Study carried out for the Ministry of Social Affairs
and Health of the Walloon Region

**New Information And Communication Technologies Serving Deaf People**

# Online Sign Language Interpretation System

Christophe Ponsard and Jean-François Molderez

Final Report
September 2003

# Table of Contents

## List of Figures

## List of Tables

# 1 Introduction

This document presents the deliverable of the project initially entitled "Study of an Online Sign Translation System", commissioned from CETIC by the Minister of Social Affairs and Health of the Walloon Region. The study was carried out during the first half of 2003.

The main objective of the project is to improve access for deaf people to administrations and public services. A more detailed presentation of the problem context and of the methodology followed is given in chapter 2. Based on a review of the key facts of the deaf world (chapter 3), this deliverable presents in detail the objectives of the project in its context (chapter 4). It then proposes three classes of alternative solutions, based on various combinations of human resources and existing or emerging technologies. These solutions are explored from the most realistic to the most forward-looking:

- in chapter 5, a system of **remote interpretation by videoconference** is studied. Interpretation is performed by a person; technology is only used to transmit the visual or auditory information. This part is grounded in reality and a feasible set of specifications is drawn up.

- in chapter 6, recent progress in **automating interpretation** is examined in terms of the many processes involved: speech recognition, gesture recognition, speech synthesis, animation of synthetic characters and machine translation into sign language. Taken as a whole, this part is forward-looking, but several partial combinations can already be envisaged and will be examined.

- In chapter 7, extensions in terms of **mobility** are explored, in particular with regard to high-speed wireless technologies and specific services for deaf people.

This document concludes by offering some directions and recommendations on the possibilities for short-term implementation, based on the needs expressed and the skills identified.

Given the field and its essentially visual nature, we have tried to illustrate our points as much as possible. The companion CD-ROM contains in particular the videos we recorded during interviews and tests, as well as collected videos illustrating some existing products and systems. A glossary of the main domain and technical terms is also available in the appendix.

References of the form [Abc12] refer to the bibliography in chapter 9.

*Note: this report is an English translation of the original French report, produced in October 2026 with AI assistance.*

# 2 Problem Context and Methodology

The general framework of this study is improving the communication possibilities of deaf people by relying on the use of new information and communication technologies (ICT).

Deaf and hard-of-hearing people face numerous communication problems every day with those around them, mainly with the hearing world. To adapt to their hearing impairment, these people of course have access to various technical solutions that make the most of any residual hearing: analogue or digital amplification, or even a cochlear implant.

These aids have their limits. To overcome them, these people and the hearing people around them have developed alternative modes of communication, the most important being lip reading and sign language. Generally, to communicate, a deaf or hard-of-hearing person will use a combination of modes that depends on the history of his or her hearing loss (in early childhood, in adolescence, in adulthood, sudden or progressive, etc.). Abilities therefore vary greatly within the deaf population and can only be characterised precisely on a case-by-case basis.

Care must therefore be taken when attempting to classify the deaf population into various "categories". Culturally, however, people with deafness can be divided into two main groups:

- people whose first language is French. These include people who became deaf at an age when they already mastered French. Besides these "late-deafened" people, this group also includes a significant percentage of people with congenital deafness who were educated using an oralist method.

- people whose first language is sign language. We will call them "signing deaf people", or simply "deaf people" when the context is sufficient.

The boundary between these groups is not clear-cut; it is rather a continuum of people who master one or the other language to varying degrees, with perfect bilinguals at one extreme and, at the other, people who have lived in isolation and have developed almost no means of communication.

Quantitatively, there are more people in the "late-deafened" category than among "signing deaf people" (about 10 times more). However, in terms of communication problems, this minority category is more affected. The main cause is their lack of command of French, which is almost a "foreign language" for these people. This makes it impossible for them to decode displayed information or to use writing in interactions.

**Without underestimating the real problems faced by other deaf and hard-of-hearing people, this study focuses mainly on solving the communication problems of "signing deaf people" in their dealings with hearing people, and more specifically in the context of administrations and public services.**

As soon as a "signing deaf person" needs to interact with hearing people outside his or her close circle, an intermediary who masters both French and sign language is often required. This is the case in particular for any administrative procedure. These intermediaries may be relatives or professional interpreters. The latter have the advantage of being perfectly bilingual and completely neutral, and of leaving the responsibility with the deaf person, who makes decisions himself or herself. Unfortunately, the current organisation of interpretation services suffers from a major problem: the chronic lack of availability of interpreters. Their small number (about twenty in French-speaking Belgium) means that meetings must sometimes be planned several weeks in advance and makes it impossible to deal with emergencies.

**The chronic lack of availability of interpreters is the main problem to be solved in this study**. It will become even more pressing since the possibility of sign language interpretation for any procedure with a service under the French Community is a **direct consequence of the recognition of sign language in French-speaking Belgium**. The solutions considered are intended for implementation within public services.

This problem was addressed using the KAOS goal-oriented analysis method [Dar93], [Lam00]. The main steps of the method are detailed below. The general structure of the report (in italics) mirrors them:

- the method is based on a preliminary study of the application domain in order to identify the actors, their characteristics and the environment in which they operate

  *In our case, this meant understanding the world of deaf people, their mode of communication, their relationships with hearing people and the basic principles of sign language. This is presented in chapter 3.*

- the method is guided by the discovery and refinement of the goals underlying the problem to be solved, which makes it possible to place the problem within a broader framework.

  *In our case, these are goals of well-being, social integration, and access to culture and information, which should in fact be guaranteed to every citizen. These goals are identified and described in the first part of chapter 4.*

- the method makes it possible to derive the solution to the problem in the form of a set of requirements forming its specifications. These are formulated not only in technical terms but also with respect to their organisational impact. Indeed, responsibilities are assigned not only for the IT to be implemented but also for the constraints imposed on the various actors who must interact.

  *In our case, one possible technical solution is to set up a sign language interpretation system by videoconference. This system would significantly improve the availability of interpreters by*

*saving their travel time, and it would also make it possible to set up an emergency interpretation system. This solution is presented in chapter 5.*

*However, even if technically possible, the system can only work if it is well suited to, and accepted by, the various people interacting through it: deaf person, interpreter, employee. We first studied the organisation of similar systems tried out abroad. We then built our own prototype and carried out a number of tests in order to validate the technical and organisational aspects and to refine the requirements.*

- the method followed is also characterised by its ability to highlight alternative solutions and to provide the relevant criteria for comparing them.

  *In our case, an alternative technical solution considered is the complete automation of interpretation, replacing the human interpreter with a chain of software components based on speech and gesture recognition, machine translation, speech synthesis and the animation of 3D virtual characters. This alternative is detailed in chapter 6. We can say straight away that, overall, it is not feasible with the current state of technology, but that automated subsystems have been identified and can be implemented in specific circumstances that we will explain.*

  *We will also explore an extension of the system suggested by the prospect of high-speed wireless communication technologies. Based on the study of these technologies and of the new services that could be offered, we will examine the prospects for improving the daily lives of deaf people.*

- finally, the method makes it easier to produce explanations and summaries, in particular by justifying solutions in terms of goals.

  *We will end the report with a summary of some implementation recommendations based on the main requirements identified and on the skills of the various actors we met.*

# 3 Domain Study

The purpose of this chapter is to highlight the salient characteristics of the world of deafness in general, and in French-speaking Belgium in particular. This description does not claim to be exhaustive, which would be impossible with the time and resources available. It aims to be representative enough to identify the needs of deaf people in the context of the problem at hand and to propose suitable solutions. For a more complete picture, we refer to other sources.[1]

We will first examine what deafness is at the medical and individual level, and the consequences it has on a person's development and place in hearing society. We will present the medical solutions to the problem without dwelling on them. Indeed, in the context of this study, the medical aspect is assumed to be adequately addressed. The deaf person is considered "as is", with the means of communication he or she has been able to develop over time. It is these means that we will seek to understand, and in particular sign language. Various organisations revolve around deaf people; we will examine their objectives and the services they offer. Many of them were consulted as stakeholders or for the validation of this study.

## *3.1 Medical Aspects of Deafness*

Physically, sound is a vibration of the air perceived by the outer ear, transmitted and amplified in the middle ear and encoded as nerve impulses on the auditory nerve in the inner ear. The information is then finally interpreted by the brain as speech in a given language, music, noise, etc. (see Figure 1)

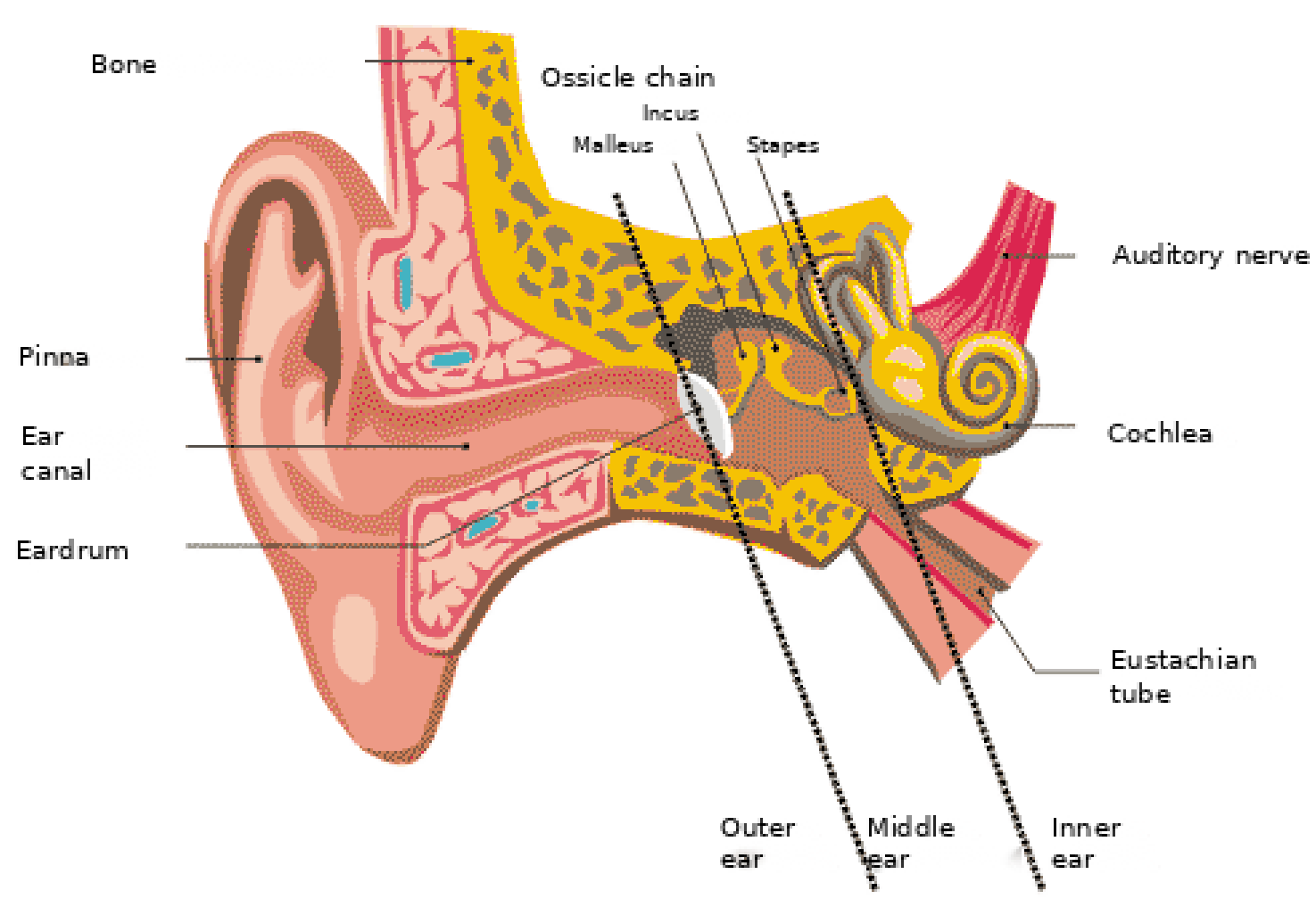

*Figure 1. Schematic cross-section of the human ear.*

The perception of sound intensity is not linear but logarithmic: that is, perception increases linearly while the intensity of the vibration increases in particular the study [Rec03], completed recently (March 2003) and conducted jointly by the UBL and Marie-Haps in the context of the recognition of sign language in the French Community exponentially. This allows the ear to handle intensities covering a very wide range, from its sensitivity threshold up to the pain threshold. Intensity is measured on this logarithmic scale, called the decibel (dB).

The ear is also characterised by a limited sensitivity that varies with frequency, between 16 and 20,000 Hz (one Hertz being one vibration per second), with maximum sensitivity between 500 and 8,000 Hz. Figure 2 shows, hatched, the perception area for speech at normal intensity, together with some typical sounds.

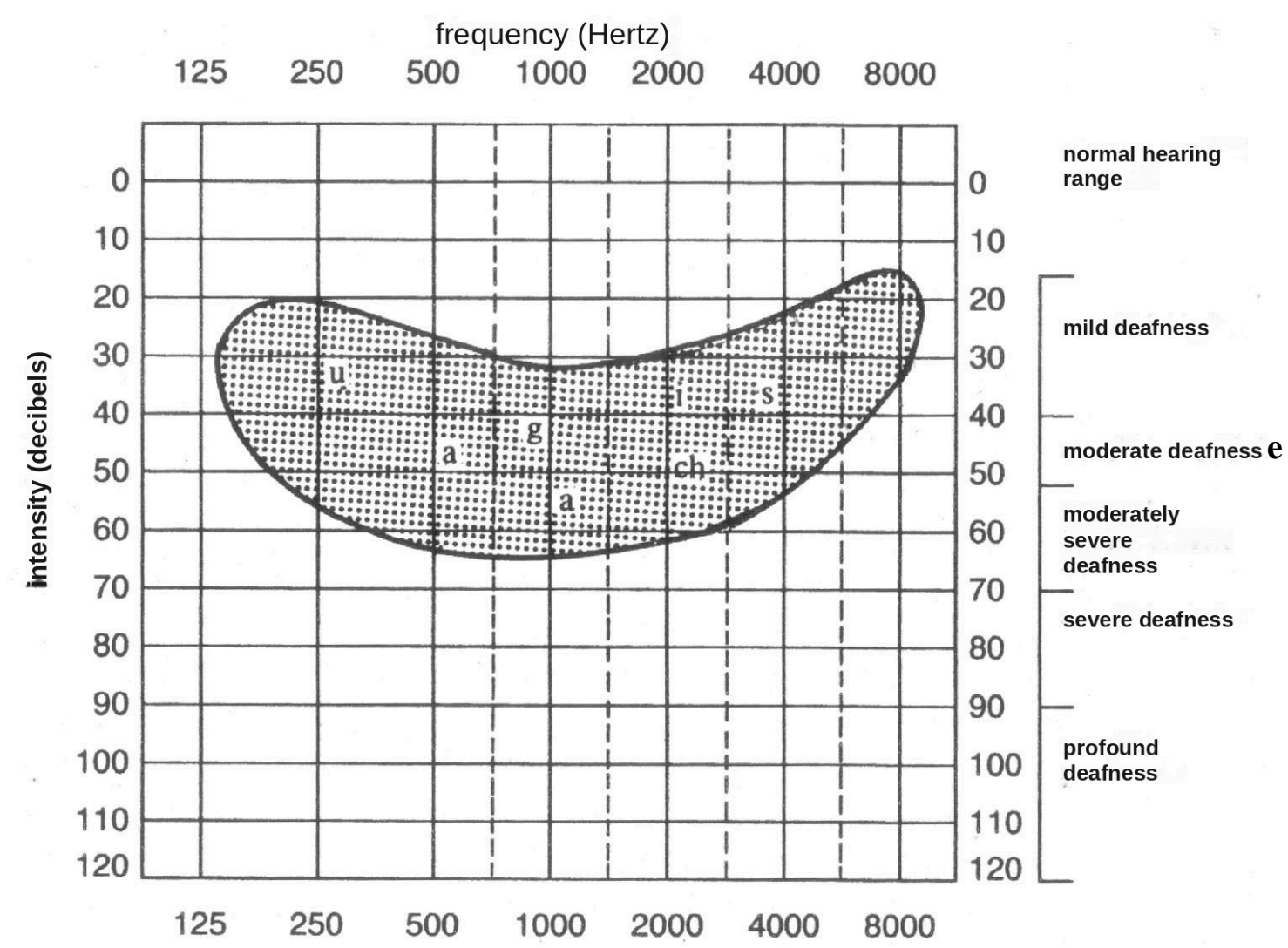

*Figure 2. Frequency/intensity diagram showing the normal communication area (hatched) and the scale of the different degrees of deafness.*

Deafness is characterised by a more or less severe degradation of hearing abilities. It is quantified by a standardised measurement of the shift in the hearing threshold. The right-hand part of Figure 2 shows a scale of the degrees of deafness, from mild deafness (from -20dB) to profound deafness (beyond -90dB). The threshold between hard of hearing and deaf is generally placed at around -70dB, beyond which even a shouting voice can no longer be perceived.

Physiologically, the problem can be located:

- either in the outer ear (wax plug) or the middle ear (amplifying chain of the ossicles, eardrum). This is called **conductive deafness**. It generally does not exceed 60dB and can usually be fully corrected.

- or in the inner ear. This is called **sensorineural deafness**. Possible causes are the loss or malfunction of the hair cells of the cochlea, or damage to the auditory nerve.

It can affect only one ear (unilateral deafness) or both (bilateral deafness).

Deafness can have a **genetic cause**. A number of hereditary prenatal syndromes have been identified (Waardenburg, Pendred, Usher, etc.). There are also degenerative post-natal forms. The genetic dimension implies that deafness can be passed on from generation to generation. This phenomenon is increased by the fact that, socially, deaf people tend to form communities and to marry each other, increasing the probability of having a deaf child. Currently, about 5% of deaf children are born to deaf parents.

There are also **non-genetic causes** of deafness. These are linked either to a problem during pregnancy (maternal illness such as rubella, Rhesus factor incompatibility), to problems during childbirth, to trauma suffered during life (head injury, exposure to noise causing irreversible damage), or simply to age (progressive "presbycusis").

Medical and technical means can reduce hearing impairment [Gou01], [Dum96]

- **surgery** on the outer, middle or even inner ear (cf. cochlear implant)

- **hearing aids (amplifying in nature):**
    - **analogue or digital**. The former already make it possible to adjust the frequency sensitivity of the amplification in order to best target the important frequencies (those of speech) and to take into account the frequency sensitivity of the impaired ear. The latter are real computers that can carry out more advanced processing on the received sound. They are much finer in terms of frequency and discriminate noise more intelligently. Several "programmes" are usually available depending on the listening environment.

    - these aids usually come in two forms: **in-the- ear** (of various sizes) **or behind-the-ear**. The latter are the most common.

    - these aids also generally offer the possibility of connecting to an external source, independent of the environment. Usually this is a special "T" mode for picking up information transmitted by magnetic induction. This system is easy to install and inexpensive. However, it is very rare in Belgium. Installing it in collective facilities (theatres, churches, public services, telephone booths, etc.) is recommended.

- **cochlear implant**: this is a genuine bionic ear whose core is a set of electrodes positioned in the cochlea. Each electrode electrically stimulates the auditory nerve according to the frequencies of the processed auditory signal. The auditory nerve and the cochlea must be intact. These electrodes work according to a strategy imposed by a processor and a decoder during fitting sessions. These two elements (processor and decoder) are installed under the skin (between the scalp and the skull) above the ear. The link between the processor and the outside world is provided by an antenna fixed close to the processor. It transmits the auditory information picked up and processed by a behind-the-ear unit according to one or more personal profiles developed during the adaptation phase. This type of aid is very expensive and of course requires surgery. Until recently, it was only considered for one ear and for severely deaf people who could no longer benefit from amplifying aids. For some years now, broader categories of deaf people and bilateral implants have been considered. On the latter point, an impact assessment study is under way in Belgium.

A fundamental point is that **no technical aid can restore intact hearing.** In the case of the cochlear implant, the adaptation phase takes several years and continues throughout life. It is therefore essential to examine the other aids to be implemented, mainly in terms of communication.

### *3.2 Consequences of Deafness*

Beyond these medical aspects, deafness is mainly characterised by the **communication problems and the complications they cause** in daily activities and personal development. These are fundamentally different depending on whether deafness occurs:

- in the **prelingual** phase, i.e. at birth or before the age of 3. This period is crucial for cognitive and language development, which involves communication with the mother, babbling, acquiring the rhythm of the language, repeating words, combining them, etc. This is not only a purely oral matter but also involves non-verbal communication (attitudes, facial expressions, clapping, etc.), which babies in fact use much earlier than their voice.

- in the **postlingual** phase, i.e. after language acquisition. There is obviously no clear boundary: while a minimum of language is acquired by around age 3, some sounds are only pronounced correctly at around age 5 and grammatical acquisition continues between 5 and 12. It is therefore only really in adolescence and adulthood that we can speak of a "late-deafened" person, and this term must always be related to the linguistic skills acquired by the individual concerned.

Unfortunately, there are no reliable official figures on the deaf population in Belgium. Figures quoted in the press are often highly variable and sometimes contradictory. The figures given here have been cross-checked from several sources (CMAP, Espace Sourd, SISW, study [Rec03]) and compared with those of neighbouring countries.

It is thus estimated that between 80 and 100 deaf children (total or profound deafness) are born in Belgium each year, which, for about 114,000 births per year, gives a ratio of around 1‰. A further 1‰ is detected during the first year. The proportion of deaf young adults is estimated at 3‰, which would represent around 2,000 children in the French Community of Belgium. This figure is lower than the frequently cited figure of 9,000 children.

The number of adults seriously limited by deafness in Belgium is around 100,000. In the French Community, it is around 30,000. Of these, according to [Rec03], it can reasonably be estimated that about 25,000 are concerned by the use of sign language.

The figures for the hard-of-hearing population are much higher: it is estimated to concern 8% of the population, the main group affected being the elderly.

By comparison, in France, more than 4 million people (about 6% of the French population) have some degree of hearing impairment: 2 million people with mild hearing loss, 1.8 million with moderate deafness and 200,000 profoundly deaf people (i.e. 4‰). Worldwide statistics report 4‰ deaf people and 4% hard-of-hearing people.

### 3.2.1 Consequences of Prelingual Deafness

These can be dramatic: language impairment and delayed cognitive development, withdrawal, emotional disorders, leading in the longer term to a real social and cultural handicap (illiterate individuals who are not integrated and not autonomous).

These people will never have a real experience of the world of sounds. For them, the natural language is sign language, at least if they are given the opportunity to acquire it. As we will see in section 3.3, sign language is a genuine language, entirely based on the visual modality and with its own syntax and grammar. This language is passed on naturally within deaf families, in a way quite similar to an oral language.

Most deaf children (about 95%) are born into hearing families who are unfortunately not equipped to pass on this language. Moreover, in order to integrate, these children will one day have to communicate with the hearing world and be able, to some extent, to master written and spoken French (through speaking and lip reading). Historically (cf. section 3.3.1), this argument was pushed to its extreme and underlies the "oralist" educational movement, which banned sign language for a century and led to its fragmentation and impoverishment. We have only gradually moved away from this way of thinking over the last twenty years or so. Many deaf adults still bear the marks of this education, which gave them no real language: they are not very good signers and their knowledge of French often remains basic. A large proportion are estimated to be functionally illiterate, that is, unable to grasp the meaning of written text.

**For deaf children born into deaf families, sign language is acquired as a mother tongue in a way quite similar to an oral language and allows harmonious development. Researchers [Rec03] have also shown that early learning of sign language by the child supports later intellectual learning, in particular mastery of French. From this perspective, French/sign language bilingualism appears essential.**

This learning can only begin if systematic and early screening is developed, and it takes place alongside the implementation of a suitable medical solution, whether amplifying or cochlear. This will have been decided with full knowledge of the facts by the parents. Cultural aspects may come into play here, particularly for deaf children born to deaf parents.

All this of course supports the recognition of signing deaf people as a linguistic minority. Sign language has already been recognised in French-speaking Belgium as a language of instruction since 1998. A process of institutional **recognition of sign language** is currently under way and will have practical implications, in particular regarding:

- early access to sign language for deaf children. We have just mentioned its importance, but we will not examine this educational aspect further as it is not directly related to our study. It is covered in more detail in studies such as [FFSB97] and [Rec03].
- the use of interpretation services. It is mainly this aspect that will concern us.

### 3.2.2 Consequences for Late-Deafened People

These are also significant. The sudden or gradual onset of the impairment can be disabling: difficulty communicating with colleagues, listening to the radio, using the telephone, ensuring one's own safety or that of others. It can lead to job loss and to depression. In practice, a functional definition of the boundary between hard of hearing and deaf is the loss of the ability to use the telephone.

These people are, however, still able to express themselves and usually gradually develop very good **lip-reading** skills, more easily than deaf people, because this ability was latent and already used unconsciously. Lip-reading training and improvement courses are also organised by the Ligue Belge de la Surdité.

Lip reading leaves a large number of ambiguities that must be resolved from the context. It also depends on the speaker, who must articulate properly and must not have any visual obstruction on the lips. Moreover, it is not always possible to maintain this visual contact, for example in group discussions or, during a class, when the teacher turns around to write on the board.

Sometimes the hard-of-hearing person uses the services of an oral interpreter, who rephrases the speech clearly, at a steady pace and at an appropriate volume.

### 3.2.3 Deafness and Disability

A delicate issue is whether deafness should be regarded as a disability. Although it is clearly disabling, deafness is often rejected by those affected and poorly understood by the public:

- deafness is not visible. It can easily be concealed, particularly by late-deafened people who find it hard to accept that their world is suddenly or gradually collapsing. They tend to work around or deny their disability, behaving as if they could still hear

- signing deaf people see themselves as a community with its own language and culture
- elderly people are sometimes deaf without being aware of it.

Deaf people do not readily accept the "wheelchair" logo used to indicate facilities for disabled people. In this respect, a finer classification, the "passe-partout" index, is emerging in Belgium and seems to interest quite a few other countries (see the website: http://www.ipp-online.org). This presentation has the merit of not lumping disabilities together (they are classified into 6 categories represented in Figure 3) and of presenting a progressive gradation of aids for each of them.

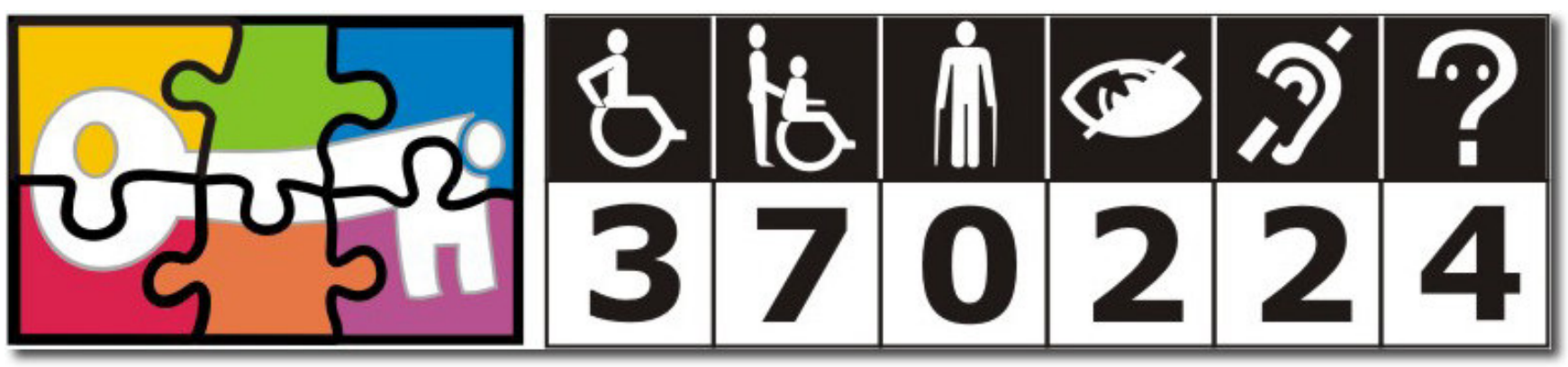

*Figure 3. "Passe-partout" index. From left to right: wheelchair user alone, accompanied wheelchair user, pedestrian with aid, blind/visually impaired, deaf/hard of hearing, frail person*

For deafness, the gradation of possible aids is as follows:

0. Nothing is provided!
1. Main information is written
2. **Main information is written and shown as diagrams (or illustrated)**
3. Information given by sound can be amplified
4. The above criteria, plus specific safety devices (e.g. flashing lights together with an alarm)
5. Possibility of being informed (e.g. subtitling of audiovisual information)
6. Presence, on request, of a person with basic knowledge of sign language
7. Permanent presence of a person with basic knowledge of sign language
8. Presence, on request, of a person with in-depth knowledge of sign language
9. Permanent presence of a person with in-depth knowledge of sign language.

The recommended minimum level is level 2, given the difficulties that signing deaf people may have with French.

In terms of general policy on assistance for disabled people, Belgium has a tradition of individual rather than collective aid. For a blind person, for example, the preference is to help finance a braille display that he or she will carry around, rather than to equip libraries with them. The individual approach makes it possible to tailor the aid better but does not always ensure optimal accessibility of public places for disabled people.

## 3.3 Sign Language (SL)

As mentioned above, sign language is a natural language used by many deaf people in their daily lives, but also by hearing people who regularly communicate with deaf people. It is estimated that for every signing deaf person, on average 2 hearing people around them make the effort to learn this language.

It is a fully-fledged language with its own vocabulary, syntax and semantics. As with oral languages, it also has regional variants (dialects) and different registers (everyday, formal, jargon, etc.).

Contrary to some beliefs, it is not universal: each country has its own sign language. However, some sign languages show similarities linked to the geographical and historical circumstances of their development: French Belgian Sign Language (LSFB[2]) and French Sign Language (LSF) are fairly close, and are related to American Sign Language (ASL) because the latter was "imported" from France. Deaf people of different nationalities generally adapt quickly to each other's sign languages: for everyday conversation, a few days are enough. Note that there is also a form of international sign language (ISL, previously called "Gestuno"), which is the "Esperanto" of sign languages.

**Sign language is a very lively and rapidly evolving language**: it is being reborn after a century of oppression (see below) that left it wounded. It needs to be reunified and enriched.

### 3.3.1 Brief History

This section gives a brief overview of the main milestones of sign language and is taken from [Gerday]. The modern history of sign language mainly took place in Belgium.

The idea that gestures could be used to express thought was put forward by Abbé Michel de L'Epée (18th century). The system he developed was, however, far from a natural sign language. He fought to impose the idea that deaf people are normal people. Following in his footsteps, Bébian proposed a truly bilingual education. The recognition of sign language as a language of instruction led to the recognition of the deaf community.

The mid-19th century saw a tremendous development of the "deaf movement" in France. Many associations were created. Ferdinand BERTHIER, a deaf teacher, was the "mobiliser" of the deaf community.

However, in the background, a dispute between schools of thought continued between, on one side, educators who used gestural teaching and, on the other, those who excluded it in order to focus all their efforts on teaching speech. Several reasons led the oralists to prevail: the majority of hearing teachers, Jules Ferry's compulsory education which called for the standardisation of teaching

---

the acronym LSBF, for Langue des Signes de Belgique Francophone, is also sometimes used. Here we use the same convention as [Rec03]: LSFB, which highlights the final B of Belgium.

methods and the suppression of minority languages, and finally technical progress, which, with the first hearing aids, gave deaf people the hope of becoming hearing again. The use of signs was gradually abolished. The Milan congress in 1878 marked the beginning of a ban on sign language that lasted almost a century. It was sustained in particular by the illusion that the emergence of hearing aids, their miniaturisation, surgery and speech therapy could turn all deaf people into hearing people. During this period, deaf culture was denied and sign language became impoverished and fragmented.

A revival appeared after May 1968, which awakened sensitivity to cultural diversity and the right to be different, leading to a collective awareness of sign language as a source and instrument of culture [Del96].

Between 1980 and 2000, more and more methods were developed and structured: unified sign language (by the CFLS), AKA, LPC, signed French, lip reading. Sign language became a language of communication and sometimes of instruction. Means of communication developed: Minitel, subtitling (1980s), Internet and GSM (1990s). Recently, more and more countries have recognised sign language and enshrined it in their constitution.

### 3.3.2 Syntactic and Grammatical Components of Sign Language

Like any language, sign language has its vocabulary, syntax and grammar. Vocabulary is the easiest to describe; it is notably the subject of the CFLS lexicons for unified LSFB. At present, the grammar is still only partially described. Here we give a few basic elements, mainly based on [Ron97].

Sign language involves not only the movement of the arms and hands but also the torso, the shoulders, the direction of gaze and facial expression. More precisely, a sign can be described by 5 main spatial parameters:

- the **handshape** corresponds to the shape of the hand; the handshape reproduces part of the shape of the object or action.
- the **orientation** indicates the directions in which the hand and the palm point. It is used to conjugate certain verbs or to specify the orientation of objects
- the **location** refers to where the signs are made. The location of the hand gives the place where the action takes place, or a generic place. It may be on the torso and head or in the surrounding space within reach of the arms.
- the **movement** covers the movements of the arms, wrists, hands and fingers. It represents an action performed by or on the object, and its dynamics can indicate verb moods.
- the **facial expression** conveys the mode of discourse (interrogative, negative, etc.) or adverbials of manner (joy, anger, etc.)

Unlike French, which expresses things sequentially, the elements that make up a sign are produced in parallel. Thus, although producing an elementary sign takes longer than pronouncing the "corresponding" word, sign language has a rate similar to that of spoken language.

Some sentences can even be reduced to a single sign, as illustrated in Figure 4.

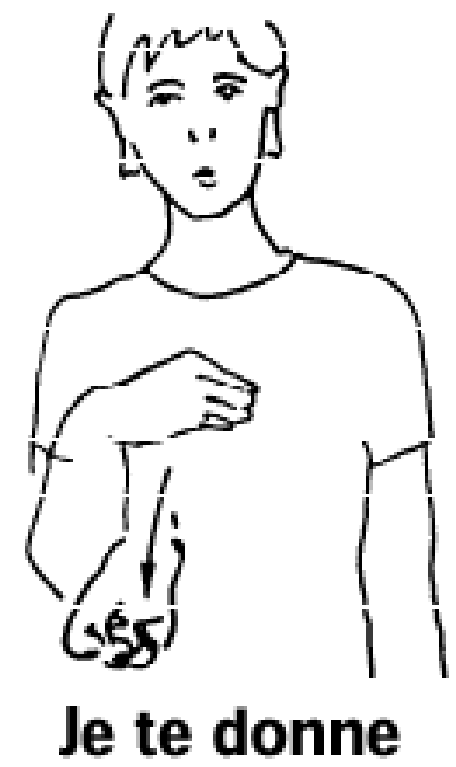

*Figure 4. Sign for "I give you".*

Sign language has many grammatical features:

- verb tenses do not exist: only the moment at which the action takes place is specified. To do so, it is placed on a time line perpendicular to the body: the future is in front and the past behind (in most sign languages)

- classifiers, which are generic signs representing classes of objects of similar size, shape or thickness, and which are found in some oral languages such as Chinese
- the placement of characters and things in the signer's space, to which he or she can refer, for example through the direction of gaze.
- the order in which signs are produced is less important than their visual arrangement in space. Generally, adverbials of time and place come first and the verb at the end.

### 3.3.3 Sign Language vs Other Forms of Gestural Communication

Sign language should not be confused with other forms of gestural communication.

**A. Fingerspelling**

This is a way of spelling out the letters of the alphabet using simple, static handshapes.

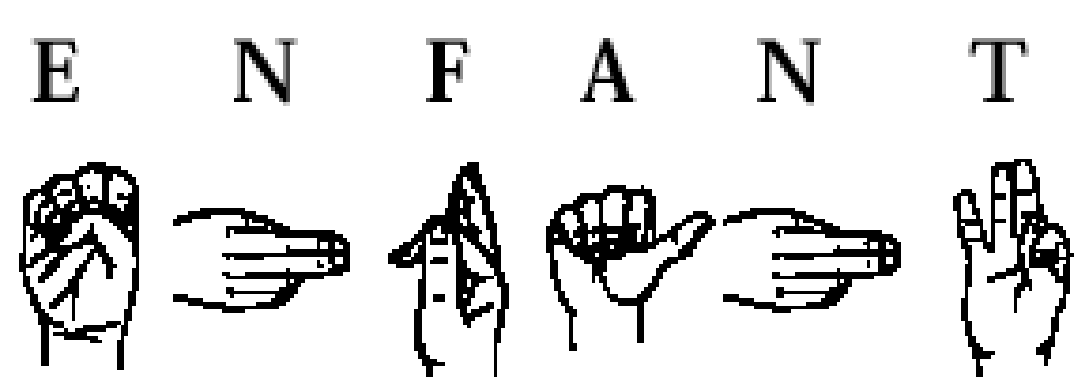

*Figure 5. Fingerspelling (source: [Bon98]).*

**B. Signed oral language**

This is a purely syntactic use of signs in the context of the grammar of the oral language (here French, but the same principle applies to other languages). It often serves as a bridge language between signing deaf people and hearing people who do not master sign language.

Signed French tries to map signs onto French words and forces sign language to follow French word order. The speaker also generally speaks at the same time (bimodal communication). Nuances, the use of space and facial expressions are generally absent, and to compensate, words are spelled out using fingerspelling signs. This makes sentences very long, visually overloaded and tiring (see Figure 6).

Other forms of communication support, such as lip reading or writing, are also usually used in addition.

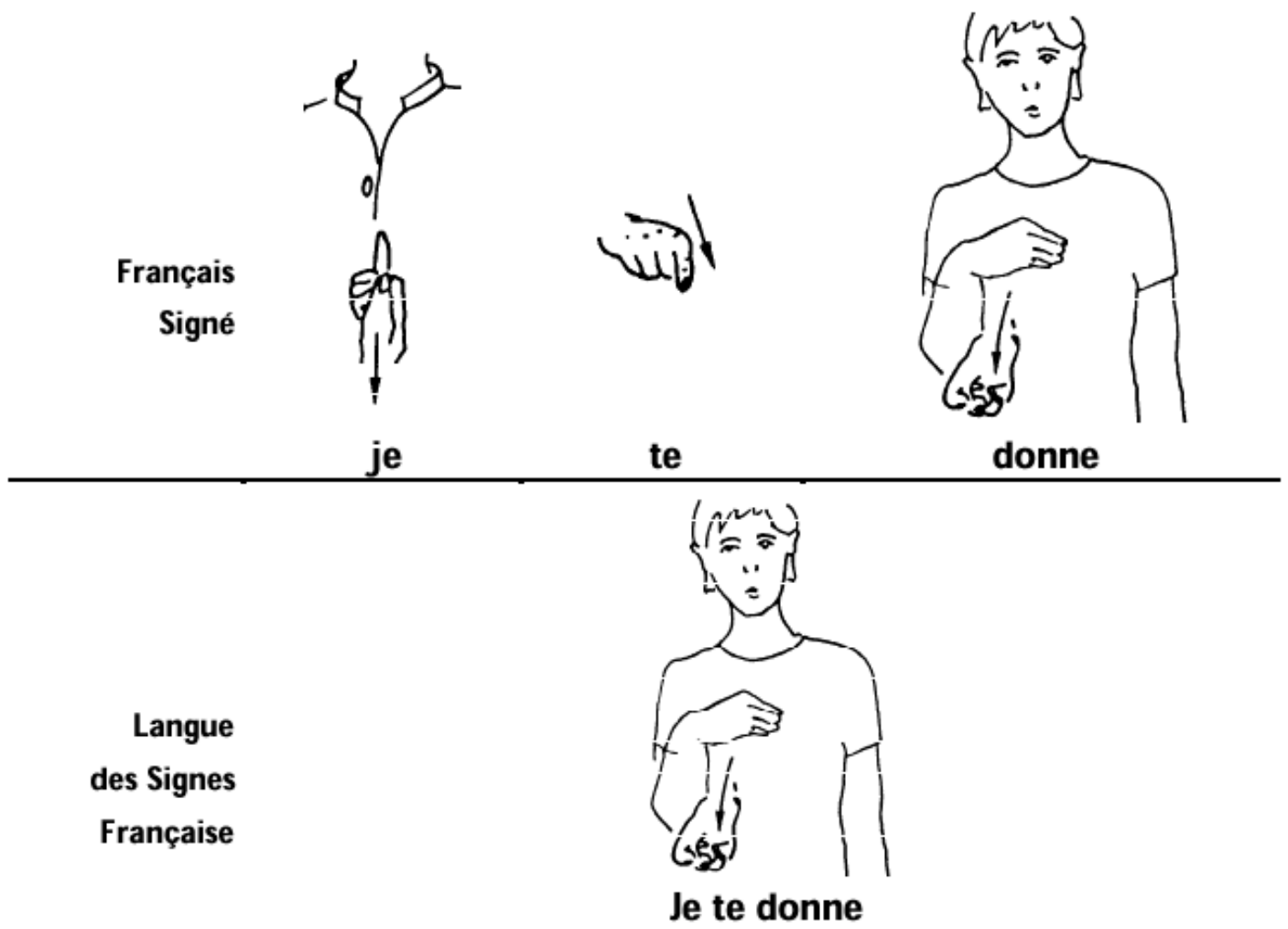

*Figure 6. Distinction between sign language (LSF) and signed language (SF) (source: [Bon98]).*

**C. Co-verbal gestures**

These complement the oral channel with visual information of a phonetic nature and carry no meaning of their own. Examples are **LPC** or **AKA**, which we will not detail here. Unlike these techniques, sign language makes it possible to communicate without the help of another communication modality.

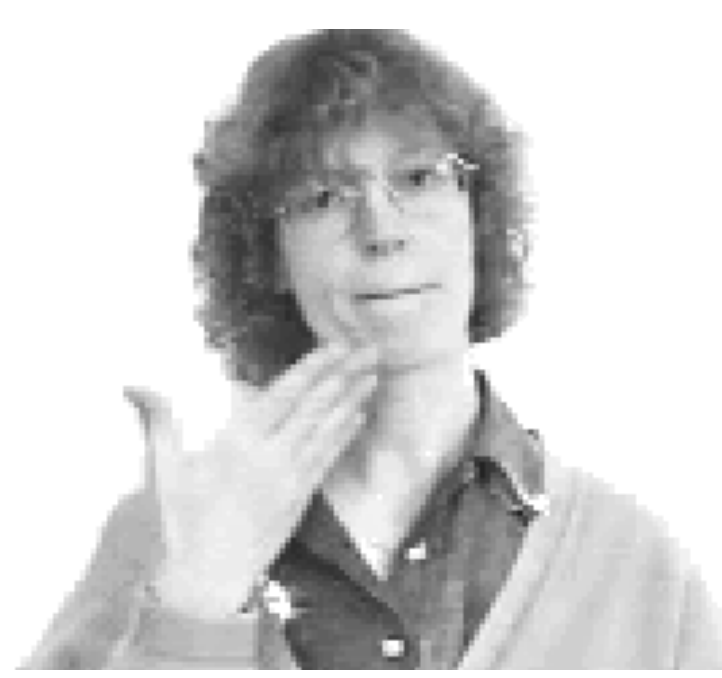

*Figure 7. LPC code for the M of "Main" (hand) (source: [Los00a])*

### 3.3.4 Means of Disseminating Sign Language

**A. Broadcast media**

Sign language is broadcast on television in two forms:

- **Sign language interpretation** of a normal programme, with the interpreter shown in an inset (see Figure 8). For the RTBF, this mainly concerns the 7.30 pm news and the children's programme "les Niouzz". Note that this programme, thanks to its simplified approach to the news, attracts a wider audience than deaf children.
- **Programmes specifically for deaf people**, made for them (and sometimes by them). For the RTBF, this was the case of the programme "Tu vois ce que je veux dire", which was recently discontinued.

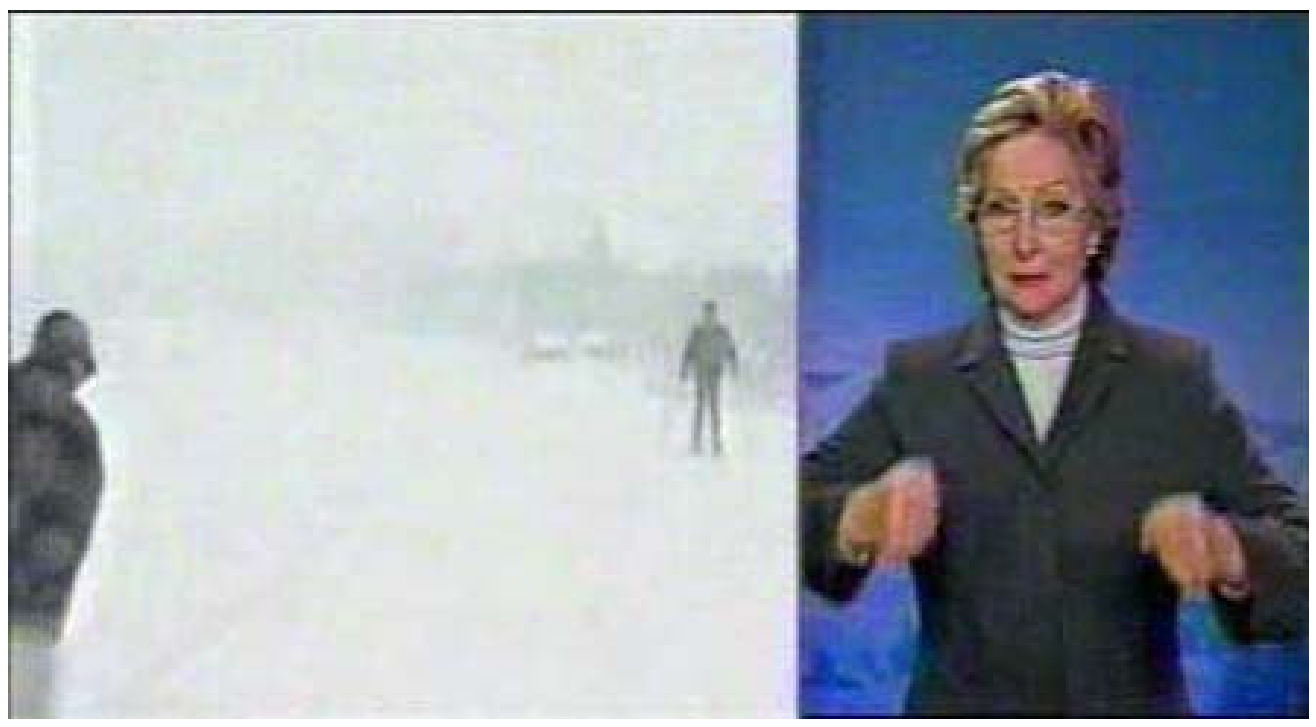

*Figure 8. Sign language interpretation of the RTBF TV news.*

Note also the **availability of subtitling** for some programmes (with, here, the obstacle of the written oral language). It is broadcast via Teletext on a special page (often page 777 or 888) and uses specific conventions: colours for the speakers, the narrator and background noises. Some subtitles are positioned spatially close to the speakers. The number of subtitled programmes varies depending on the channels and their financial resources. For the RTBF, these are mainly documentaries (such as "le jardin extraordinaire") and information programmes ("Contact": highway code, "Pulsation": medical magazine, etc.). The RTBF news is signed but not subtitled. Regarding subtitling, note finally a technical problem linked to the disappearance in Belgium of video recorders capable of recording them.

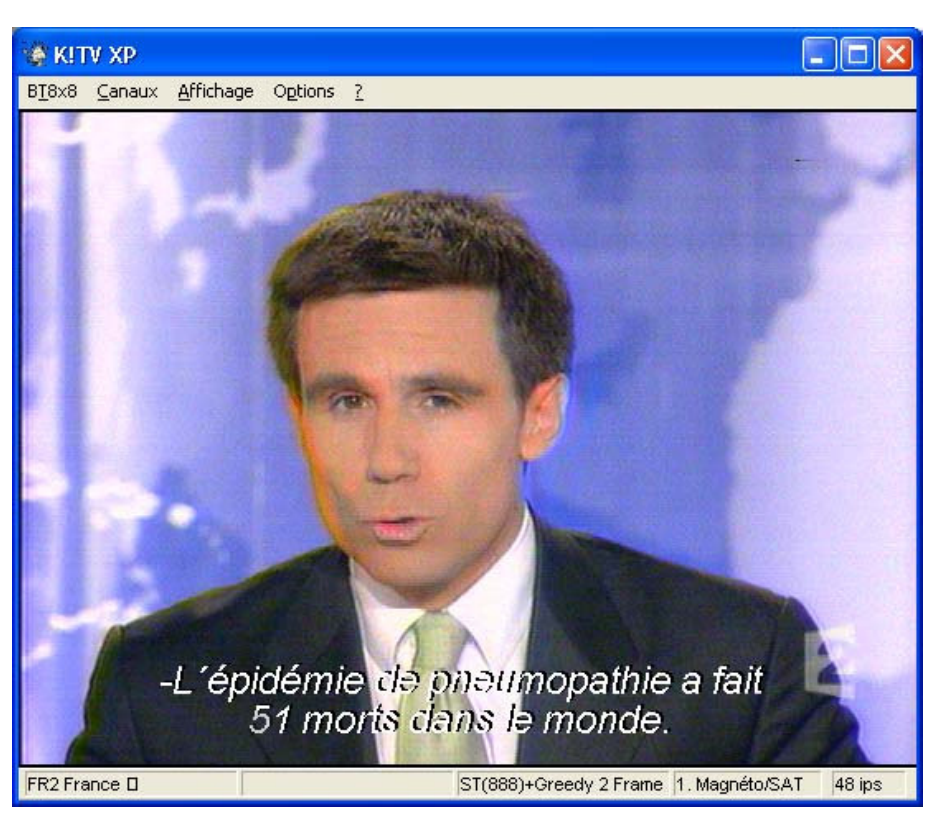

*Figure 9. Subtitling of the France 2 news.*

**B. Physical media**

Languages generally have oral and written forms and can be disseminated through various media: written, audio and/or video. Of these, audio is obviously useless for deaf people.

Unlike conventional languages, sign language has no written form equivalent to the signed language and in general use. Various written options exist but have specific uses:

- the Hamburg notation (HamNoSys) is used by linguists. It is reminiscent of hieroglyphics (see Figure 10)
- descriptions of signs using drawings with arrows (see Figure 4) are used by sign language dictionaries and are not always easy to decode.
- notations based on French but supplemented with annotations are used as a means of transition towards French. They are being investigated by literacy centres such as Alpha Signe.

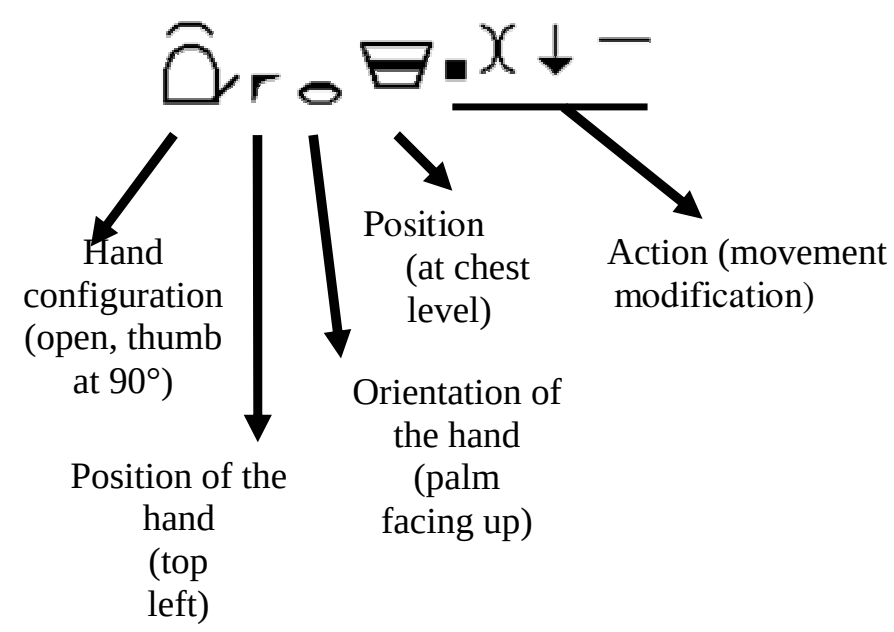

*Figure 10. HamNoSys notation for "tired"*

In fact, before the advent of **VHS video**, sign language could only be passed on from person to person. Video makes it possible to watch stories but has certain drawbacks, linked to the single camera angle and to tape storage, which does not allow easy access (in the case of a lexicon, for example).

**DVD**, a more recent technology, allows easier navigation and also supports several viewing angles. This possibility is little known because it is rarely used, owing to the difficulty and cost of production. DVDs also offer subtitles that are sometimes specific to deaf people (but of course with the obstacle of the written language for signing deaf people).

The **multimedia CD-ROM**, another recent technology, makes it possible to combine various complementary representations of the same information: image, textual description and video (see Figure 11). This is useful for learning. It also allows easy, intuitive navigation through multiple indexes (according to the parameters of sign language, or in alphabetical order for French). However, storage is limited to a few hundred signs per disc, and only with small videos of fairly average quality (otherwise the number drops further: quality and size require a lot of space). The **DVD-ROM** pushes this limit back by an order of magnitude and can offer several thousand signs, but remains insufficient for complete dictionaries (several tens of thousands of signs). In addition, this is only digitised video, which also suffers from the problems of viewing from a fixed angle and at a fixed speed.

A promising technology that will be considered in this study is the use of **3D synthetic characters** (called "avatars"). They advantageously replace video by offering a much higher compression rate: the information stored is a description of the signs, which can be almost linguistic in nature. Its volume is comparable to that of the equivalent written text. In addition, these characters can be viewed from various angles and at variable speed, which is very useful for learning. It is also possible to adapt the synthetic character to the audience (child, adult, etc.) and even to give it the appearance of a real person. The main restriction concerns the ability to render all the nuances of sign language, in particular facial expressions and certain specific signs. The use of this technology will be considered in detail in chapter 6 of this report.

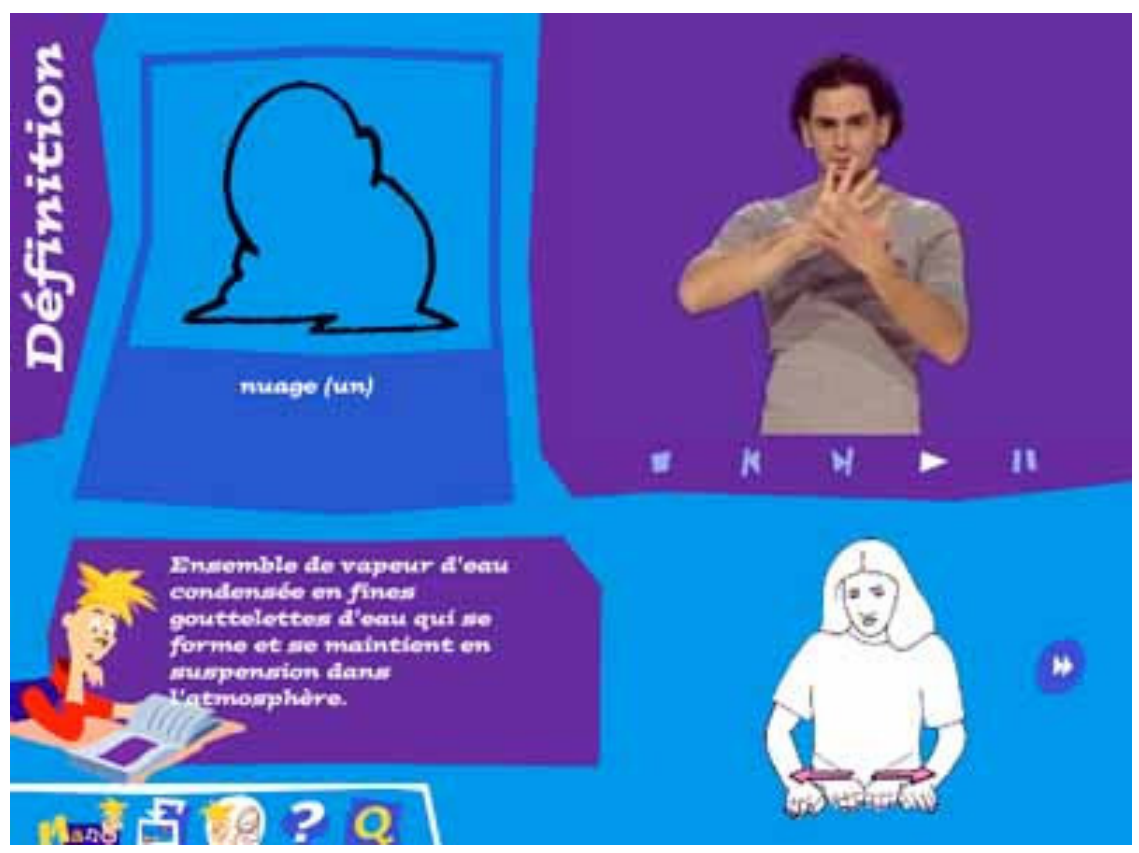

*Figure 11. Explanation of "Cloud" in the multimedia CD-ROM "Les signes de Mano"*

The main qualities of the various media considered are summarised in the table below.

| **Medium** | **Book** | **VHS** | **CD-ROM** | **DVD** | **[Avatar]** |
|---|---|---|---|---|---|
| Storage volume | - (fixed) | - | - | + | ++ |
| Ease of production | - | - | - | + | ++ |
| Flexibility of use | - | - | + | + | + |
| Rendering of movement | - | + | + | + | + |
| Navigability | + | - | ++ | ++ | ++ |
| Multimedia content | - | - | ++ | ++ | ++ |
| Expressiveness | - | + | + | + | -/+ |
| Uncluttered look | + | - | - | - | + |

*Table 1. Comparison of the qualities of different media (key: - = poor, + = good, ++ = very good)*

### 3.3.5 Means of Communication in Sign Language

**A. Communication with a hearing person**

Since the aim is to get a message across, every means is used to achieve it. Several situations can arise:

- the hearing person has learnt sign language or has partial knowledge of it. This is generally the case for most hearing people around the deaf person. Communication can take place in sign language or, more likely, in signed French.
- the hearing person does not master sign language, but the deaf person can manage with lip reading and written French.
- the hearing person and the deaf person have no common language. It is then advisable to involve a third person who knows both languages. This will be either a family member or a professional interpreter. The latter offers more guarantees in terms of neutrality, accuracy and confidentiality. By not getting involved in the deaf person's decisions, she also leaves more responsibility with the deaf person.
- if, in addition, the deaf person needs support for the procedures in which he or she is involved with hearing people, the help of a social worker who masters sign language will be considered.

As for **sign language interpreters**, only about twenty are active in French-speaking Belgium, either self-employed or working for one of the interpretation services (SISW, SISB) or social services (L'Epée, InfoSourd, SARE). The profession is mainly made up of women. Sign language interpretation, like all interpretation, is a complex process involving many cognitive tasks in parallel (listening, understanding, searching for equivalents, reformulating, monitoring, etc.; see [ElOu03]). **The interpretation process is mentally exhausting and its quality degrades on average after about twenty minutes**. For long, uninterrupted interactions, several interpreters taking turns must therefore be planned.

**B. Remote communication**

At present, there is no operational equivalent of the telephone for deaf people: **they cannot communicate remotely in sign language**. This requires the transmission of images, and videoconferencing systems are not yet widely operational. We will consider them in this study.

Deaf people have to use written French to communicate remotely. The various means available to them are:

- **fax**, very widespread and easy to use, which also allows drawings.
- the **Minitel**, imported from France, was long one of the only means of remote interactive communication. A directory service was set up by TeleContact. The development of the Internet has made it obsolete.
- **Internet**-based options: e-mail, chat and instant messaging. However, this requires overcoming the obstacle of using a computer and the investment it entails.

- **SMS**, which has the advantage of mobility and does not require very formal written language. Special rates have been negotiated for deaf people.

No **relay service is available in French-speaking Belgium** (see Figure 12). Such services operate in the United Kingdom, the Scandinavian countries and the USA. This service allows a deaf person to contact an intermediary, who relays the call to a telephone subscriber who does not have the deaf person's means (fax, Minitel, etc.). Permanent services can be set up, in particular for access to emergency services. Two recent developments make this gap less critical than before:

- the GSM boom, which, thanks to SMS, provides a text-based means of communication available to everyone (deaf and hearing) everywhere. A deaf person can therefore use a hearing person close to them as a relay.
- the opening of emergency services to other means of communication: 100 and 101 can be reached by simple fax. It is also possible to contact Touring Secours by SMS.

The multiplication and diversity of means of communication do not only have positive aspects. Organising widely accessible services also becomes more complex. The volatility of Internet communication software can also discourage people (deaf and hearing alike) from investing in it.

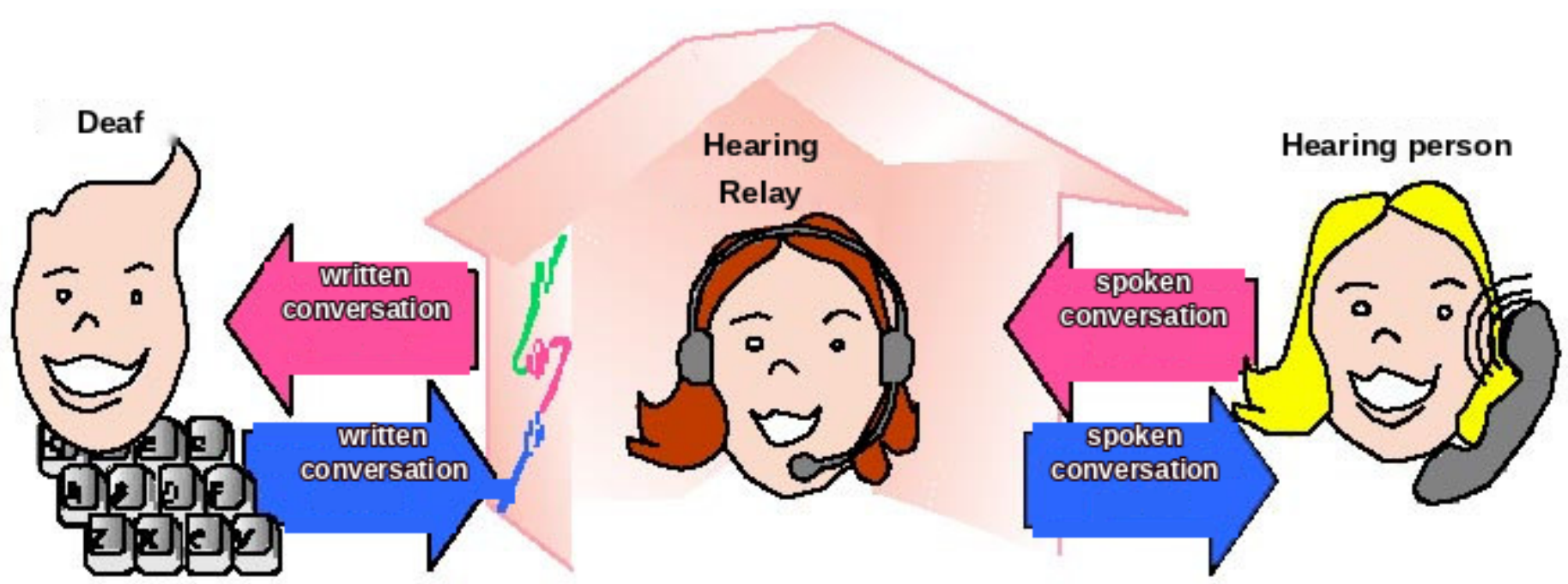

*Figure 12. Text relay system (source [MCC99])*

## *3.4 Organisations Related to Deafness*

### 3.4.1 Overview of the Deaf World in French-speaking Belgium

The deaf world is mainly a world of associations formed by the deaf families of a region. These associations are relatively independent of each other. There are about 80 of them in French-speaking Belgium.

These associations organise various recreational, sporting or cultural activities that allow deaf people to meet. They also typically handle educational activities in sign language for children or for parents of deaf children. Within some dynamic associations, more ambitious projects can emerge, such as the design and publication of sign language material (books, videos, CD-ROMs).

This mosaic of associations makes it difficult to join forces to express a common vision of the problems to be solved and to develop global projects. This coordination is provided by the FFSB (Fédération Francophone des Sourds de Belgique), which also manages links with other bodies: the interpretation services of Wallonia (SISW) and Brussels (SISB), social and job-seeking services (L'Epée, InfoSourd, SARE, etc.), and contacts at national level (the Flemish federation: Fevlado) or international level (European: EUD, or worldwide: WFD).

In education, an aspect less relevant to this study, there are about ten establishments such as "La petite école" in Montegnée (CMAP), the Ecole Intégrée (comprendre et parler) and "Ecole et surdité". They offer adapted teaching with personalised follow-up (audiologists, speech therapists, etc.) and use the various means of communication mentioned above in various ways (internally and/or through integration into mainstream education). For a more detailed overview, in particular regarding the use of sign language, see [FFSB97] and [Rec03].

In research related to sign language, we have researchers in sign language linguistics (Institut Marie Haps, the psychology faculty of the ULB, the faculty of philosophy and letters of the ULg), in technical aids (CRETH, at the psychology faculty in Namur) and in the use of computing (faculty of applied sciences of the UCL).

### 3.4.2 Stakeholders Consulted

This section details which of the above bodies were consulted and involved in the process of preparing this study. We also give a brief overview of their main skills.

| **Organisation** | **Contact persons** | **Skills** |
|---|---|---|
| ABILS<br>Association Belge des Interprètes en Langue des Signes | Annie Gillet<br>Anne-Marie Wauquiere | Representative organisation of interpreters in French-speaking Belgium. SL interpretation services. |
| Alpha signe | Jeannine Pawlack | General training courses for deaf people |
| APEDAF<br>Association de Parents d’Enfants Déficients Auditifs | Véronique Gernay | Information and support for parents, advice on technical aids. |
| AWIPH<br>Agence Wallonne pour l’Intégration des Personnes Handicapées | | Policy for disabled people<br>Management of individual aids<br>Approval/subsidising of services<br>Information, training, awareness-raising<br>Documentation centre |
| CMAP<br>Medical Centre of Montegnée | Sonia Demanez | Audiophonology<br>Integrated and special education (including sign language) |
| CFLS<br>Centre Francophone de la Langue des Signes | Caroline Ahn<br>Max Rasquinet | Unification of LSFB<br>Publication of lexicons and information brochures |
| CRETH<br>Faculté de psychologie<br>FUNDP | Pr. Michel Mercier<br>Geneviève Bazier<br>Vincent Collin | Technical aids and psychology |
| Espace Sourds | Caroline Heretinsky | Production of SL learning materials.<br>Basic SL courses.<br>Activities for deaf people |
| FEVLADO<br>Federatie voor VLAamse Doven | André Lathouwers | Flemish equivalent of the FFSB<br>Coordination, projects, technical aids |
| FFSB<br>Fédération francophone des Sourds de Belgique | Martine Fraiture<br>Jean-Louis Brassiène | Coordination, projects, technical aids |
| I3D<br>Université de Lilles | Jean-Marc Toulotte<br>Olivier Losson | Use of NICTs |
| InfoSourd | Yves Van Minnenbruggen | Social support and interpretation (Brussels) |
| INGI/TELE<br>FSA/UCL | Yves Deville<br>Benoît Macq<br>Xavier Wielemans | Use of NICTs |
| L'Epée | Sylviane Sciorre | Social support and interpretation (Liège, Namur, Eupen) |
| Marie Haps | Damien Huvelle<br>Rachid El Ouaghli | Linguistic studies on SL.<br>SL interpreting courses. |
| SARE | Pascale Van Der Belen | Job-Seeking Assistance Service |
| Signe3D | Pierre Beghin | Use of NICTs |
| Télécontact | Marc De Muynck | Directory of deaf people and their friends (minitel, etc.) |

*Table 2. Stakeholders consulted.*

### 3.5 Scope of the Study

This description of the domain shows that the population with hearing impairments of varying degrees is large. It is also very heterogeneous in terms of the impact of hearing loss on the lives and abilities of deaf people.

In order to limit the scope of the study to the relevant aspects, we recall and clarify the framework within which we will work:

- **the context is improving accessibility of public services. If the solution can have applications in other contexts (such as education), these will be mentioned.**
- **the study focuses mainly on the minority of signing deaf people. We will also mention whether some aspects of these solutions can benefit late-deafened people or even other classes of people.**
- **the study focuses on an adult audience and does not consider aspects specific to education.**
- **medical aspects and hearing aids are assumed to be dealt with.**
- **the case of people who master neither French nor sign language is not considered. They require social services and specific communication techniques.**
- **the case of people with other disabilities affecting their ability to communicate is not considered.**

# 4 Requirements Analysis

In this chapter, we set out the main goals that a communication support system for deaf people must achieve. The KAOS method for analysing and structuring needs will be used (see [Dar93][Lam00]). It was already outlined in chapter 2 and will be progressively detailed throughout this chapter.

Some goals are obvious, since their scope is universal: **everyone's well-being, autonomy, access to culture and education**. In order to refine these very general goals and to examine whether and how they are actually achieved for deaf people, we will rely on the domain study carried out in the previous chapter.

We will then arrive at a **tree structure of goals**. A very general example of a goal tree is shown below (Figure 13). General goals are found at the root, technical means at the leaves, and the various people and languages used to interact are represented in the environment outside the tree. Moving through the tree corresponds to asking "why" when moving towards the root and "how" when moving towards the leaves. In the following pages, we will present this diagram more symbolically.

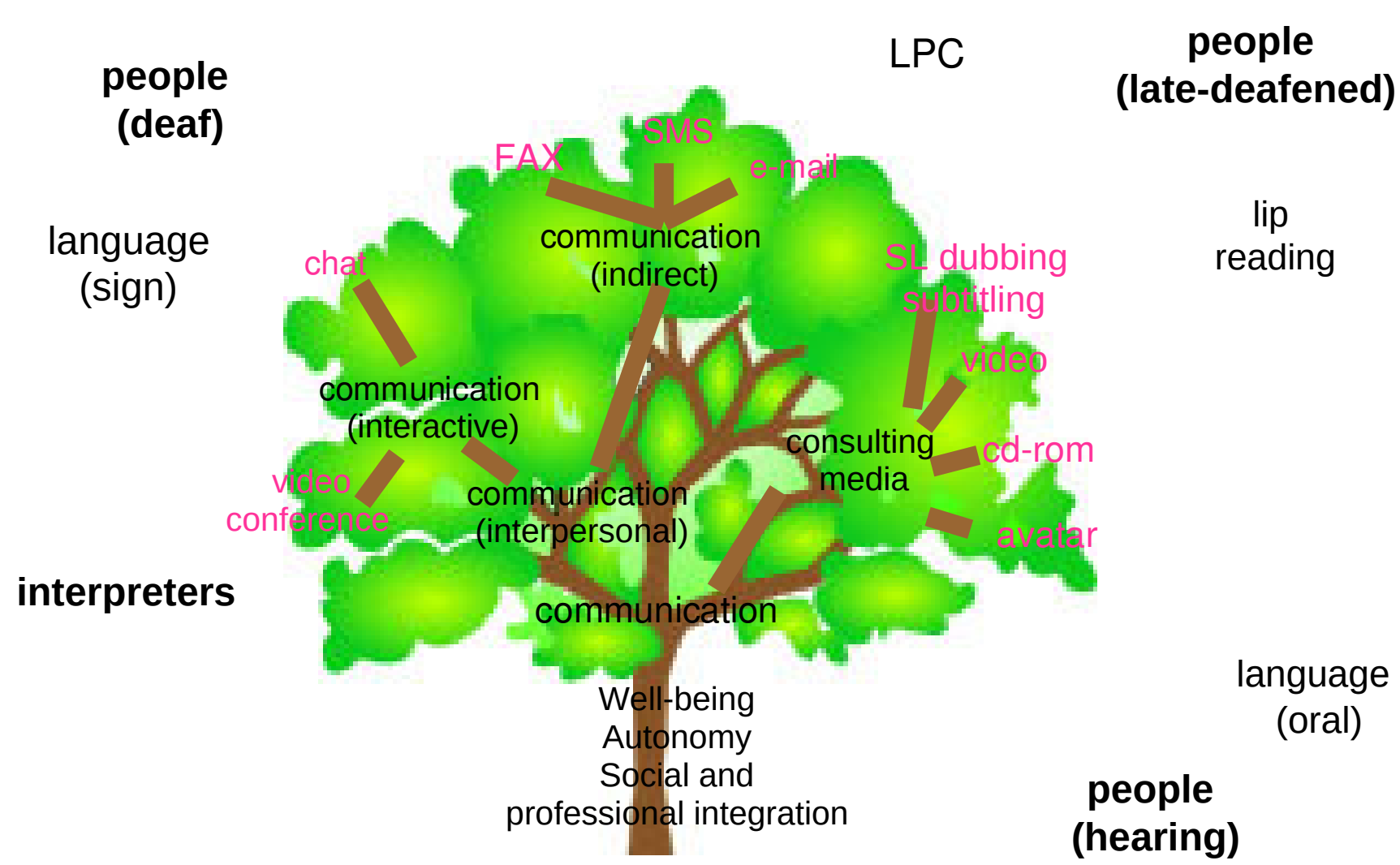

*Figure 13. Concept of a goal tree*

This goal tree will provide the general framework within which we will examine the problem. The problem will then be analysed on the basis of the concrete scenario of a current interpretation procedure. An analysis based in particular on the identification of obstacles will be carried out, and alternative solutions will be proposed in a framework involving not only organisations but also technological elements. Each of these alternatives will be detailed and assessed in the following chapters. They will ultimately be validated against the main functional and non-functional goals described in this chapter.

## *4.1 Inventory of Needs*

The goals were collected during numerous interviews. We also examined the skills of the various organisations as well as existing technical solutions, and asked ourselves why they exist. Various requirements of the daily life of a deaf person were identified and linked to the general goals already mentioned.

From here on, we will present goal diagrams more symbolically:

- the root, corresponding to the general goals, will be placed at the top
- the various nodes at the intersection of branches are shown using various geometric shapes: parallelograms for goals, hexagons for agents.

We will also use the following additional notions:

- the notion of **responsibility** assigned to an agent (written <<resp>>). The agent may be human or software. In the simple example above, the responsibilities are all assigned to technologies (FAX, SMS, etc.). In this section, they will also be assigned to organisations (SISW, SARE, etc.)
- the notion of **requirement** (in the sense of a specification): a goal placed directly under the responsibility of an agent that is part of the system to be developed. For example, in this study, the requirements will concern aspects related to interpretation but not to education.
- the notion of **obstacle**, represented by a parallelogram symmetrical to the goal and marked with a cross. It represents a barrier to be overcome in order to achieve a certain goal, linked to it by an **obstruction relationship** (written <<obstruction>>). Additional goals introduced to overcome obstacles are linked by a resolution relationship (written <<Resolution>>).

### 4.1.1 General Goals

Following the approach above, we obtain the diagram in Figure 14. The general goal of well-being is made up of three sub-goals: improving education, autonomy, and access to culture and information.

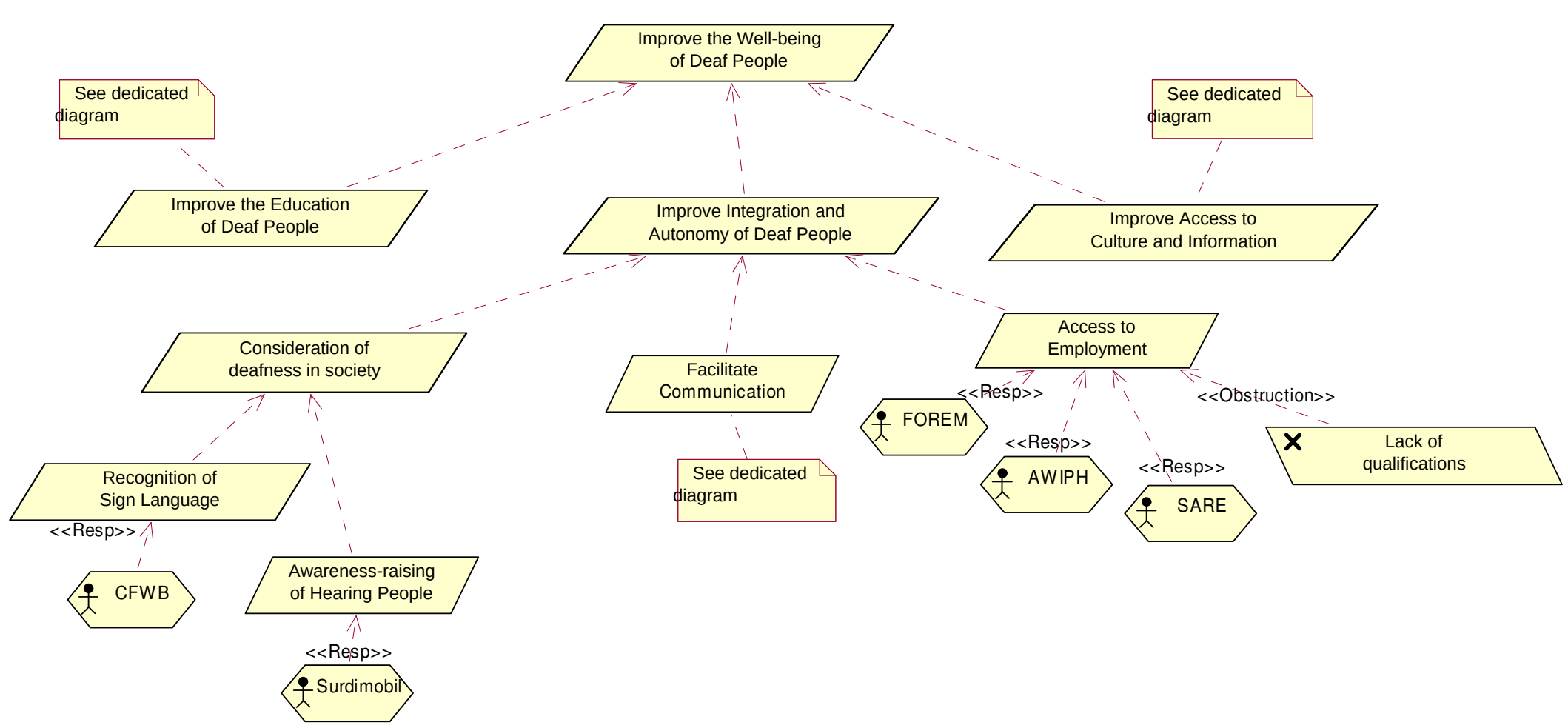

*Figure 14. Main goals and responsible organisations.*

The education goal is outside the context of this study and falls mainly within the competence of the French Community. It will nevertheless be detailed for information in section 4.1.3.

Integration falls mainly within the competence of the Walloon Region. Three sub-goals contribute to ensuring integration:

- First of all, the problem of deafness must be taken into account in the hearing world. Two aspects come into play here:
  - at the institutional level, sign language needs to be recognised.

  This point has been the subject of a study and has just been achieved.
  - at the level of the general public, which is relatively unaware of the issue, but there are some very good initiatives, such as the SurdiMobil. Events such as the World Deaf Day or announcements related to the recognition of sign language also contribute to raising awareness.

- next, the ability to communicate is crucial. We have highlighted huge gaps in this area. This goal will be detailed in section 4.1.2.

- finally, integration also involves access to employment. Several organisations work together towards this goal: FOREM as the general job-seeking body, AWIPH for the general policy on employment of disabled people, and specialised social services such as SARE for deaf people in particular. One obstacle is the lack of qualifications, which justifies introducing goals for training deaf adults.

Finally, the goal of improving access to culture and information will be described in section 4.1.4.

### 4.1.2 Communication Goals

Communication problems are the major problems for deaf people. We can distinguish communication problems with hearing people and between deaf people. These goals are detailed below and described in Figure 15.

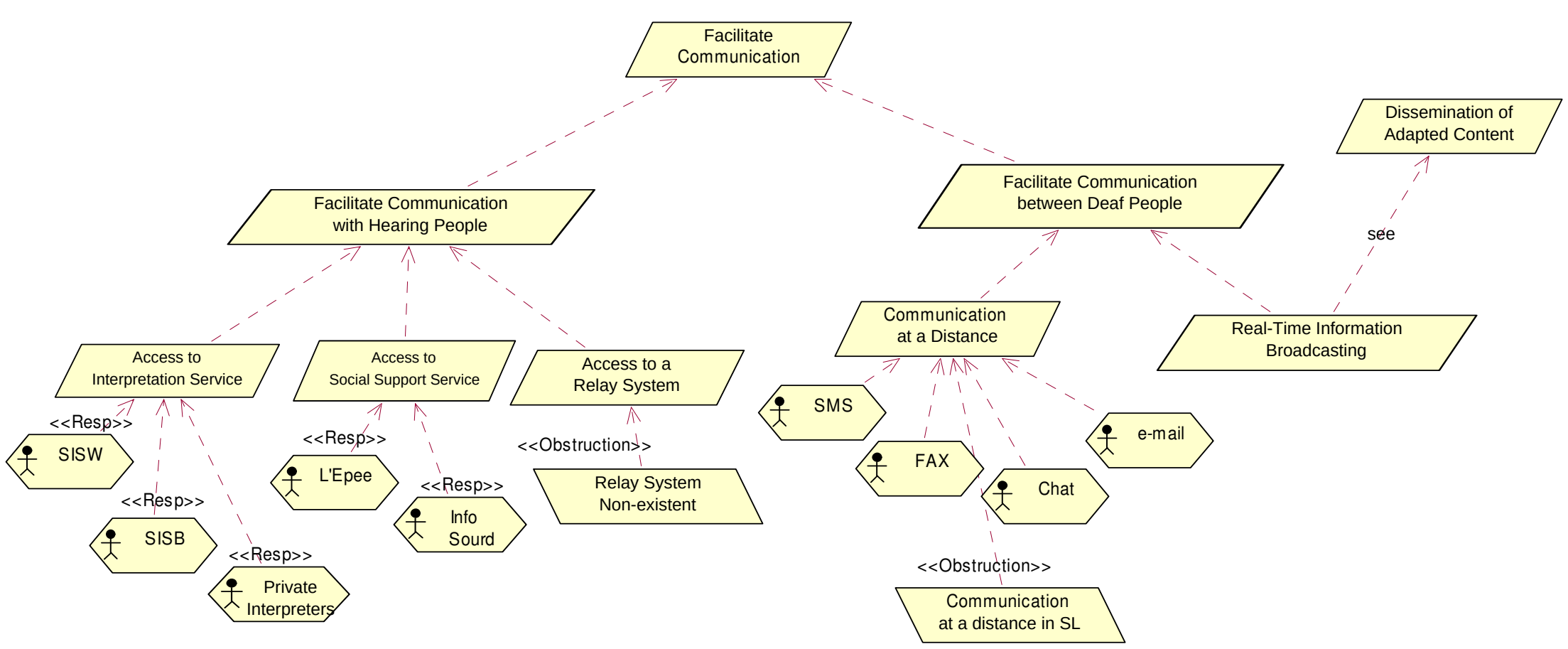

*Figure 15. Communication goals.*

Regarding **interaction with hearing people**, we have highlighted the need for interpretation services (such as SISW and SISB) and social services (such as L'Epée and InfoSourds). We have also mentioned the absence of a relay service. We will explore this goal in more depth in the second part of this chapter, as it is the key need that this study aims to address.

Regarding **interaction between deaf people**, various solutions based on written language exist. The gaps lie in means of remote communication in sign language. The solution involves the use of videoconferencing, a technology that will be considered for setting up a remote interpretation service and which could in the longer term be extended to all deaf people. The dissemination of content in sign language is also very poor and allows little indirect dissemination through the media. We will examine this point when refining the goal of access to culture and information (see 4.1.4)

### 4.1.3 Goals for Improving the Education of Deaf People

We give this refinement for information and to enrich the context. A first refinement considers the educational path on the one hand and the available material on the other. The latter goal also contributes to access to culture and media and will be considered in that context (see next section)

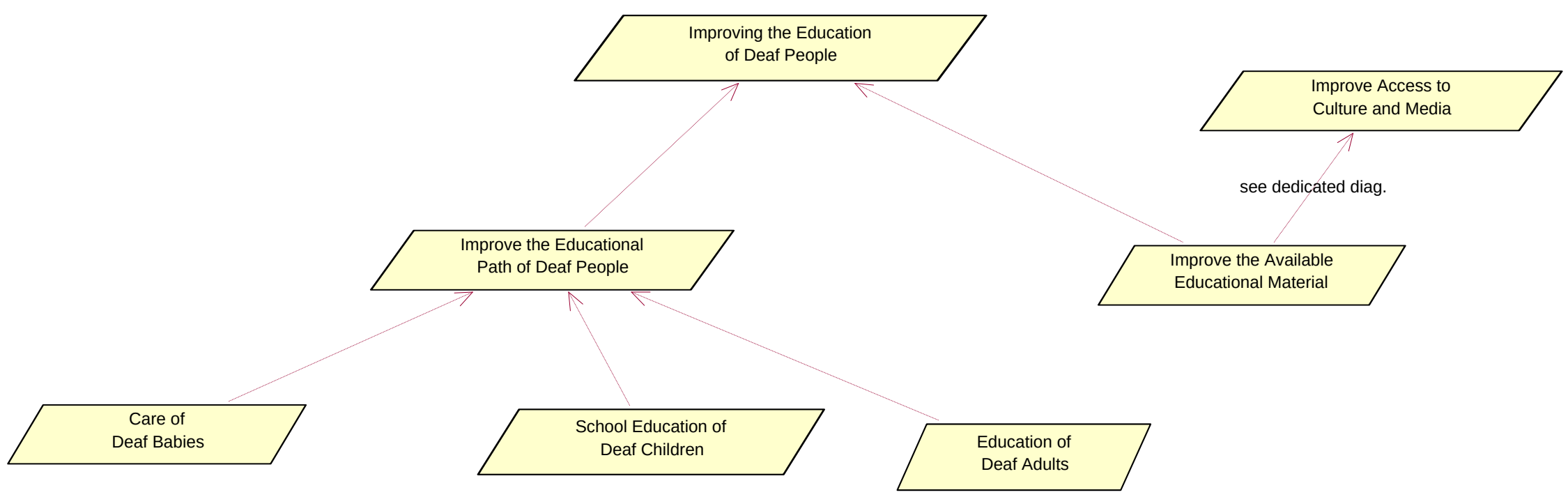

*Figure 16. Goal of educating deaf people*

The refinement of the educational path is a breakdown by age group. It is only outlined in the three diagrams that follow.

**For very young children**, as highlighted in the domain study, it is important to provide care as early as possible. This involves means of detecting deafness (typically at the ONE), medical care by audiologists, and also linguistic follow-up enabling the acquisition of both French and sign language, with a view to total communication. This last aspect is not yet fully achieved.

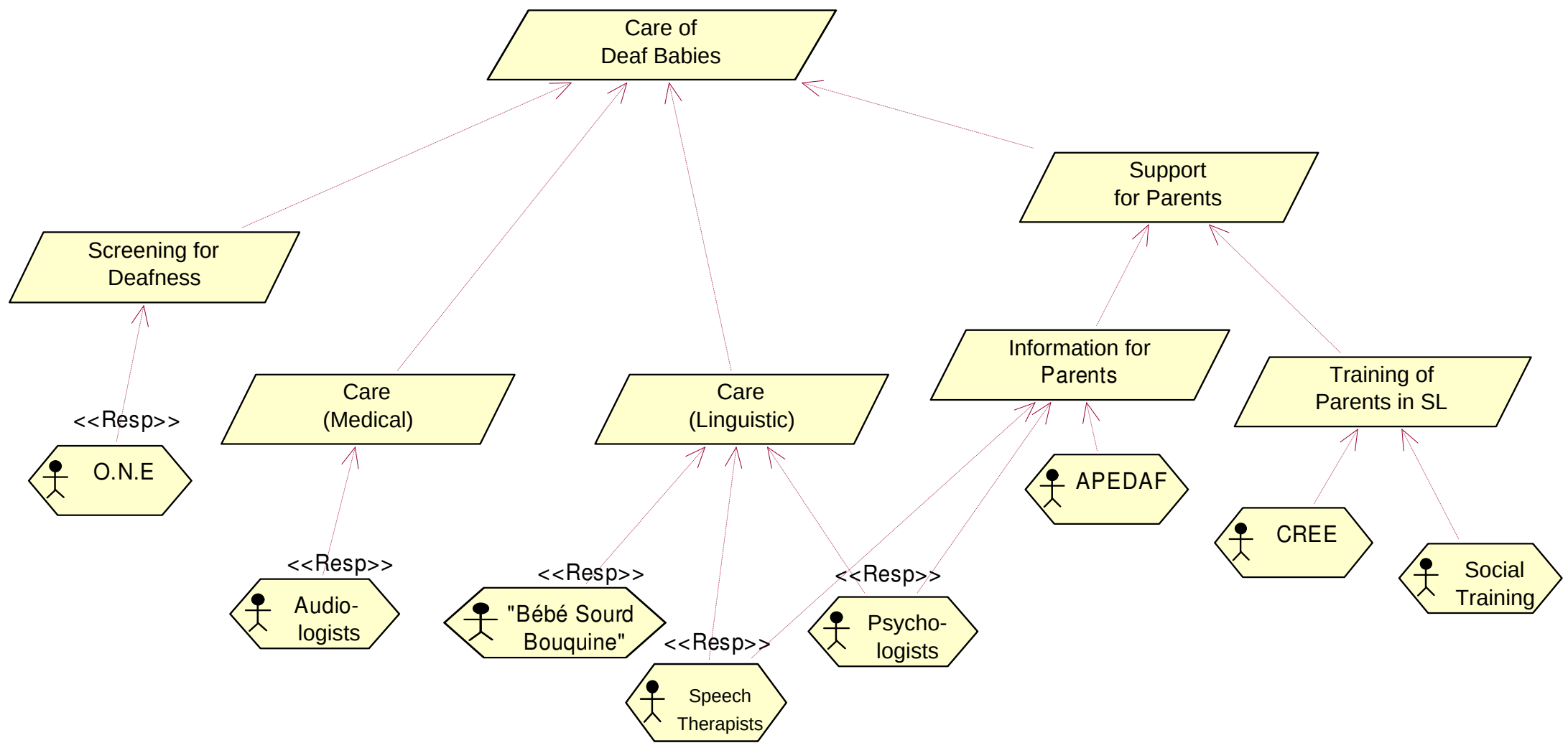

*Figure 17. Goal of caring for deaf babies.*

**For school-age children**, we find the various types of education described in the domain study, as well as multidisciplinary follow-up.

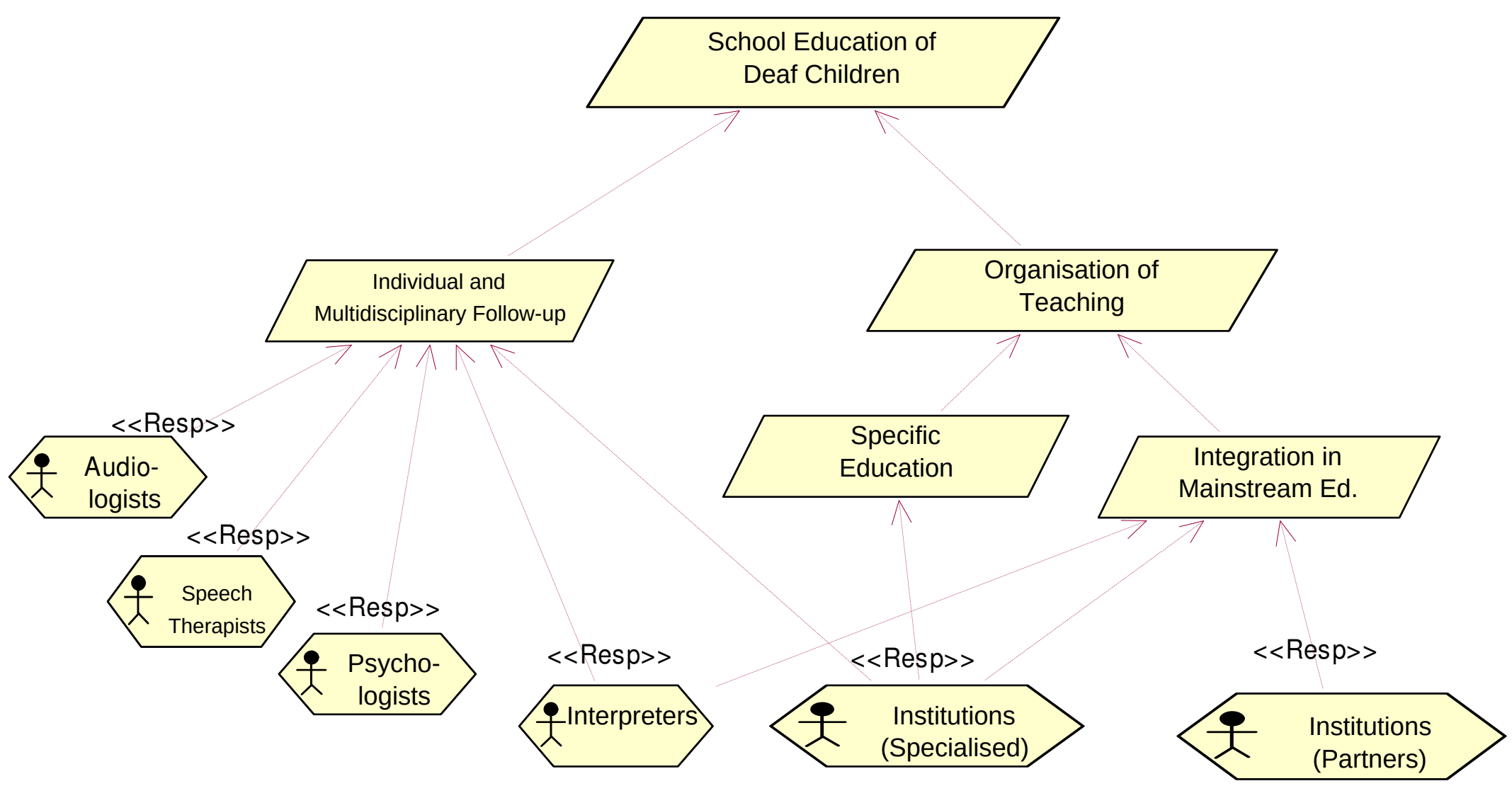

*Figure 18. Goal of teaching deaf children.*

**For adults**, there is a goal of improving command of French, which is a problem for many. It is provided in particular by courses offered by Alpha Signe. There is also a goal of training deaf people for employment. This is addressed in particular through integration into a mainstream educational path with appropriate support (notably interpreters). In practice, only a small percentage of deaf people follow it. Other solutions are apprenticeship-based training, which can take place in collaboration with partner companies (for example, the Ateliers du Monceau, which practise French-SL bilingualism).

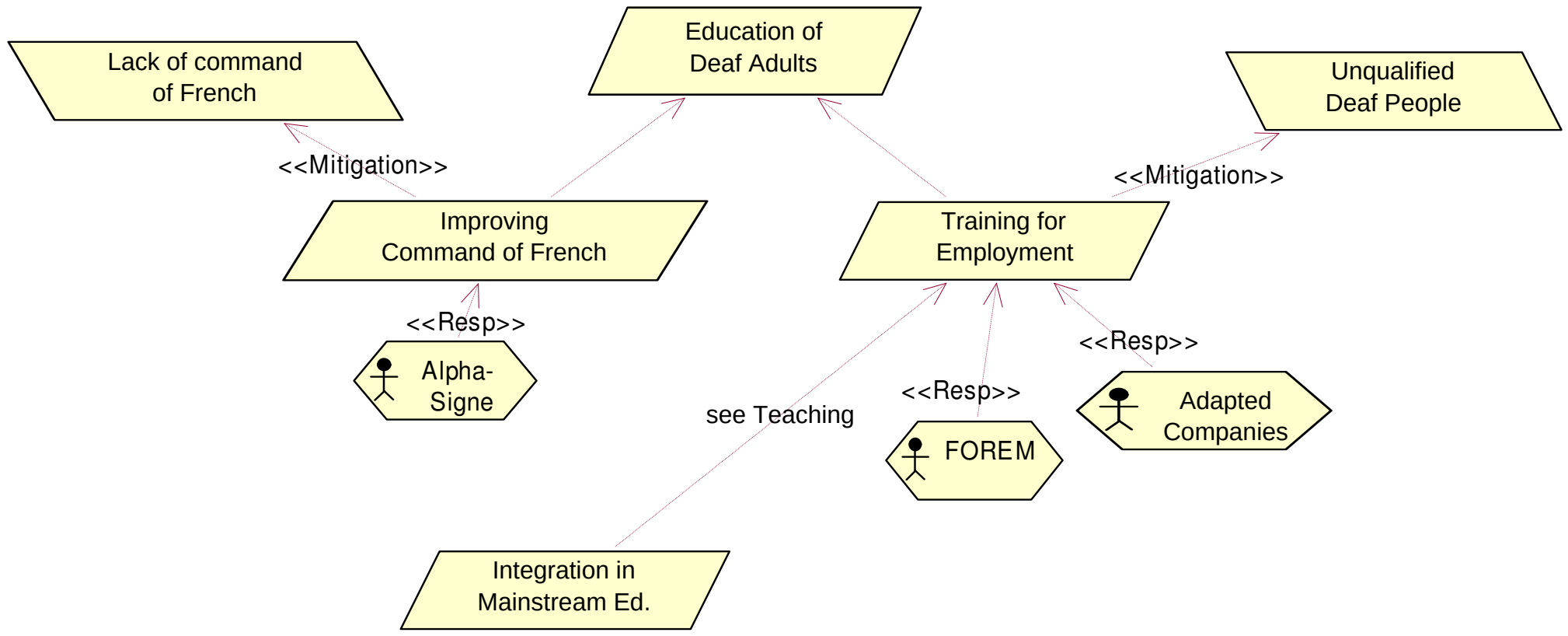

*Figure 19. Goal of educating deaf adults.*

### 4.1.4 Goals for Access to Culture and Media

Access to culture and media can be refined (cf. Figure 20) by considering, on the one hand, the availability of content adapted to deaf people and, on the other, the dissemination of this content. This dissemination can take place either on one of the physical media already widely discussed in the domain study, or through real-time broadcasting. We will examine these three goals in more detail.

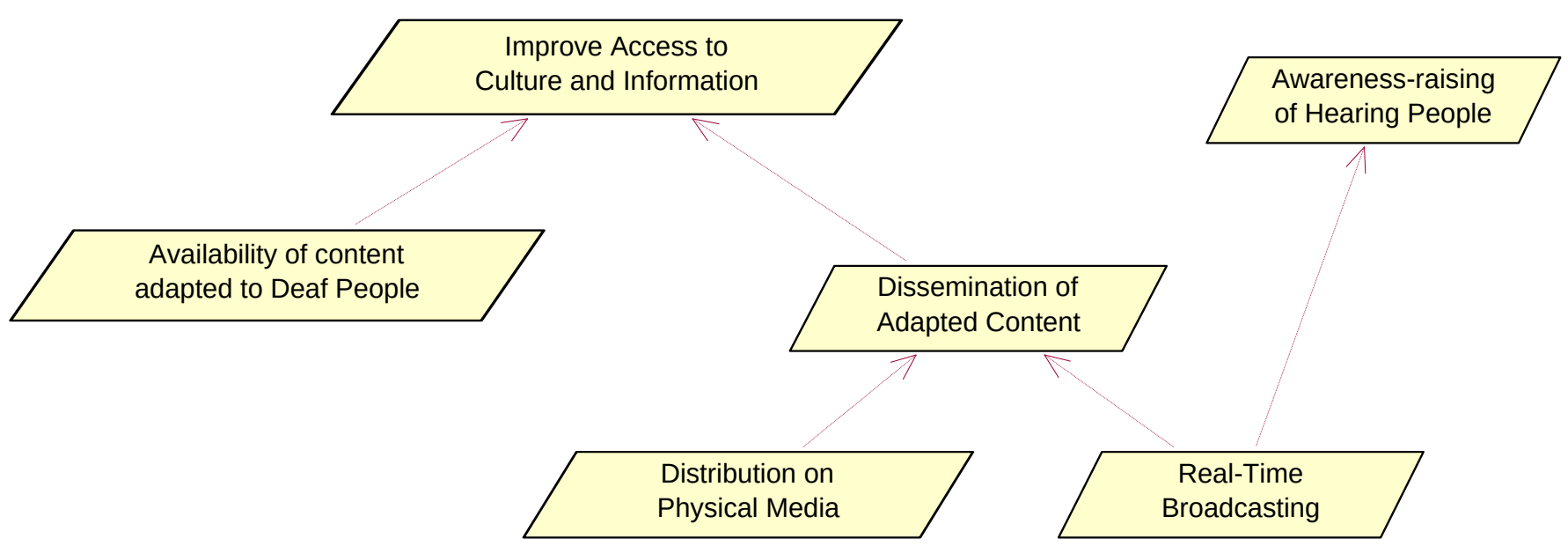

*Figure 20. Goal of access to culture and information.*

**The availability of content** adapted to deaf people breaks down mainly into three sub-goals:

- the availability of content ABOUT sign language, typically learning material: courses, lexicons, etc. Various organisations for deaf people have developed such material:
  - the CFLS has produced paper lexicons of unified LSFB
  - associations in the Tournai region have produced a CD-ROM version of part of this lexicon (2 CD-ROMs) as well as children's books combining drawings, simple sentences in French and representations in LSFB (e.g. the "Woopy" series)
  - LSF material is also available, for example the "DILS", a dictionary with very complete indexes in sign language and in French, and "Les Signes de Mano", which makes remarkable use of multimedia

- the availability of content IN sign language, which requires dubbing by an interpreter
- the availability of subtitled content. For signing deaf people, this solution suffers from the language barrier. It is much better suited to the larger class of late-deafened people. Implementation is complex and costly because very good synchronisation with speech must be ensured. The workload is estimated at 40 times the broadcast time. For real-time broadcasting (see next point) it is even more complex.

Cost constraints unfortunately severely restrict the amount of content disseminated and lead to painful choices about content. The priority is to focus on important subjects, health information for example.

- regarding subtitling: very few commercial DVDs have subtitles specifically for deaf people. The number of subtitled programmes is also very limited in French-speaking Belgium (RTBF: a few programmes per week, no news). It lags far behind other European countries (about 80% on the BBC, 20% on TF1 and A2). The programmes are mainly documentaries (such as "Jardin extraordinaire") or information (on health, road regulations)

- regarding sign language material broadcast on the RTBF: there are two news programmes: the main TV news and the "Niouzz", intended for deaf children but also followed by a wider audience given its very clear and simple approach to current affairs. Published material is mainly produced on the initiative of deaf associations when they find the means. It mainly concerns information on crucial subjects such as AIDS, pregnancy, etc.

**Alternatives** that have not been explored for content dissemination are:

- the use of virtual characters, for their advantages in terms of learning (viewing angle, speed), adaptation to the audience and storage volume on a given medium.
- online dissemination on the Internet, which makes it possible to keep the medium up to date with the evolution of this living language.

Combining these alternatives is also recommended: online video dissemination on the Internet remains bandwidth-intensive, and the use of synthetic characters would greatly reduce it. We will develop this point in chapter 7.

These alternatives have been studied with the CFLS, which could make use of them for its process of unifying Belgian sign language, at the level of:

- the process itself: weekly working meetings could be more efficient if discussion could start with encoding and consultation via the Internet. The community could also be more involved in the process.

- the product: the lexicon itself could be online.

However, various obstacles have been identified:

- the limits of expression of virtual characters

- the use of Internet technologies in the deaf community, and in particular among older people (as among hearing people)
- pricing: the current lexicons are sold and are a significant source of income for the CFLS.

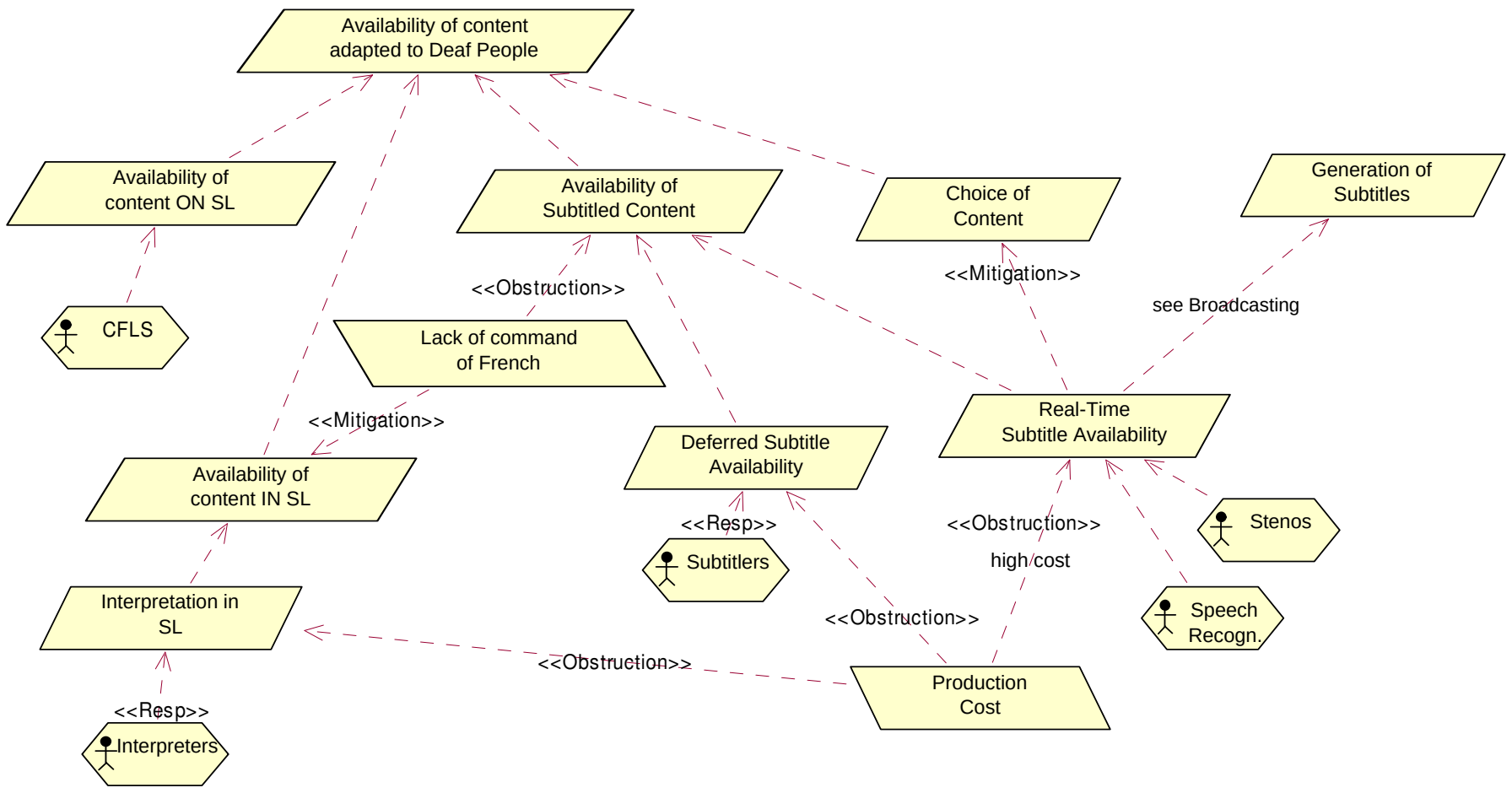

*Figure 21. Goal of availability of content adapted to deaf people.*

**The goal of "real-time" broadcasting** is described in Figure 22. This mainly concerns analogue broadcasting over the air or by cable, but online digital broadcasting can also be envisaged in the future.

- for sign language, the organisation is not very complex, as interpretation is always done in real time. For fairly long programmes, several interpreters must be planned (ideally taking over about every 20 minutes).

- for subtitles, the process is much more complex and costly, because it cannot be prepared in advance for "live" interviews. It requires highly specialised people (stenographers, for example on France 2) or very advanced technologies (speech recognition, for example at VRT). Perfection is not achievable here and an error rate of a few percent is tolerated.

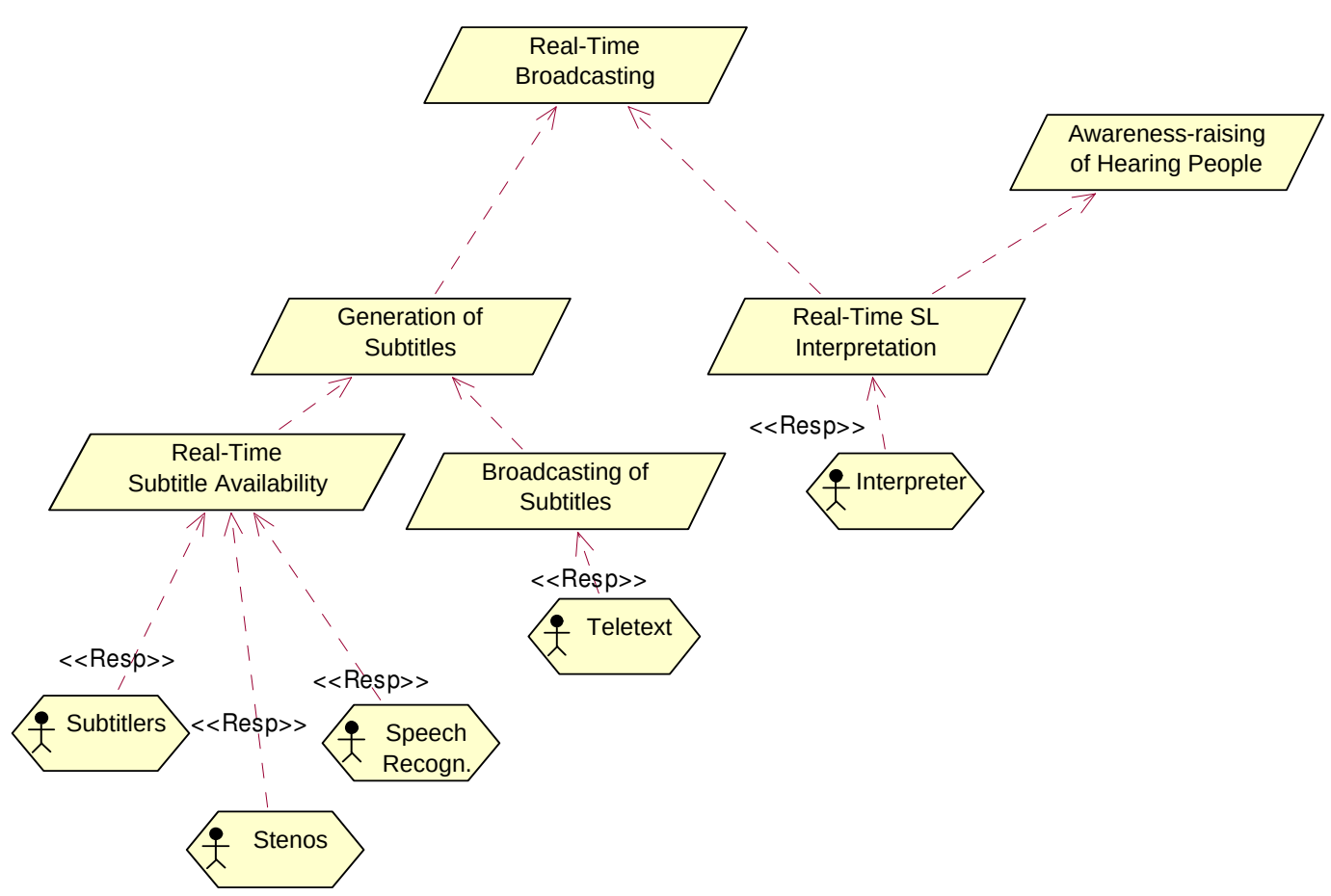

*Figure 22. Goal of "real-time" broadcasting.*

Regarding distribution on physical media, we again find the need for suitable media, already discussed in the presentation of the domain. Another goal is the presence of competent publishers. For content relating to sign language, this implies the participation of deaf people in validation and possibly in production. Achieving this last goal also helps to improve access to employment.

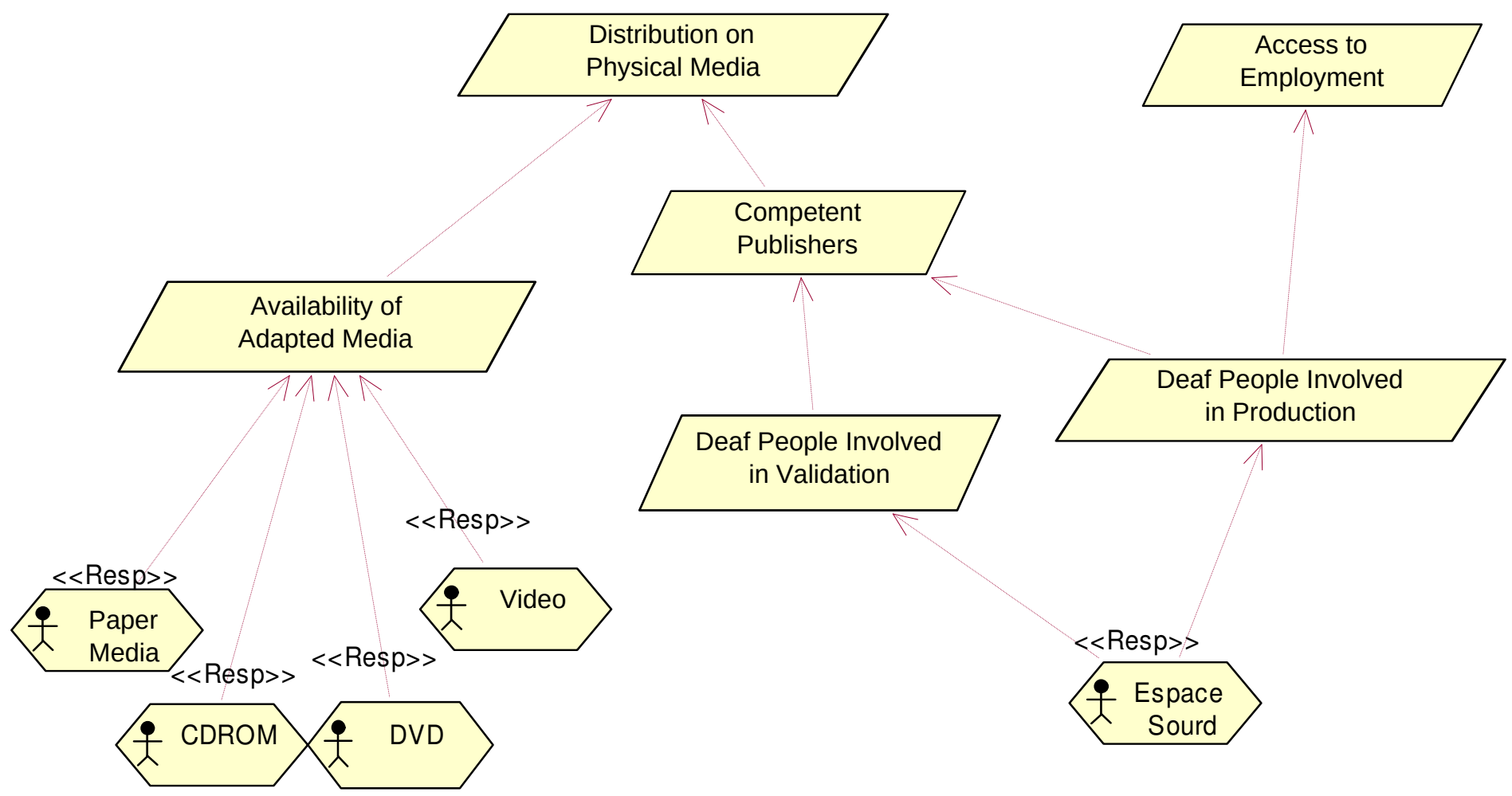

*Figure 23. Goal of distribution on physical media.*

### 4.1.5 Non-functional Requirements

The requirements elicited so far have mainly described functionalities that improve the daily lives of deaf people: interpretation service, dissemination of information, etc. They are complemented by non-functional requirements, which correspond to qualities of their implementation. We have already mentioned several of them; others were collected during interviews or from the literature. Here is a summary of the main ones:

- **Personalisation**: systems must be able to adapt to the very diverse skills and situations of deaf people; the proposed solutions must be flexible in their organisation and technical aspects.
- **Performance**: conversational systems must by definition be provided in real time. They must allow good-quality conversation. More detailed criteria (image resolution, fluidity, etc.) will need to be established.
- **Affordability**: access costs for users must remain acceptable. Support policies may need to be provided.
- **Reliability**: the technical solutions chosen must be proven and offer the best guarantees of availability.
- **Interoperability**: for technical solutions, preference will be given to those based on a standard that does not tie the solution to one manufacturer and allows equipment from different manufacturers to work together. This also ensures greater durability of the system.
- **Security**: the systems chosen must guarantee the anonymity and privacy of their users.

## *4.2 Detailed Analysis of Access to Interpretation*

The general organisational analysis in the previous section gives a good overview of the problems affecting deaf people and shows the importance of sign language interpreters, who appear many times. We will now examine in more detail individual access to an interpretation service in the context of a procedure at a public service. We will examine a typical scenario, identify the main obstacles and propose various alternatives to resolve them.

### 4.2.1 Communication Situations

As mentioned in the domain study, contacts between deaf and hearing people are often problematic (for example between a deaf person who has difficulties with French and a hearing person with little awareness). We have identified a number of typical situations:

- administrative: at the municipality (renewal of an identity card, passport application, etc.), at the Walloon Region (housing grant, etc.)
- with a public service in the broader sense: health insurance fund, bank, insurance, electricity (Electrabel), water (SWDE), etc.
- legal: at the police, at the lawyer's, in court
- medical: at the general practitioner or specialist, at the hospital, at the psychiatrist or psychologist
- cultural: associations, conferences, guided tours, masses, etc.
- professional: meetings, training, job interviews, trade unions
- private: shops, parents' meetings, etc.

Communication requires the presence of a person proficient in French and sign language. This role can be filled by someone close to the deaf person, but this comes up against several obstacles:

- the sign language skills of these people are not always perfect. They may unintentionally alter the content of the messages exchanged.

- neutrality is not guaranteed: the content of the information passed between the deaf and the hearing person may be deliberately altered (censorship, unwanted additions, etc.)
- lack of confidentiality: the deaf person may have to reveal aspects of his or her private life
- responsibility: the deaf person may be completely excluded from the decision-making process, which does not encourage him or her to take charge.

These obstacles justify **access to a professional interpretation service**, ensuring **accuracy of the message, neutrality and professional confidentiality**. It also gives the deaf person greater autonomy, as shown in Figure 24. We will briefly examine how such a service works.

A typical illustration that we had the opportunity to witness is a deaf person being summoned to the police. It was a demonstration sketch during the inauguration of the Namur branch of the social service "L'Epée" (see Figure 25)

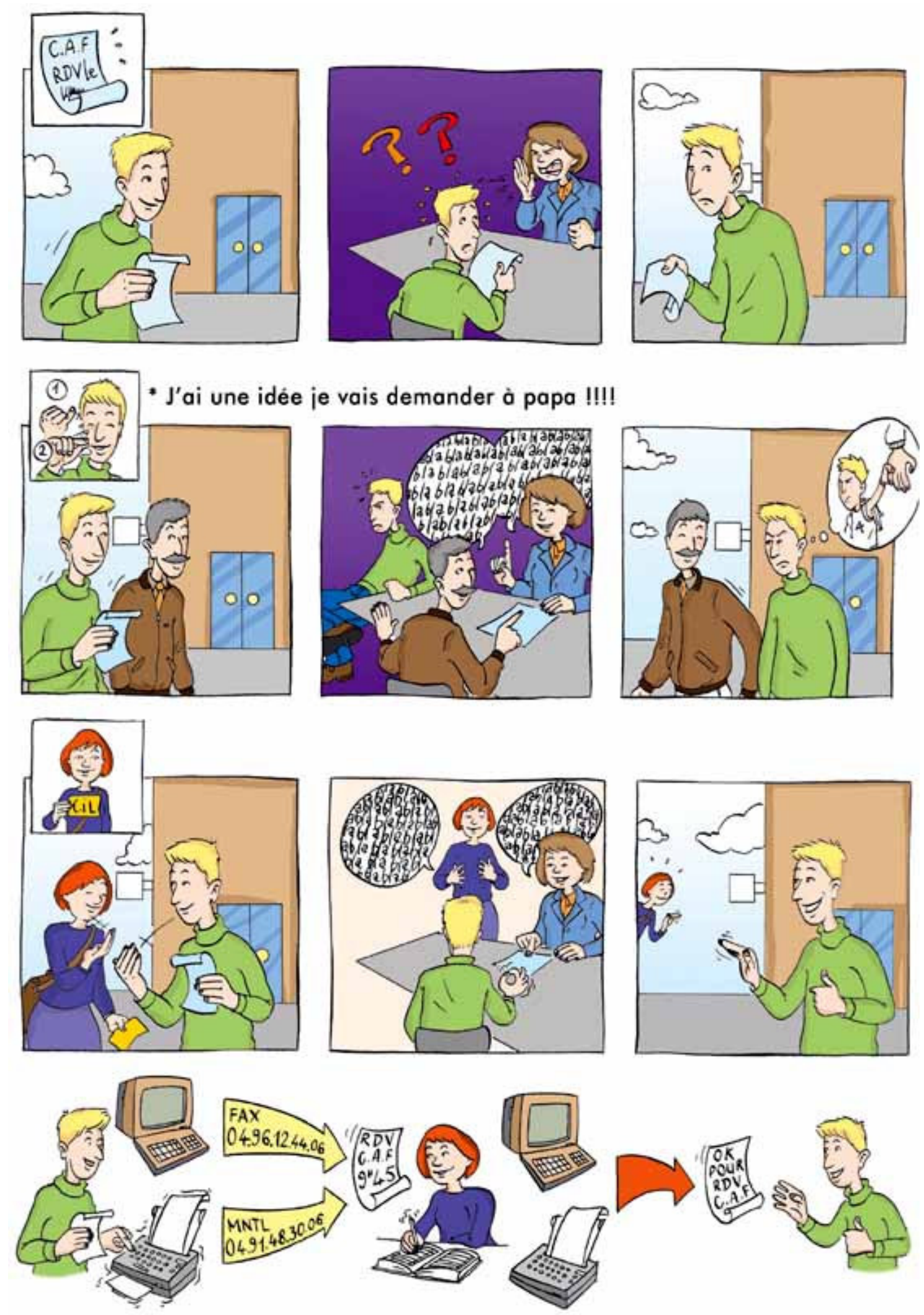

*Figure 24. Sign language interpretation service.*

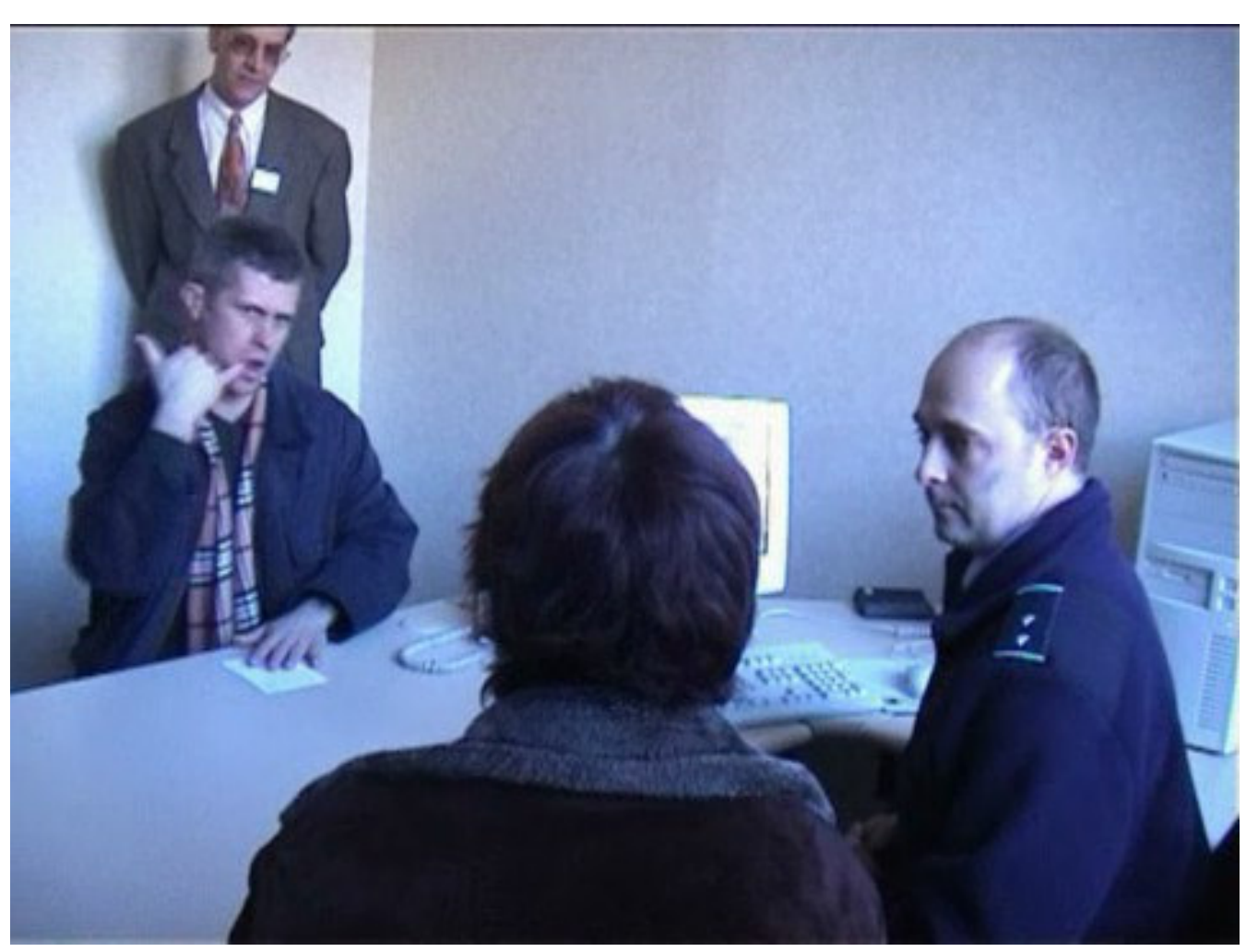

*Figure 25: Deaf person and her SL interpreter at the Namur police*

The scenario presented was as follows:

- *the deaf person was caught speeding (deaf people can obtain a driving licence)*

- *the "pro justicia" notice received was thrown away by the deaf person, who did not understand what to do with it (the person may have difficulty understanding French, in particular the vocabulary used in this type of document)*

- the deaf person then receives a summons to appear in person at the Namur police.
- This time, the deaf person has understood what to do and contacts an interpretation service (more precisely the social service in this case). Technically, this contact can be made by a means accessible to the deaf person: fax, SMS or in person to explain the problem. The latter is more suitable here, since it is a social service.
- the service schedules the appointment according to the availability of the person and the interpreters and the constraints of the summons.
- the deaf person and the interpreter meet at the agreed place and time.

A simplified version of this scenario is shown in Figure 26 as a sequence diagram (UML notation). This diagram shows, with horizontal arrows, the various interactions between the users of the system. Each user corresponds to a vertical line, with time flowing from top to bottom.

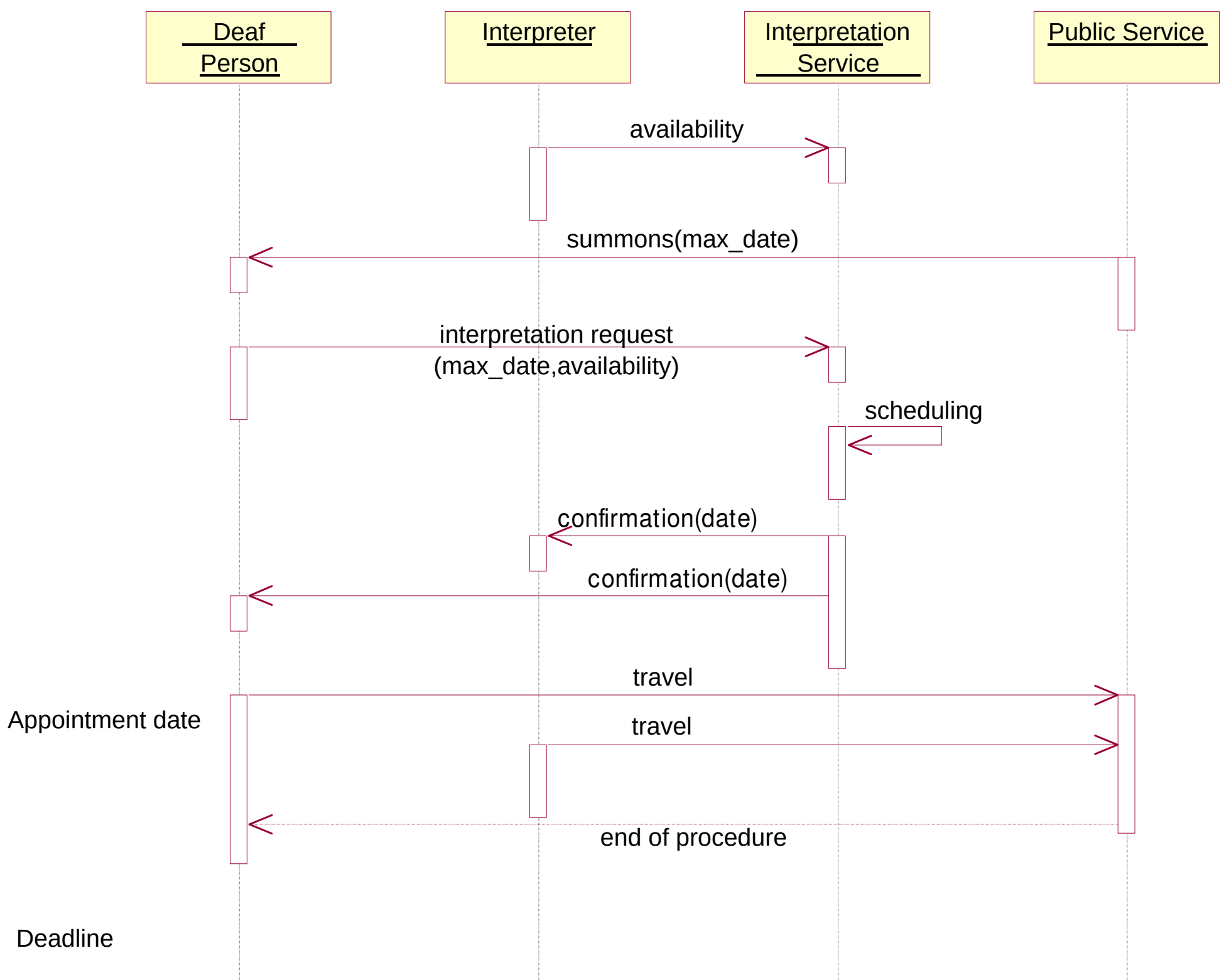

*Figure 26. Scenario for carrying out an interpretation.*

Regarding the cost of the service, deaf people have free interpretation hours in the form of "tickets", which are reimbursed to the interpretation services. For professional needs, this allowance is larger but often insufficient. Deaf people also exchange tickets between intensive and occasional users. For a comparison with the situation in other European countries, see [EU00].

### 4.2.2 Analysis of an Interpretation System

The goal of offering an interpretation service can be refined into 2 sub-goals: availability of interpreters and scheduling. We will examine the many obstacles to these goals and identify ways to resolve them (see Figure 27)

- the **small number of interpreters**, about twenty in French-speaking Belgium, lengthens waiting lists at the scheduling level: an appointment must be planned several weeks in advance. Improvements here require training more interpreters and, more generally, improving their working conditions, in particular through recognition of their status.
- a large part of the **time is lost travelling**: sometimes two hours on the road for a few minutes of interpretation! At the scheduling centre, this is taken into account when allocating requests to interpreters, but many factors make it difficult to satisfy: for example, the urgency of some requests or the lack of interpreters in some regions (such as the province of Luxembourg)

- the two previous points make it impossible to **handle emergencies** in the current system.

In the following sections, we examine 3 ways of mitigating these obstacles.

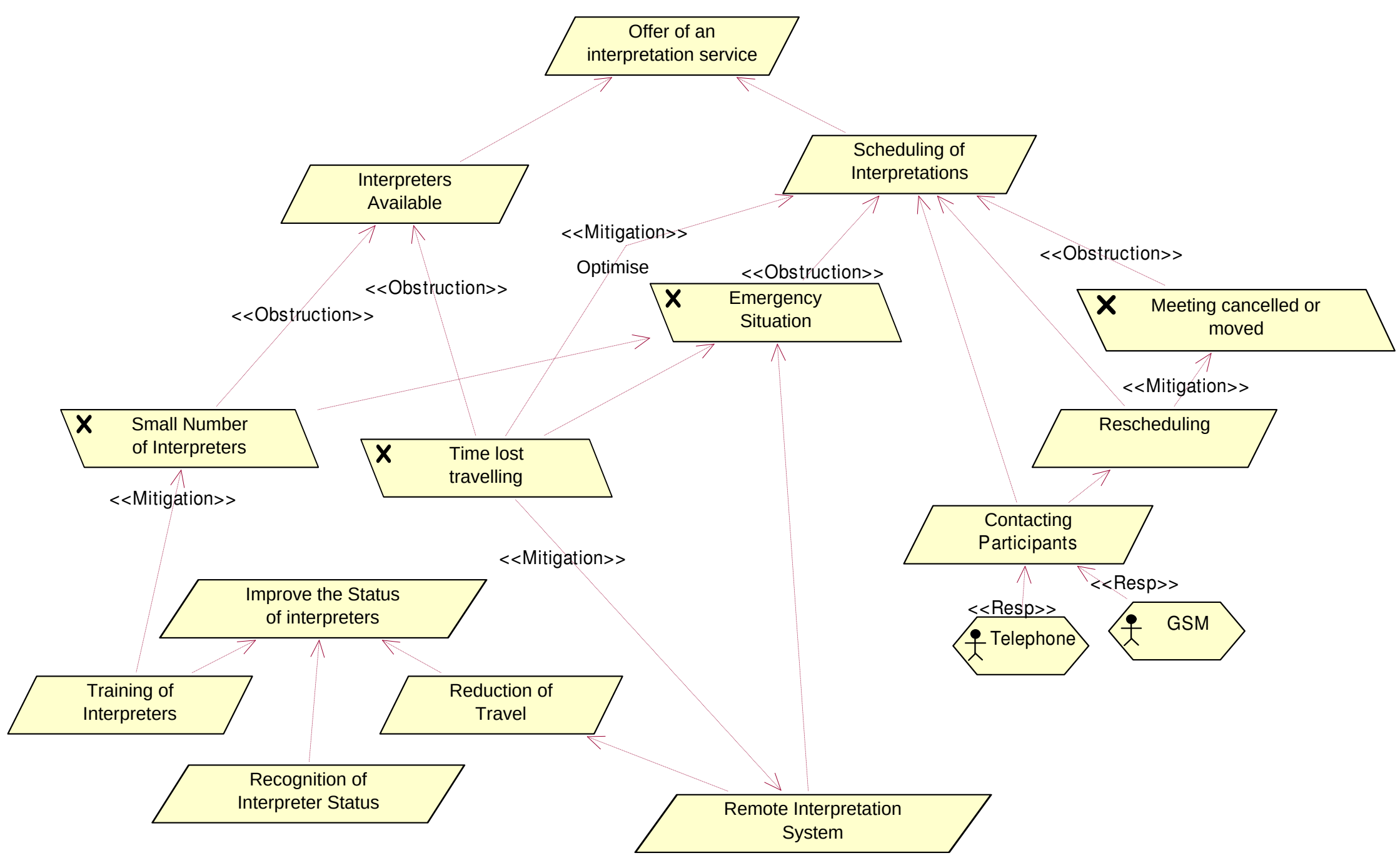

*Figure 27. Analysis of obstacles to interpretation.*

### 4.2.3 Solution 1: Remote Sign Language Interpretation System

A first possible solution is to implement a remote interpretation system, which no longer requires the interpreter to travel. This system must use technology to capture and transmit images and speech between the deaf person, the interpreter and the hearing person. A priori, the system offers several advantages:

- improved availability of interpreters by saving travel time, and therefore shorter waiting times to satisfy a request

- improved working conditions for interpreters, as travelling is tedious. One could even envisage interpreters carrying out interpretations from home.

- the possibility of handling emergencies, by organising an on-call service.

However, new obstacles also appear. They were identified during interviews, in particular with the interpretation services. A survey of deaf people was also carried out. The main ones are:

- the physical absence of the interpreter requires the deaf person to face the hearing person alone at first.
- the quality of remote communication: interpretation on screen, image quality in terms of resolution and fluidity, appropriate framing, etc. This requires a careful examination of the technical limitations.
- intimidation related to the technological aspect.
- the notion of confidentiality given the presence of a camera.

The interpreters interviewed consider that the system is useful but should be reserved for fairly simple procedures and neutral situations. Routine operations in an administration seem appropriate. The system is not recommended in crisis situations or those with an emotional dimension. The medical and legal fields therefore seem less suitable. Nevertheless, the urgency of a situation may justify its use. By comparison, when a hearing person who speaks only a foreign language is admitted to a hospital as an emergency, an interpreter can be called by telephone.

The survey was a multiple-choice form available on the Internet from the FFSB website. This method has already been used for other surveys. It is obviously very biased in that only "connected" deaf people could respond. They have a very active profile and are a priori in favour of such a system. The results (presented in the appendix) do point in that direction: most of the deaf people who responded have no preconceptions and accept the absence of the interpreter. Concerns were expressed about image quality. Among the places that should be equipped with such a system, the most requested are hospitals, the police and municipal offices.

Note also that this **remote interpretation system** should not be confused with a **video relay system**, in which the hearing person is not beside the deaf person but in telephone contact with the interpreter. The second system is, however, a trivial extension of the first and could also be considered. In the context of contact with a hearing employee within an administration, however, the question arises of where the videoconferencing station used by the deaf person should be located.

Chapter 5 studies the use of videoconferencing in more depth.

### 4.2.4 Solution 2: Automated Interpretation System

In the previous solution, the system still relied on a human interpreter. One option to consider would be a system based on machine translation, insofar as it can be implemented. We mentioned the complexity of the interpretation process in the domain study. Automating it requires a chain of complex components using advanced techniques:

- for communication from the hearing to the deaf person: speech recognition, machine translation from French into sign language, animation of synthetic characters
- for communication from the deaf to the hearing person: recognition of gestures and facial expressions, machine translation from sign language into French, speech synthesis.

We will study these techniques in more detail in chapter 6. We will see that while some are very mature, others are still in the realm of pure research.

An interpreter therefore cannot yet be replaced by a machine. This is not surprising, since translation systems for oral languages, which have been studied much more, are also far from effective.

However, partial systems for specific communication situations can be considered. For example, a system based on the use of predefined phrases ("formulaic language") coupled with a virtual character that interprets them in sign language was developed within the European ViSiCAST consortium and tested at the British Post Office. This system is described in detail in chapter 5. A validation phase showed that the system makes it possible to carry out simple, routine procedures at a post office, such as cashing a cheque or buying stamps. However, it makes the procedure longer and heavier. Its main use seems rather to be a safeguard for cases where the transaction does not go well with "makeshift means" (paper, gestures, digital price display, etc.)

Chapter 5 will be devoted to a more exhaustive exploration of partial systems and the specific situations they could address.

### 4.2.5 Solution 3: Extension to Mobility

In the two previous solutions, we implicitly assumed that people are at fixed locations: the deaf and the hearing person are together and contact remotely the interpreter, who is in an interpretation centre. Fixed stations are provided at both locations and a communication link connects them. Its nature has not been detailed, but with current technology it is a "wired" link.

Wireless technologies have been developing continuously since the arrival of GSM. GSM truly liberated deaf people thanks to SMS, which, although originally a "technical" service, has become one of its main features (including for hearing people). SMS has allowed deaf people to stay in touch everywhere with those around them, whether deaf or hearing.

This development is set to continue towards total high-speed connectivity. This is the ultimate goal of UMTS, an expensive technology that has been the subject of much speculation. **Eventually, the availability of bandwidth everywhere will finally give deaf people the equivalent of GSM for sign language.**

We are not there yet. At present, a lighter and more targeted technology is developing instead: WiFi, the wireless equivalent of local computer networks. It provides Internet access at equipped sites (called "HotSpots"). "PC" and "GSM" technologies will certainly converge in the coming years: the arrival of personal digital assistants, GSM phones with multiple communication functions (such as the Nokia 9210i) and "tablet PCs" are current signs of this.

For the remote interpretation system, this opens up new possibilities:

- contacting an interpreter remotely from any "HotSpot", not just from a specially equipped location.
- the terminal belongs to the deaf person and not to the administration (personal rather than collective equipment). These terminals should be easy to carry, for example PDAs or "tablet PCs", and must be fitted with a camera.

Communication infrastructure and equipment costs would be reduced, provided that the available infrastructure is suitable (guaranteed quality)

We will discuss these options in more detail in chapter 7. In particular, we will examine the objectives of the European WISDOM project.

### 4.2.6 Summary of the Operational Alternatives Considered

To meet the need for remote interpretation, three alternative solutions have been discussed. Each draws a different boundary between the "automated" system and the actors interacting with it. Figure 28 gives an overview of the design of each alternative, in particular the human agents involved and the technologies used.

In the next 3 chapters, we will explore these alternatives in more detail, focusing on feasible systems and their requirements, in particular for implementation within a public service.

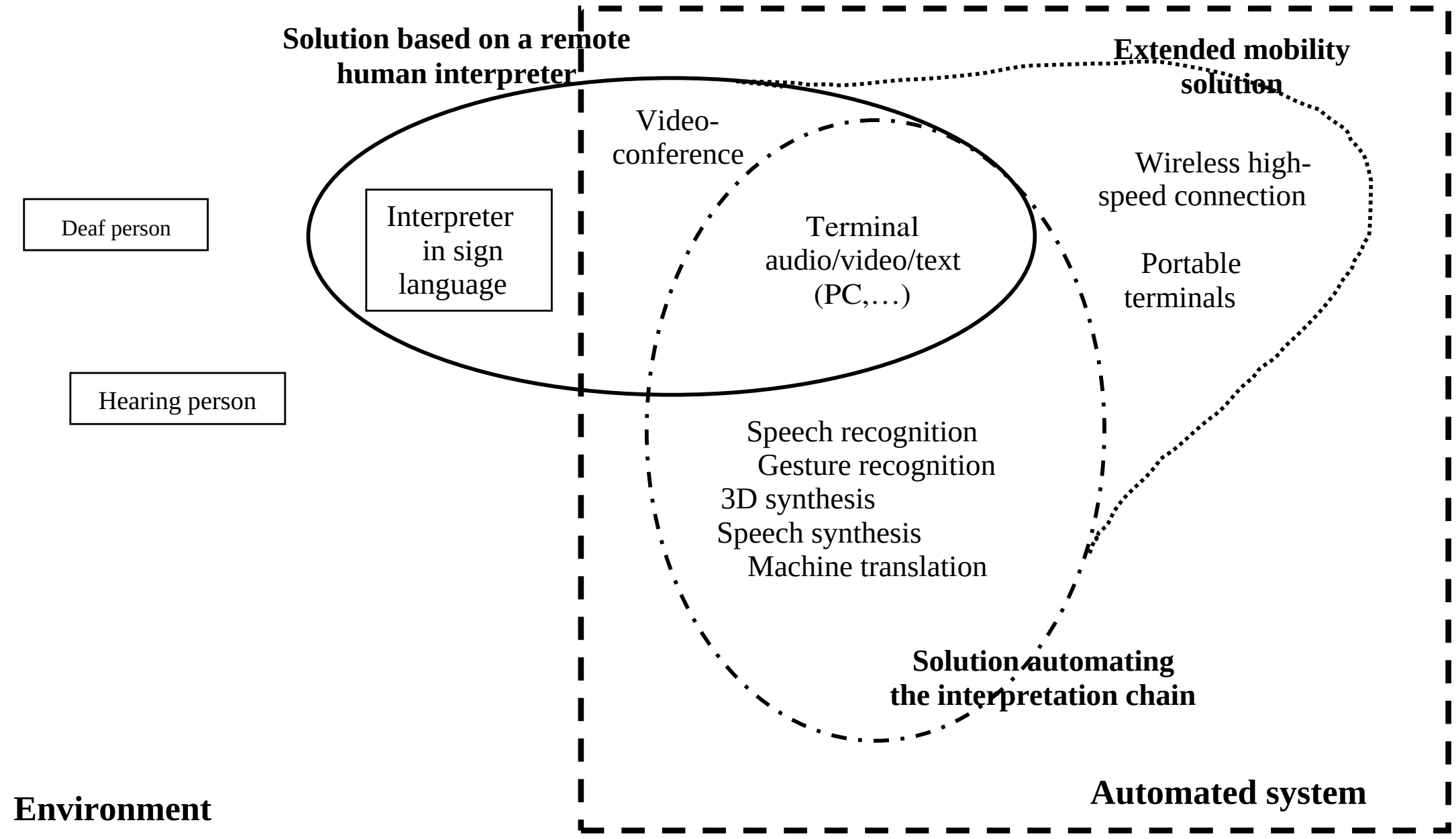

*Figure 28. Main alternatives for an interpretation system.*

# 5 Sign Language Video Interpretation

In this chapter, we examine in detail the design of a sign language video interpretation system, whose objectives were established in the previous chapter. We will first examine the **principle** of this system, then the constraints related to the underlying technology: **videoconferencing**. In order to highlight the requirements specific to a remote sign language interpretation system, we will review **existing systems and studies**. On this basis, we will propose a **summary of the requirements** for the design of a video interpretation system suited to sign language, in order to make the best possible use of videoconferencing. These requirements will be **validated** through a series of tests we carried out: we will describe the prototype deployed and the lessons learned.

## *5.1 General Principles*

Figure 29 recalls the principle of **remote interpretation**: the deaf and the hearing person are physically together and the sign language interpreter is remote. A video link allows communication in sign language between the deaf person and the interpreter, and an audio link allows communication between the hearing person and the interpreter. The latter may be provided by telephone. The main problem is carrying the video so that the important elements are transmitted with good quality: in particular hand movements and facial expressions.

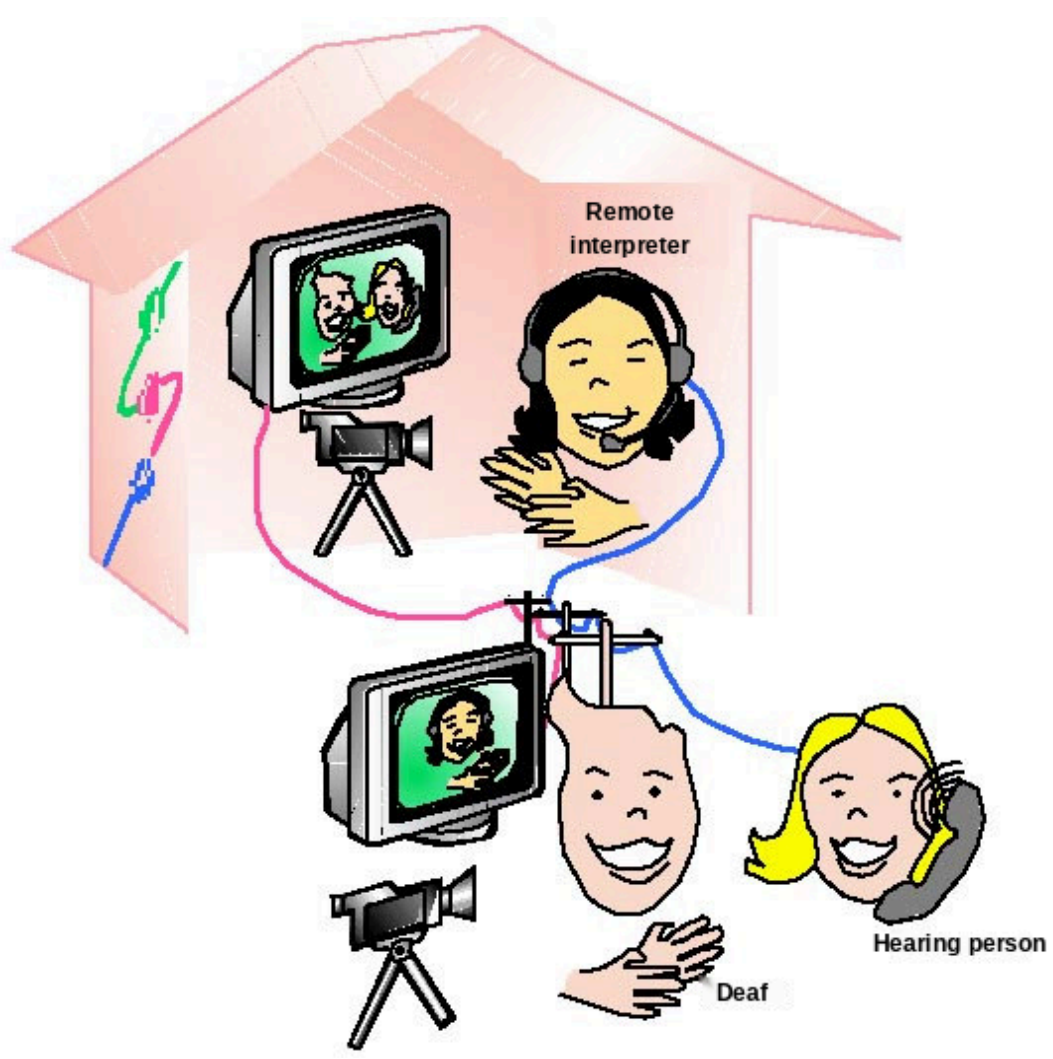

*Figure 29. Principle of remote interpretation (source [MCC99]).*

Figure 30 illustrates the **video relay system**, which is the extension of the text relay system presented in the domain analysis (see Figure 12). Compared with the remote interpretation system, it is clearly a natural extension of it: the hearing person is simply contacted by telephone. If the deaf person has personal equipment, he or she can also be spared from travelling (sometimes far). **The system we are going to design can therefore evolve to support this relay service.**

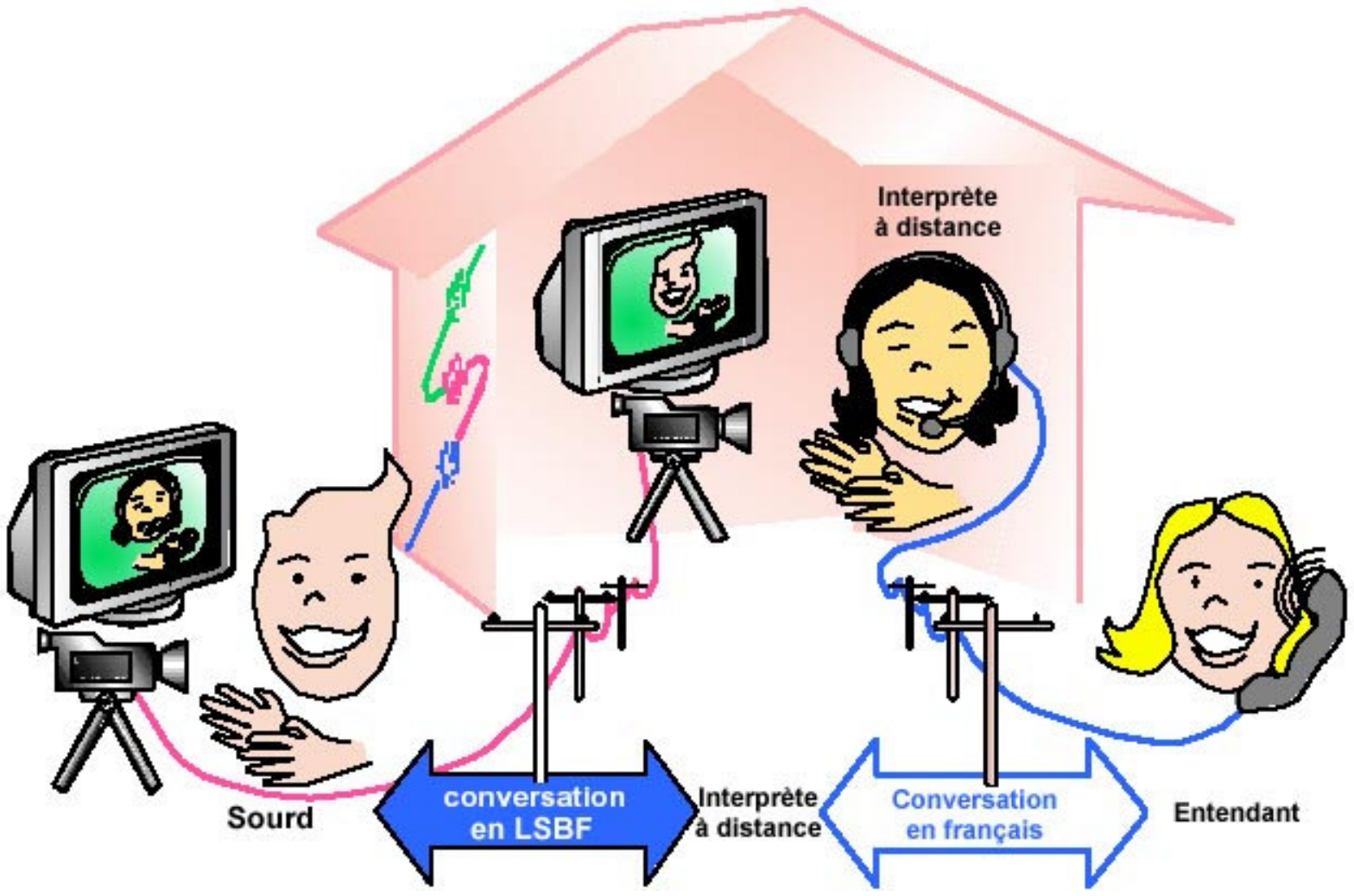

*Figure 30. Principle of video relay (source [MCC99]).*

Let us also note two other extensions of the system

- the system works independently of the oral and signed languages used: it is equally suitable for French or English, LSF or ASL. It merely transmits auditory and visual information without interacting with its content. This also makes it possible to consider **extending the system to the wider public of late-deafened and hard-of-hearing people who use oral interpreters**. These interpreters could also offer their services remotely, with the camera essentially transmitting their face and in particular the well-articulated movements of their mouth.
- on a larger scale and in the longer term, the direct, everyday use of videoconferencing between 2 signing deaf people can be envisaged.

## *5.2 Videoconferencing*

In this section we give a brief overview of videoconferencing technology.

### 5.2.1 General Presentation

Videoconferencing allows people who are far apart to communicate. It transports voice and moving colour images in real time. At present, videoconferencing is mainly used:

- by companies, to avoid costly and sometimes dangerous travel (post-September 11 syndrome, war in Iraq, SARS, etc.). This justifies major investments in sophisticated videoconferencing rooms
- by individuals on the Internet, who can thus communicate with friends or family members who are sometimes very far away, at very competitive prices.

Note that for most hearing people, the image does not necessarily add anything compared with the telephone. It constrains the behaviour of the speaker, who must stay close to the camera, whereas most telephones are cordless.

In addition, the camera is also an intrusion into the home. Internet users have overcome this inhibition: they think nothing of "multi-tasking" by checking their mail or surfing during the conversation.

Technically, videoconferencing takes place in a context of digital transmission: unlike the telephone, transmitting analogue video on a large scale is not feasible. Compression techniques are needed. Any videoconferencing system will therefore require, at each site:

- one or more video cameras
- one or more microphones
- a codec (COder-DECoder), which performs several functions: conversion of analogue video and audio signals into digital signals; compression and bit-rate reduction of these digital video and audio signals; multiplexing of these signals with various remote control and signalling signals; electrical interface with the digital link and call management; decompression of the received digital signals; analogue conversion of the received digital signals

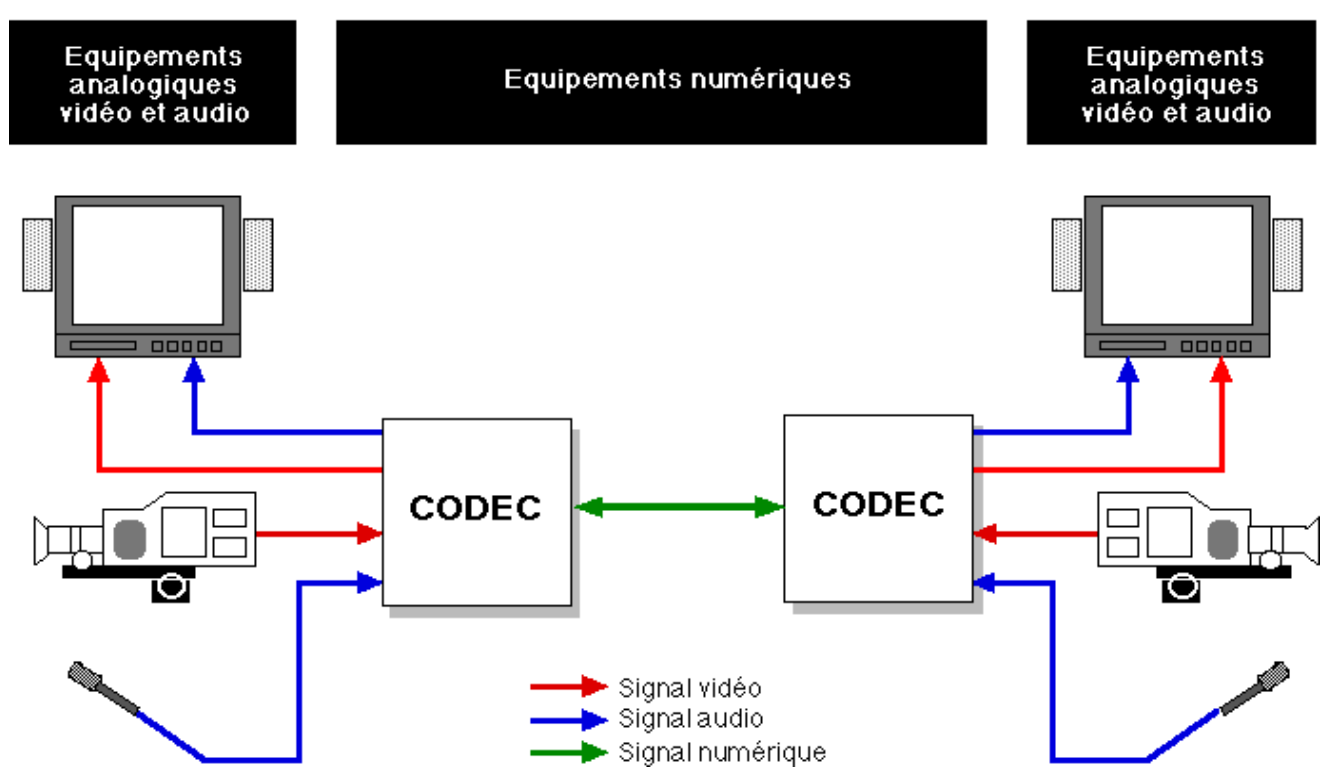

*Figure 31. Principle of videoconferencing.*

The setup can be supplemented with collaborative work tools such as instant messaging, a whiteboard, application sharing and synchronous file transfer.

There may be 2 or more sites. For more than two sites, this is called multipoint videoconferencing. It is more complex to implement and requires a "bridge" to exchange video streams between all sites. In this work, we will not need to consider this case and will restrict ourselves to point-to-point videoconferencing.

### 5.2.2 Digital Networks

Videoconferencing works over digital transmission networks. Several types of network can be considered:

- ISDN, a circuit-switched network mainly offering 2 communication channels (64Kbps) that can carry voice or data. They can be combined into a single data channel with double the bit rate (128Kbps).

- the Internet, a packet-based network. Various connection modes are possible: very low speed ("56K" modem over a telephone line) or high speed (cable, ADSL)

Two important notions related to network capabilities will be needed in what follows:

- the notion of **bandwidth**: this is the maximum information rate available for communication. It is measured in bits per second (Kbps: kilobits per second, or Mbps: megabits per second). A distinction must be made here between **downstream** (towards oneself) **and upstream bandwidth** (towards the network), as they are generally not symmetrical: for example, downstream bandwidth is larger for ordinary surfers, whose main activity is consuming information from the Internet. It is more symmetrical for companies that offer services.
- the notion of **quality of service (QoS)**: this is the minimum guaranteed information rate. This notion is necessary for good-quality real-time transmission and involves notions of priority and resource reservation. It is present in "switched" networks, and ISDN in particular. On the Internet, there is currently no strong guarantee.

For videoconferencing:

- **symmetry is desirable**: for example, the availability of huge downstream bandwidth on ADSL is not very useful if the other party only has a small upstream bandwidth and is therefore very limited in the video quality they can offer.
- **quality of service is desirable**: without it, the link is likely to regularly suffer significant drops in quality and interruptions, which will undermine the use of the system and lead to its rejection.

Here is a table showing these characteristics for the main solutions for individuals and companies:

| **Network** | **Bit rate** | **QoS** | **Cost (indicative)** |
|---|---|---|---|
| **Private use** | | | |
| Modem on phone line | 56 Kbps (not duplex) | No guarantee | Free (local call) |
| ISDN | 2x64 Kbps | Guaranteed | €30/month |
| Cable | Up to 4 Mbps down<br>128 to 256 Kbps up | No guarantee | €35/month |
| ADSL | 1 to 3 Mbps down<br>128 to 192 Kbps up | No guarantee | €35/month |
| **Business use** | | | |
| Multiple ISDN | 3x128 Kbps | Guaranteed | 3x€30/month |
| ADSL | 256 to 512 Kbps up | 128 to 256 Kbps "guaranteed" by ISP | €150-500/month |
| SDSL | 1 to 2 Mbps up and down | 256 to 512 Kbps "guaranteed" by ISP | ?? |

*Table 3. Communication infrastructure for videoconferencing.*

### 5.2.3 Videoconferencing Standards

Videoconferencing is standardised by the ITU through the following recommendations:

- H.320 for ISDN (and more broadly for networks with quality of service up to 2 Mbps)
- H.323 for the Internet at reasonable bit rates
- H.324 for very low-speed Internet networks (56K modem over a telephone line).

These recommendations mainly specify how the videoconference proceeds and refer to many other recommendations concerning audio coding (G.7xx series) and video coding (H.261, H.263). Collaborative work tools are standardised by the T.120 standard. These two standards are thus nicknamed "umbrellas" (see Figure 32) and guarantee the interoperability of equipment developed by the various manufacturers. This is crucial for the development of such a system. For greater flexibility, these standards provide a mandatory part and an optional part.

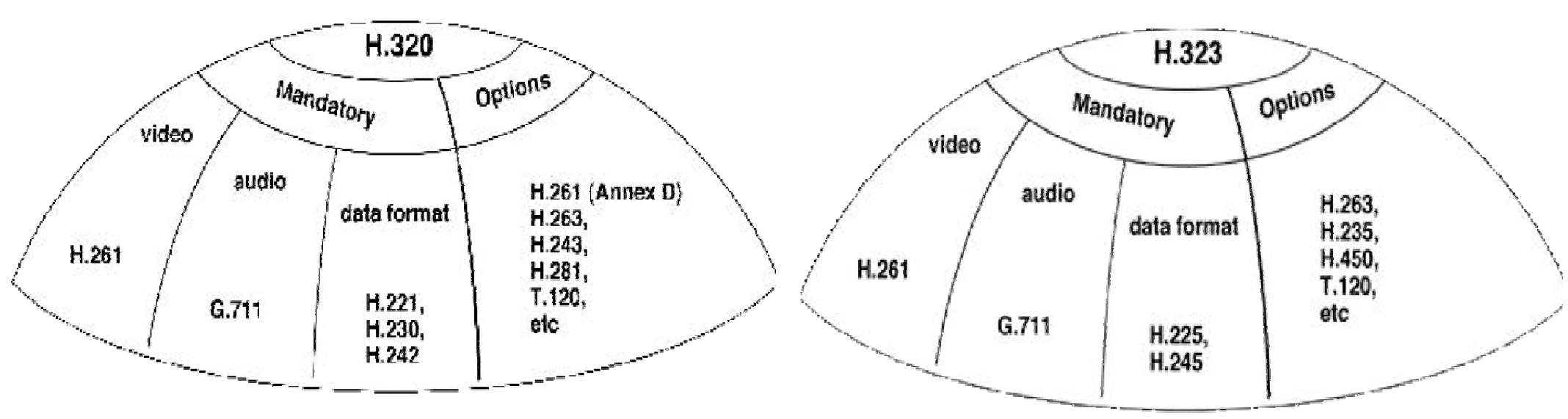

*Figure 32. "Umbrellas" of the H.320 and H.323 standards.*

**A. Audio quality (G.7xx)**

There are many audio codecs. They offer compression rates between 16 and 64 Kbps and cover voice frequencies from 3.1 KHz (normal telephony) to 7 KHz (extended quality). The basic codec is G.711, also known as PCM (Pulse Code Modulation); the others are optional. The audio codec is therefore usually negotiated dynamically when the connection is set up, depending on the codecs available on each side and on preferences.

**B. Video quality (H.261 and H.263)**

Table 4 shows the supported image formats. They are generally multiples and sub-multiples of the CIF format (352x288). They have a width/height ratio of 4/3. As an indication, CIF resolution is slightly below television resolution and 4CIF (704x576) corresponds to DVD-type resolution. QCIF (176x144) is the highest resolution that must be supported by the H.261 and H.263 codecs. CIF is optional.

H.261 applies spatial compression within an image (similar to JPEG images) and temporal compression between images (based on motion compensation). H.263 improves on H.261, notably motion prediction. These principles are the same as for the MPEG coding used in VCDs and DVDs, but one fundamental constraint is different: the system must cope with a variable bit rate.

| **Image format** | **Width** | **Height** | **H.261** | **H.263** | **Uncompressed bit rate (Mbps)** | | | |
|---|---|---|---|---|---|---|---|---|
| | | | | | 10 FPS | | 30 FPS | |
| | | | | | B/W | Col. | B/W | Col. |
| SQCIF | 128 | 96 | N/A | Yes | 1.0 | 1.5 | 3.0 | 4.4 |
| QCIF | 176 | 144 | Yes | Yes | 2.0 | 3.0 | 6.1 | 9.1 |
| CIF | 352 | 288 | Opt. | Opt. | 8.1 | 12.2 | 24.3 | 36.5 |
| 4CIF | 704 | 576 | N/A | Opt. | 32.4 | 48.7 | 97.3 | 146.0 |
| 16CIF | 1408 | 1152 | N/A | Opt. | 129.8 | 194.6 | 389.3 | 583.9 |

*Table 4. Video qualities*

The compression rate is therefore not fixed but depends on the available bandwidth. It can range from 25 (very good quality) to 150 (average quality). Having selected the quality, size and number of frames per second (FPS), Table 4 can be used to calculate the required bandwidth. For example, for colour CIF at 20 FPS, an uncompressed bit rate of around 24Mbps is needed (by interpolation between 10 and 30 FPS). A 384Kbps connection will therefore require a compression of around 62.5, which is good quality on the scale given.

Note also that these codecs offer a trade-off between quality (image degradation due to compression) on the one hand and fluidity (number of frames per second) on the other. They therefore impose neither parameter, but degrade one and/or the other depending on the instantaneous bandwidth. For example, if the user has given priority to image quality, the codec will first sacrifice the number of frames per second if bandwidth decreases.

Note already that for interpretation, both parameters matter: quality is crucial because some elements are very small (fingers, gaze), and movement must also be rendered well. In this respect, a rate of 12 frames per second seems to be the minimum acceptable. All these parameters will be studied in more detail later in this chapter.

Finally, a new codec is in preparation: H.264, which follows on from H.261 and H.263. The expected gain is a doubling of compression for the same quality. This would reduce bandwidth requirements accordingly, or rather allow much better results with current bandwidths. However, it also requires much more computing power, which is not a problem with recent processors.

### 5.2.4 Videoconferencing Terminals

Various types of terminals are available:

- "set-top boxes" are devices placed on a television, which is used for display and sound. The device contains the camera, the microphone and the encoding and decoding unit.
- "all-in-one" units look like a telephone but with a screen and a built-in camera. Note that these are poorly suited to remote interpretation because the screen is small. They can usually also be connected to a television, but this brings us back to the previous solution.

- solutions based on personal computers. The terminal consists of a multimedia PC connected to the network, equipped with a good-quality webcam and videoconferencing software. This software will be described in more detail below.

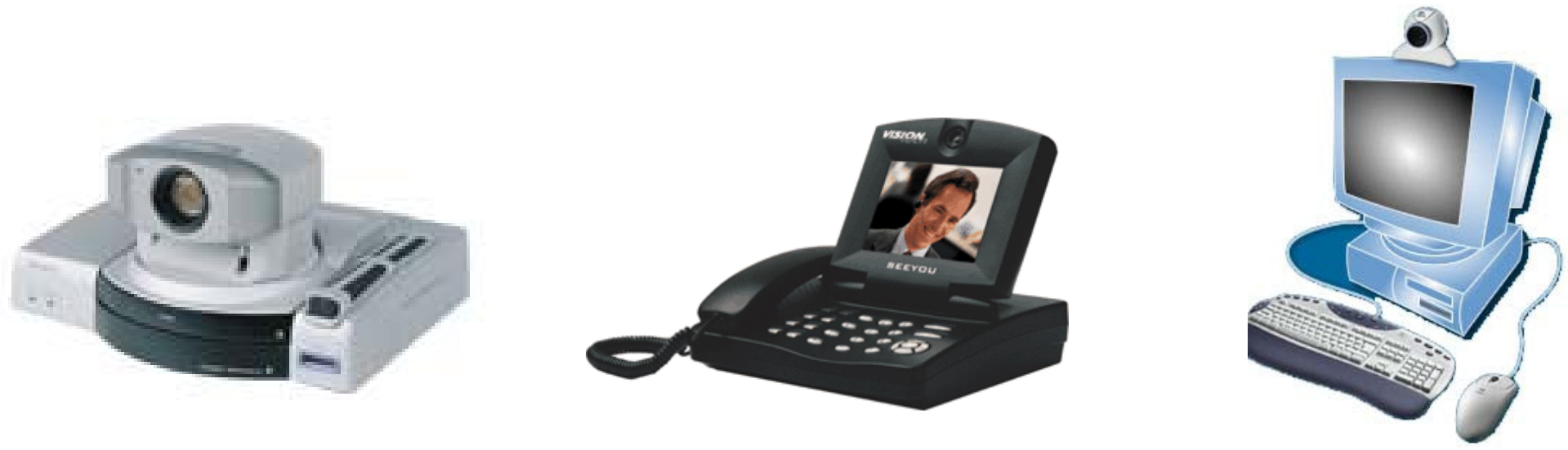

(a) (b) (c)

*Figure 33. Different types of videoconferencing terminals:*
*(a) set-top box (b) all-in-one (c) computer-based*

Most type (a) and (b) equipment is compatible with both H.320 and H.323. Some manufacturers also support proprietary protocols that require all parties to use their product. The main manufacturers are Aethra, Polycom, Sony and VCON.

We do not consider in more detail solutions based on the H.324 protocol (using plain telephone lines). Existing products cannot guarantee good-quality video over such a narrow bandwidth. As an example, we mention the recent "Beamer Personal Video System" (see Figure 34), which supports the H.324 standard. It offers QCIF resolution but only delivers 2 to 5 frames per second on average.

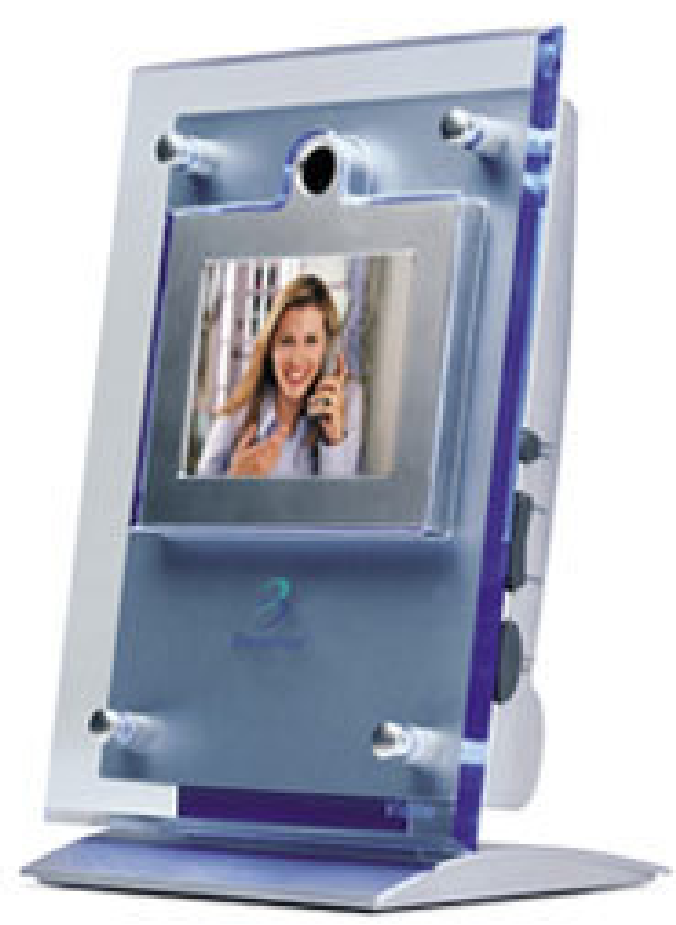

*Figure 34. Beamer PVS.*

As for prices, the cheapest solutions are based on personal computers (under €1,000); "set-top boxes" and "all-in-one" units are already available for around €2,000. Professional versions cost between €5,000 and €10,000 (or more) but include features related to multipoint conferencing and offer luxurious complete meeting-room installations with giant screens that are not needed here. To these costs must of course be added the connection costs mentioned earlier.

### 5.2.5 Videoconferencing Software

This software is intended for the personal computer solution. It is generally H.323-oriented, but some packages can work in H.320 with one or more ISDN cards. Most software works with standard hardware. Some take advantage of specific PC hardware for which they are the companion software (e.g. VCON Vigo). Most of these packages also support the collaborative functions of the T.120 standard: chat, shared whiteboard, file transfer and application sharing.

| **Logiciels** | **Platform** | **Features** | **Price** |
|---|---|---|---|
| **Génériques** | | | |
| NetMeeting<br>http://www.microsoft.com/windows/netmeeting/ | PC-windows | H323 | Supplied with Windows |
| CuSeeMe<br>http ://www.cuworld.com | PC-windows | H323 | Free entry-level version |
| Sorenson EnVision<br>http ://www.s-vision.com | PC-windows | H323, used for relay in the USA | ?? |
| VideoLink<br>http ://www.smithmicro.com/videolinkpro | PC-windows | H323 | 60$ |
| eConf (France Télécom)<br>http://www.econf-communication.com/ | PC-windows | H323, SDK available to adapt the interface for deaf people (e.g. Sweden) | ?? |
| Marvin (Envilogg, Suède)<br>http ://www.envilogg.com/ | | | |
| OpenPhone<br>http ://www.openh323.org | PC-windows<br>Linux | H323<br>Basic, more of a demo | Gratuit<br>(open-source) |
| GnomeMeeting<br>http://www.gnomemeeting.org/ | Linux | H323, based on OpenH323 | Free<br>(open-source) |
| **More specific** | | | |
| Allan eC (Omnitor, Suède)<br>http://www.omnitor.se | PC-windows | H323 | ?? |
| X2X Visual Meeting (VCON)<br>http://www.xenex.fi | PC-windows | H320, H323<br>(specific to Finland) | ?? |

*Table 5. Main videoconferencing software.*

Among the software packages, some are more specifically adapted to the needs of deaf people. We will examine them in more detail in the next section, in the context of existing online interpretation systems.

## 5.3 Existing Systems and External Evaluations

This section briefly reviews the main systems that exist around the world and the studies carried out to evaluate them. The aim here is not to be exhaustive but to give a representative view and to highlight the requirements related to the use of videoconferencing for sign language interpretation.

### 5.3.1 Existing Systems

We will examine the situation in the following countries:

- **Scandinavian countries and the USA**: they are further ahead in addressing the problem of remote interpretation for various reasons:
    - the problem of distance and accessibility has arisen in a more critical way (large countries, winter isolation, etc.)
    - social legislation is very advanced in the Scandinavian countries. Sign language is also already officially recognised there
    - the USA did not experience the ban on sign language
- **Closer countries and regions**: the Netherlands, France and Flanders. Here the aim is to look at developments among our neighbours, who are more or less at the same stage as we are.

In countries that already had a text relay system, the video system (relay and/or remote interpretation) was set up more easily. The following countries already had a text relay system:

- in the USA: it is in fact a legal obligation imposed on the states by the "Americans with Disabilities Act" (1990).
- in Europe: Denmark, Norway, Sweden, Finland, France, the Netherlands, Switzerland.

**In the USA**, many relay networks are active and several offer remote interpretation. They may be organisations linked to a US state, to a telephone operator, or a national organisation. Some examples:

- USAVRS network: http://www.usavrs.com
- CAN network: http://www.caninterpreters.com
- SignOnASL network: http://www.signonasl.com
- ATT operator: http://www.relaycall.com
- State of Wyoming: http://www.dawyo.org/VRI.htm

**In Europe**, we know that Sweden and Finland have set up a video interpretation system:

- in Sweden, the project started in 1995 and opened to the public in 1997. The service covers about sixty terminals in the country and is open Monday to Friday from 8 AM to 8 PM. It handles about 400 calls per month (1998 figures).
- in Finland, the system (see Figure 35) was deployed between 1998 and 2000 to 22 households and one workplace in the east and north of the country (a large, sparsely populated region). The system was financed by the equivalent of the national lottery (Finnish Slot Machine Association) and then taken over by the municipalities.

Closer to home, trials or projects are being studied:

- in France: EDF has set up a service for its deaf customers (eSourds: http://www.esourds.com). A major project is also in the preparatory phase: WebSourd (http://www.websourd.org). It is a portal designed entirely for deaf people. It will include access to all relevant information for these people and to services such as explanations of procedures through LSF videos and an online interpretation service. It is due to open at the end of 2003.

- in the Netherlands: experiments have been taking place in the Netherlands for more than a year as part of the "BeeldBellen" project (http://www.beeldbellen.info)
- in the United Kingdom, there is no system in operation, but trials have been carried out, notably by the "Centre for Deaf Studies" in Bristol as part of the "SignWorks" project, and are currently continuing as part of the European Wisdom project. We will detail them below.

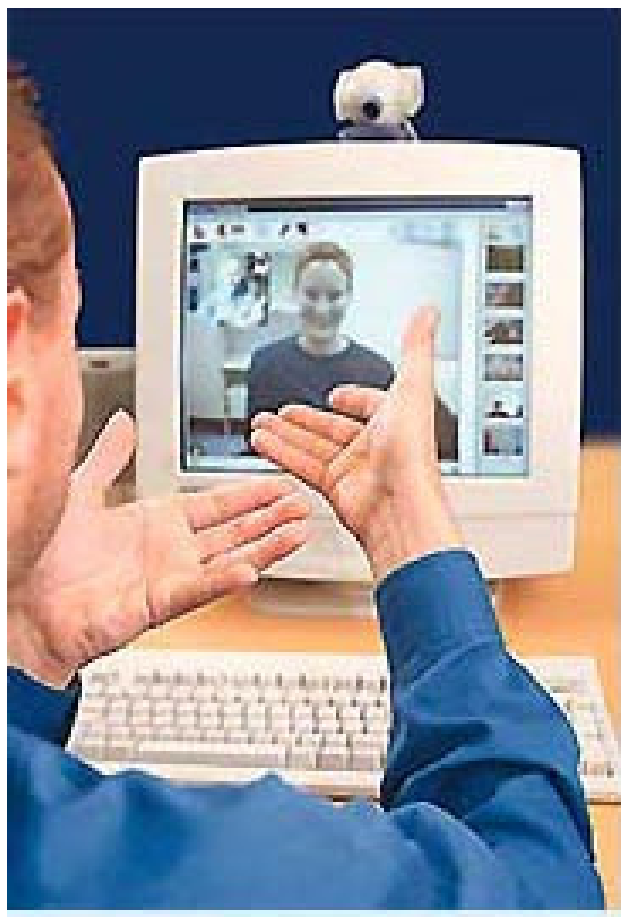

*Figure 35. The Finnish system.*

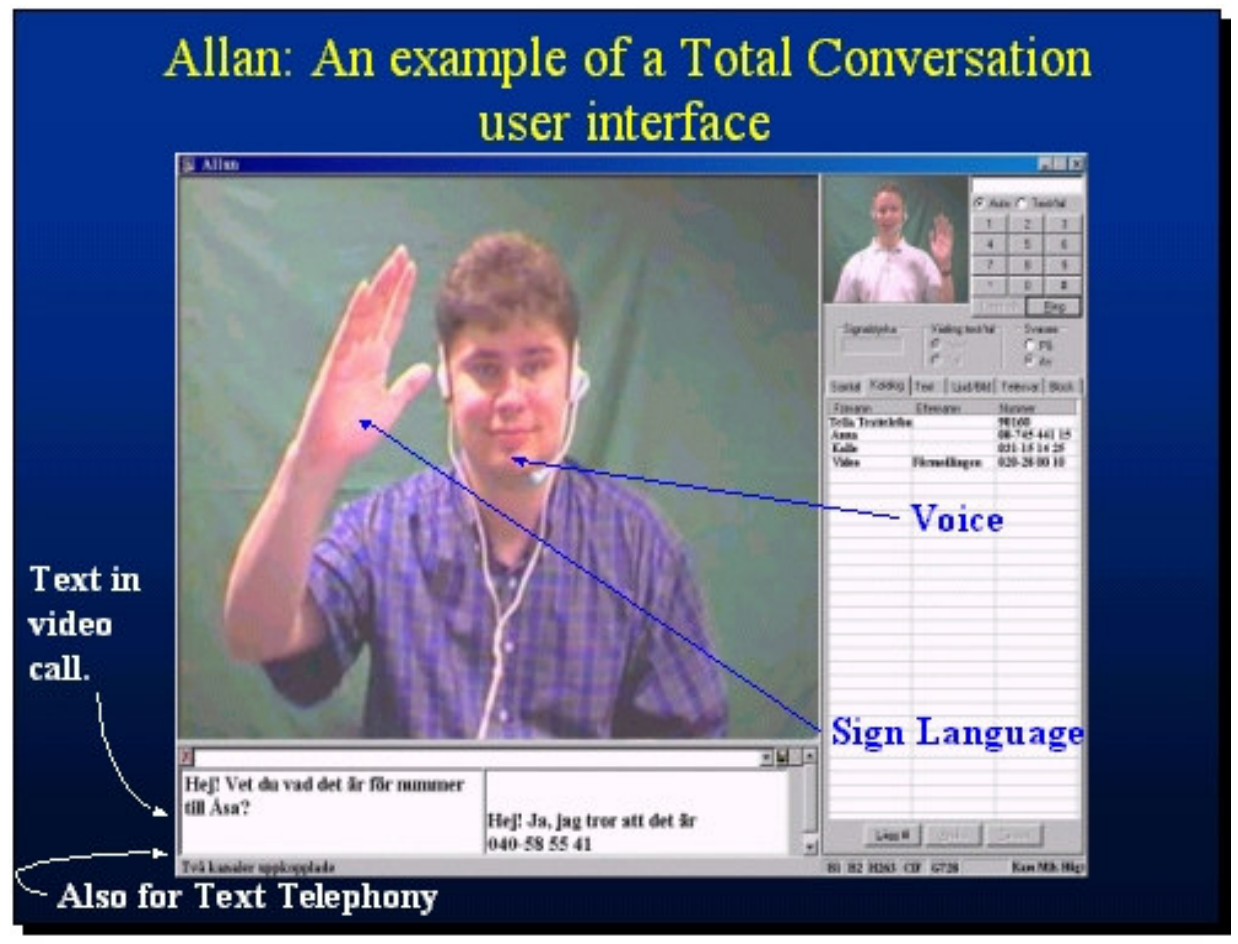

*Figure 36. Omnitor Allan eC (Sweden).*

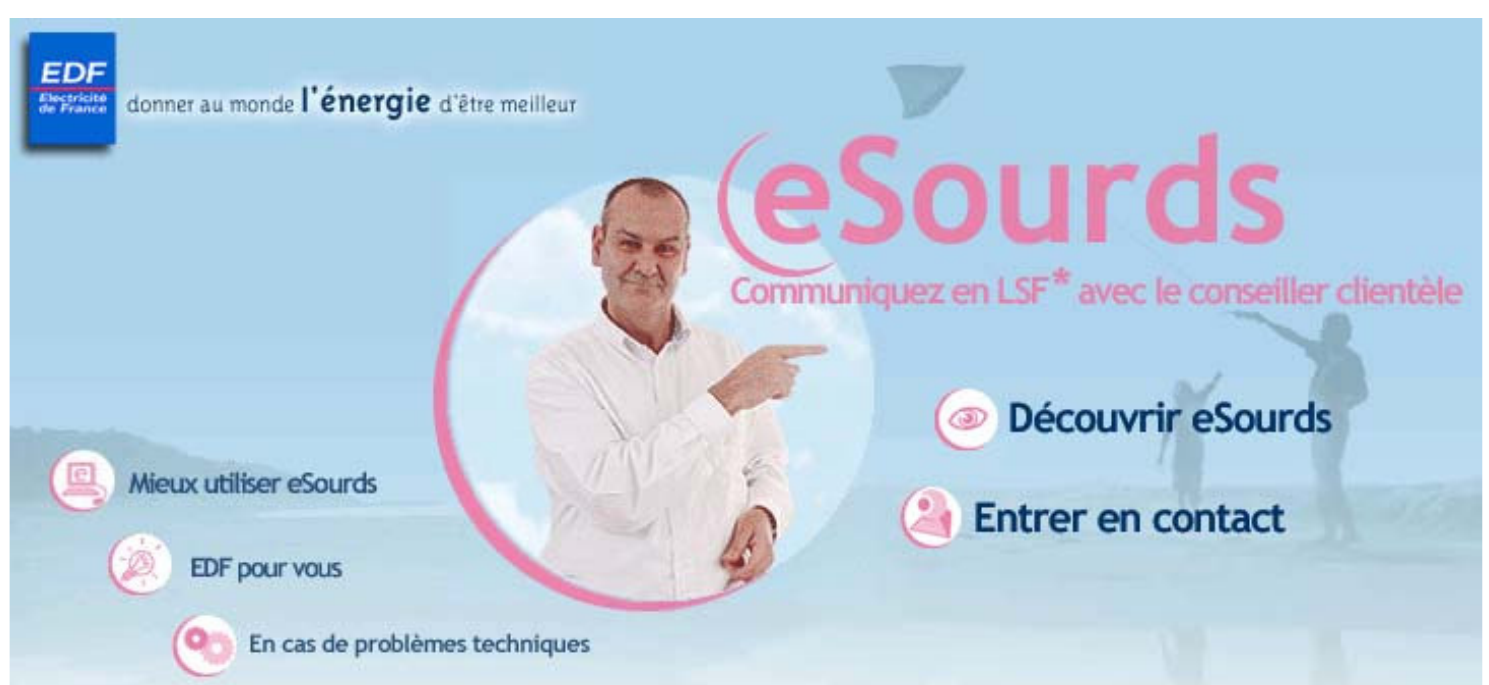

*Figure 37. eSourds, customer service for EDF's deaf subscribers.*

**Technically**, we find the various solutions presented in the section on videoconferencing:

- Triple ISDN, the only one that guarantees quality. It allows work in very good conditions but is expensive to implement. Setting up the triple connection can also take a long time and cause problems. Finland opted for this system.

- Single ISDN (2x64 kbps), considered acceptable but tiring: fast signs are jerky and some details (fingers, direction of gaze) are sometimes too blurred to be clearly distinguished. A comparative video example is provided on the companion CD-ROM.
- Broadband Internet, whose quality of service depends largely on the connection set up. Quality is generally close to that of single ISDN. This is the case in France for eSourds, which is based on NetMeeting with small "QCIF" videos within web pages.

- A mixed system accepting several systems. This is the case of the Dutch "BeeldBellen" project.

Software specific to deaf people has an interface characterised by:

- a concern for simplification:
    - omission of unnecessary features (complex directory, etc.)
    - integration within a web page (usually based on NetMeeting) with simplified controls (a single "call the centre" button)

    - or else: operation in full-screen mode, hiding any other applications on the platform to imitate a real videophone.

- a concern to integrate text-based means of communication (included in the T.120 standard) and to make them systematically present in the interface (see Allan eC above)

As for areas of use, it is not always only remote interpretation but sometimes also direct contact between deaf people (Finland). Contacts with public services are always mentioned.

As for applications in the medical field, where we saw that there were conflicting goals (urgency vs accuracy vs emotional aspect), we identified several systems in this context:

- a Canadian evaluation project at McGill University: http ://ww2.mcgill.ca/icc/canarie/signLanguage/
- an American system (University of Rochester), which has become operational: http://www.urmc.rochester.edu/strongconnections/
- an English system at Leicester University Hospital, which has two videoconferencing links with interpreters located at the Leicester deaf centre. They serve the emergency department and the cardiac care unit.

We will not go further into this aspect or the other application areas that are outside the strict scope of our study. For a review of the needs and of the various technologies that can be used by deaf people, see [EU00].

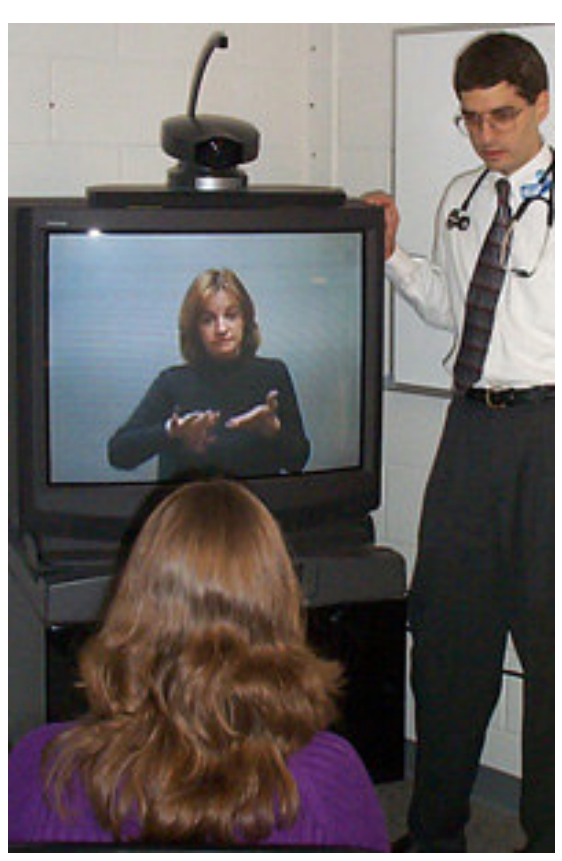

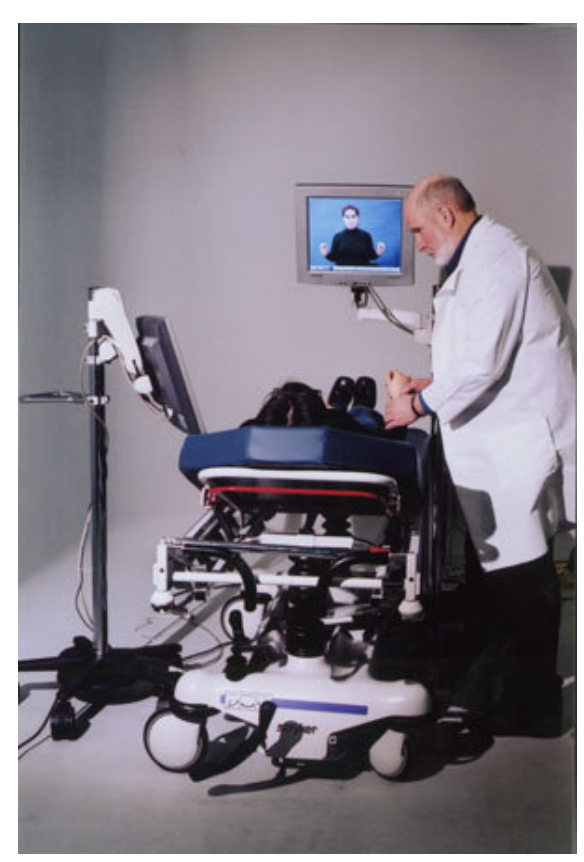

*Figure 38. Use of remote interpretation in a hospital setting.*

### 5.3.2 External Evaluations

**A. ACE study (Australia)**

The remote interpretation system was tested with 19 deaf participants and 6 Auslan interpreters, in the presence of 26 hearing people representing service providers. The main lessons are given here; for more details, see the report [Mcc99].

All participants said they were in favour of the remote interpretation system, in particular because it offers **a more natural interaction than the text relay service, especially in expressing emotions**.

Compared with a video relay system, participants also expressed a preference **for the physical presence of the service provider**. This presence allows better control of the pace of the conversation. This preference also reflects the importance of human contact, particularly in medical interventions. Nevertheless, participants recognised the **usefulness of video relay in remote areas where service providers are absent.**

An important recommendation of this study is to establish a **set of protocols** ensuring effective three-way interaction:

- a protocol for controlling the pace of the conversation;
- protocols on the interpreter's side to manage the conversation in different circumstances, for example according to the client's level;
- a protocol to determine the best seating arrangement and thus ensure the best visual contact;
- protocols to determine the best physical environments and clothing
- protocols to introduce participants to the concept of remote interpretation before starting the conversation; this includes a warm-up period allowing the deaf person and the interpreter to get to know each other's style, a protocol for replacing the interpreter if one of the two parties does not wish to continue the conversation, and guidance to lead the service provider towards an effective conversation

**B. "SignWorks" evaluation, Bristol (UK)**

The Centre for Deaf Studies in Bristol has carried out several studies [Bla98a], [Bla98b], [Bur98] on the use of videophones and remote interpretation in British Sign Language (BSL). These studies were mainly part of the "SignWorks" project, which aims to promote employment for deaf people. The main lessons learned are as follows:

- Conversation in sign language is possible but requires an effort to adapt, in particular regarding the speed at which signs are produced. Deaf people therefore sign more slowly and take on average 50% more time (for example 1 min 30 instead of 1 min)
- Fingerspelling is problematic: avoid the small screens of "all-in-one" videophones and prefer larger screens.

- Acceptance is good: at the end of the scripted tests, a free conversation started about criticism of the system itself, and it was carried out using the system.

- Three configurations between interpreter and hearing person were devised: with, on one side, respectively the deaf person alone, the hearing person alone and the interpreter alone, and on the other side the two other people. The most popular configuration was the hearing and deaf person together with the interpreter remote. This is the configuration we consider in this study.

**C. Evaluation by the State of Pennsylvania (USA)**

The evaluation confirms that traditional telephone links are inadequate, that a single ISDN line is acceptable, but that a triple line provides better comfort. Implementation is envisaged in particular in the medical field.

http ://www.cermusa.francis.edu/prototype/telemed/interpret/

**D. Omnitor evaluation (Sweden)**

The Swedish company Omnitor has established precise measurements of quality requirements for video communication in sign language [Hel97], based on measurements and on ITU recommendations [H.Supp-1]. The main conclusions are as follows:

- Regarding **spatial resolution**, taking into account the framing (medium shot) and the size of the finest elements (fingers and eyes) implies a CIF-type resolution (352x288). A QCIF resolution (176x144) still makes it possible to distinguish the fingers but not the direction of gaze; part of the message may be lost.
- Regarding **temporal resolution**, the fastest signs are those of fingerspelling (sometimes 7 signs per second). Considering that 3 frames are needed to perceive a sign, a rate of 21 frames per second is necessary. Lower rates are possible but cause signers to slow down and lead to some frustration: 12 frames per second is the minimum.

- Regarding **movement**: avoiding blur during movement is necessary to increase the chances of perceiving handshapes. Here, a good camera and good lighting are important factors.

- Regarding **delays**: below 0.5 seconds, the delay is not really noticeable; beyond 1.2 seconds, it becomes irritating.

## *5.4 Requirements for the Design of a Remote Interpretation System*

Based on the general requirements described in chapter 3 and on those highlighted in the previous section, we will now propose a more precise design of the remote interpretation system. Here we describe the requirements for the various components of the system: the organisation of the public service in terms of reception procedure and infrastructure, the organisation of the interpretation service, and the videoconferencing infrastructure. These requirements are sometimes qualified in terms of minimum required level and recommended level.

### 5.4.1 Preparing the Interaction at the Public Service

**A. Welcoming the deaf person**

Currently**,** a deaf person going to the place of the interaction first meets the interpreter in the waiting room, and communication is established between them: greetings, exchange of news and a more precise understanding of the procedure to be carried out. This welcoming step is essential for the deaf person, who is certainly reassured by meeting someone who speaks "his or her language".

**In the new system,** it is important to provide a suitable welcome for the deaf person. Among the various means that can be used:

- the use of "passe-partout" icons indicating that deaf people are catered for. The level could be equivalent to a physical interpreter available on request (level 8).

- staff training, in particular on how to behave with deaf people and some basics of sign language
- the provision of visual documents: brochures, videos, etc. explaining how the system works if the deaf person is using it for the first time
- it is also important that the same organisation is found in the various services, so that the deaf person immediately feels on familiar ground if he or she has already used the system before

**B. Familiar interpreter**

As the world of deaf people and interpreters is small, the deaf person may already have dealt with the same interpreter. When dispatching, an attempt is made to take this into account to some extent (especially if there are explicit requests: scheduling is already complex without this additional constraint). This remains valid for the remote interpretation system.

**C. Preliminary exchange**

During this preliminary exchange, communication between the 2 parties adapts, mainly on the interpreter's side, who assesses the deaf person's regional origin, general level of education and command of sign language. In the new system, it is important that this key step is preserved through a preliminary private period between the deaf person and the interpreter.

### 5.4.2 Organisation of the Premises in the Public Service

Many interactions in a public service take place at a counter, which is not a very suitable place for videoconferencing:

- **lighting** and **ambient noise conditions** are not very good for transmitting and receiving audio and video
- the deaf person may use residual hearing and lip reading: a **glass screen** is an obstacle both to the effective transmission of the employee's speech and sometimes to seeing the employee well enough (thickness, cleanliness, frosted areas, etc.)
- the **layout of the counter**: a counter is a face-to-face configuration for two people, while interpretation involves a third person, who usually sits next to the deaf person, which is sometimes already problematic with normal counters. For a remote interpretation system, the equipment must be placed appropriately: it requires visual contact with the deaf person but also sometimes with the hearing person (or with documents he or she holds). A triangular configuration around a table is more appropriate.

- a **more private location** is preferable, to avoid the feeling of being on display and for greater confidentiality

A suitable place for videoconferencing should therefore be provided, with the following characteristics:

- good lighting
- good sound insulation from the outside (closed room reserved for videoconferencing for the duration of the session)
- triangular configuration in which the deaf person, the hearing person and the videoconferencing system each occupy one corner.
- seated position around a table with any documents

Note that a specific room does not need to be reserved for this sole purpose: many services have a multimedia room with a television that can be used with a few adjustments.

### 5.4.3 Organisation of the Interpretation Centre

In principle, the geographical location of the centre is not very important. It should preferably be located centrally with respect to the interpreters and have good telecommunications infrastructure. The service can, however, be part of a larger organisation:

- a general interpretation centre such as the CIRE
- linked to services more specific to deaf people (for example, SISB/Info-Sourd/ SARE/FFSB/APEDAF are already grouped in the same building, where interpreters are permanently present)

As with traditional interpretation, a scheduling system is needed. This system can be integrated into the existing one if the system is set up within an existing interpretation service that shares field interpreters and one or more fixed interpreters for remote interpretation. Depending on demand, the number of simultaneous fixed interpreters can be determined.

**The initial system** can consist of a single station and evolve according to demand. If there are several stations, the system will nevertheless offer a single external access identifier, with calls being redirected to a free station. Despite the scheduling system, stations could still become saturated (delays, unexpected calls, etc.): the call can then be put on hold and/or redirected to a special station. For example, a station could be reserved for the secretary responsible for scheduling, who could use it to make appointments in addition to other means (SMS, fax, telephone, etc.)

**A possible extension** is to set up an **emergency** interpretation system. An incoming call must be signalled so that it is not missed (on-screen alert, external audible and visual alert). The call has priority over other calls if no station is free. A special station may also be provided. If the service operates around the clock, an on-call system will have to be ensured. The design of such a system must be examined in more detail with potential partners (100 service, hospitals, etc.). It is outside the scope of this study.

### 5.4.4 Communication Quality

Remote interpretation inevitably degrades communication, which must be minimised, because any additional noise in the exchange impairs "listening" comfort and, in particular, makes the many tasks the interpreter must perform simultaneously more difficult and speeds up fatigue. We will review a few factors.

#### A. Communication infrastructure

This point was discussed at length at the beginning of this chapter. The recommended system for sufficient bandwidth and quality of service remains the triple ISDN link. That is the price of the interpreters' comfort.

Solutions based on broadband Internet should be examined and can be considered if access providers undertake to guarantee bandwidths above 200 Kbps.

#### B. Use of the camera

An interpreter who is physically present perceives signs in three dimensions and from an angle she chooses as the most suitable (choice of seat at the start of the session). The interpreter also follows the parties with her eyes. The captured image is two-dimensional, from a fixed angle, and covers only a defined area.

For the interaction, **a fixed camera** seems sufficient, but **care must be taken in choosing the viewing angle and framing**. These will be determined by both parties at the start of the session.

**Regarding framing**, on the interpreter's side, the classic framing for sign language is used: the medium shot focusing on the upper body above the waist. On the side of the deaf person and the employee, the question arises whether or not to show the employee (so that everyone sees everyone). The employee's image is not essential but can add something. The options are:

- simultaneous display of the deaf and the hearing person. In this case, some resolution is lost and the parties must sit physically closer together. This is not always possible, for example for a job interview.
- successive display depending on who is speaking. Implementation is then more complex: it requires either an external operator to handle framing, or a two-camera system with a push button to switch the display from one person to the other. The images can be combined so that the person who is not speaking is shown in an inset.

Framing must also be maintained. A seated position does not allow major disruptions in this respect, unless the procedure requires leaving the videoconferencing table at certain times. The simplest system is to display a local control image as an inset, with each participant making sure to stay correctly framed. This task can possibly be entrusted to an external human operator or to an automatic system ("intelligent" motorised camera).

**Regarding the viewing angle**, a slightly oblique view is recommended for better observation of the signs, whereas a frontal view makes it possible to simulate eye contact.

**Regarding the two-dimensional aspect**, the (very expensive) Teleportec system makes it possible to simulate a three-dimensional physical presence (http://www.teleportec.com/)

**C. Image**

The image is transmitted with a certain quality in terms of refresh rate and resolution. The image must be fluid in order to render correctly the movements that make up signs: no blur (often linked to the camera sensor) and no jerkiness (linked to slowdowns of the link). Of the meaningful elements, the hardest to transmit appear to be the subtleties of facial expressions and handshapes. See here the quality measurements carried out notably in [Hel97] and the requirements derived from them:

- minimum temporal resolution of 12 frames per second, 25 recommended
- minimum spatial resolution QCIF (176x144), CIF (352x288) recommended
- maximum delay 1.2 s; a delay of less than 0.5 s is recommended

Videoconferencing gives its best results with very good lighting and a plain background. Good contrast must also be ensured so that the interpreter's silhouette stands out clearly. Interpreters often choose plain, dark clothing to make their hands stand out.

**D. Screen size and type**

The image received must replace the interpreter as well as possible. **A good-sized image** is therefore essential, especially on the deaf person's side:

- a small LCD screen (as offered by "all-in-one" videophones) is insufficient.

- a good-sized LCD screen (17") offers a very good, very restful image
- a large television is a good technical solution (its displayed resolution is comparable to the CIF format) and an economical one. In addition, it may already be available. Compared with an LCD screen, a TV image is more tiring, in particular because of flicker and because one sits closer than usual. A 100Hz television is better in this respect. It should be avoided on the interpreter's side.
- a video projector offers a very large image but is expensive (>€2,500) and requires subdued lighting incompatible with the camera.
- let us also recall Teleportec's "3D" system (http://www.teleportec.com/), which is prohibitively expensive.

**E. Audio transmission**

One element not to be overlooked is correct audio transmission, which is essential for communication with the hearing person. Audio is handled synchronously with the image by the videoconferencing software. For capturing and playing audio, two solutions are possible: closed (directional microphone and headset) or open (free microphone, loudspeakers). Technically, the latter can cause echo problems and possibly feedback (Larsen effect).

At the public service, the open solution seems preferable: it is less restrictive for the employee (no headset) and is essential if several hearing people are involved or if the deaf person has residual hearing.

On the interpreter's side, a headset microphone seems preferable for good listening comfort and possible isolation from the working environment.

**F. Additional means of communication**

A system offering additional means of communication should be used:

- **text chat** can be useful for spelling certain names
- a **whiteboard** for drawing a map visible to the interpreter
- **file transfer** so that the interpreter can study a form closely if an electronic version is available

These are optional in the videoconferencing standards but are offered by most software. The interoperability of the functions offered should be checked.

A system offering a user-friendly "total communication" interface (for example "Allan eC") is recommended over mainstream videoconferencing solutions designed for hearing people

For displaying documents, a special camera is recommended. Some videoconferencing systems include one as standard.

## 5.5 Validation Tests

In order to validate the design detailed in the previous section, we carried out a series of validation tests. They placed deaf people, interpreters and hearing people in typical communication situations that might be encountered in real administrations. They were carried out on a prototype for which we controlled all the parameters.

The tests were carried out at the end of May 2003 at the FFSB premises, with the participation of deaf people recruited by the federation and interpreters from both SISW and SISB. Several external observers were invited.

A secondary objective was also to familiarise the participants with the possibilities and limits of the videoconferencing concept.

### 5.5.1 Hardware Infrastructure Used

The tests were carried out under controlled conditions on a private high-speed network (100 Mbps). It was artificially throttled to simulate different commercially available bandwidths (see Figure 39).

The stations were based on laptop PCs equipped with very high-quality cameras (more than 1 million pixels, a resolution far higher than those used in videoconferencing). The advantage of the laptop is that it can simulate the other types of terminal ("set-top box" and "all-in-one") through its LCD screen and TV output.

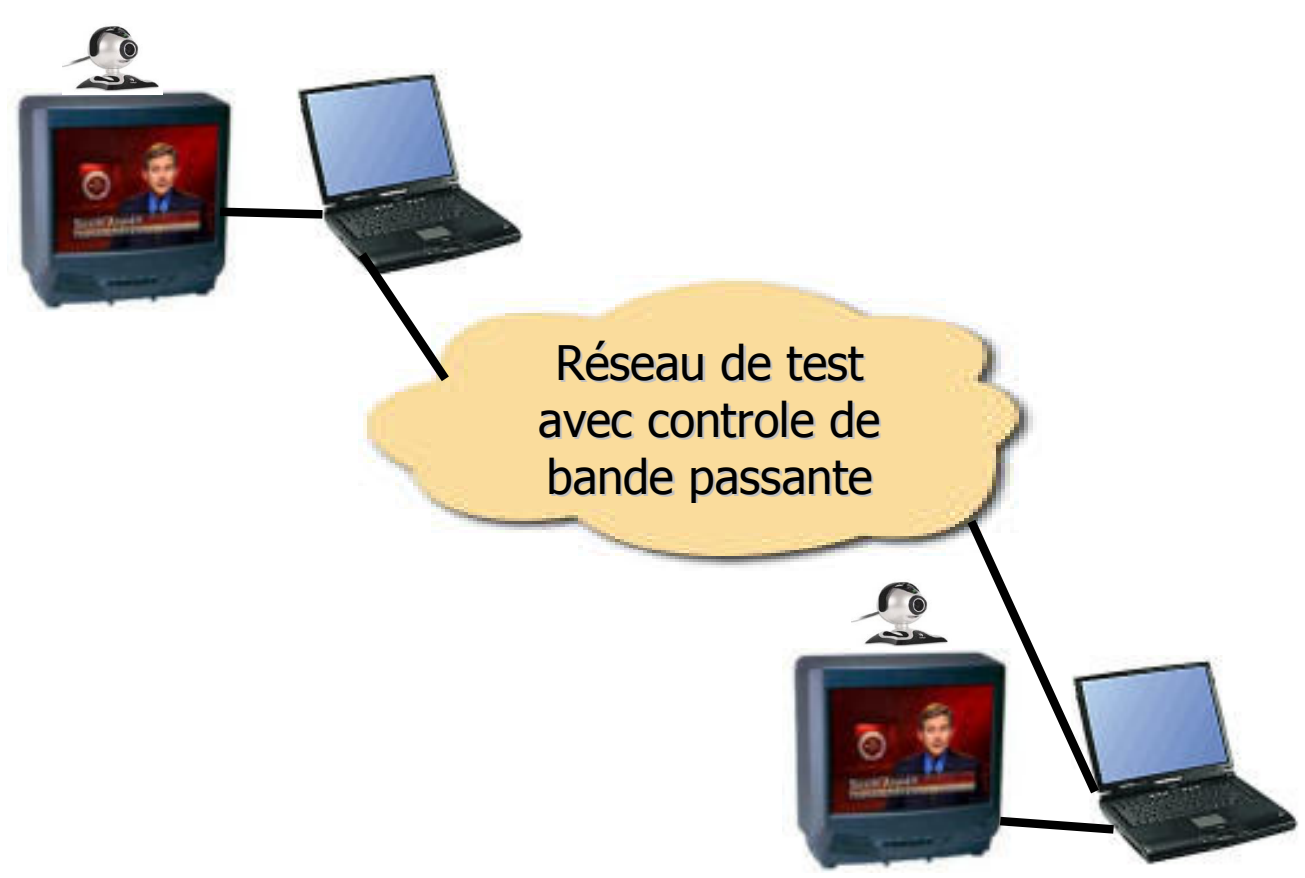

*Figure 39. Prototype used for the tests.*

The interpreter's station (see Figure 40) was located in the FFSB library. No plain background was provided, and natural and neon lighting was supplemented by a 100 W spotlight providing direct lighting. The station was equipped with:

- a Fuji FinePix 2600Z digital camera in webcam mode
- an LCD screen (14 inches)
- a semi-professional headset microphone

The station at the service (see Figure 41) was located in the FFSB meeting room, with a white wall as background. Natural and neon lighting was supplemented by a halogen floor lamp providing indirect lighting. The station was equipped with:

- a high-end Logitech Quickcam 4000 webcam with built-in microphone

- an LCD screen (14 inches) and a television (50 cm)
- additional external PC-type loudspeakers

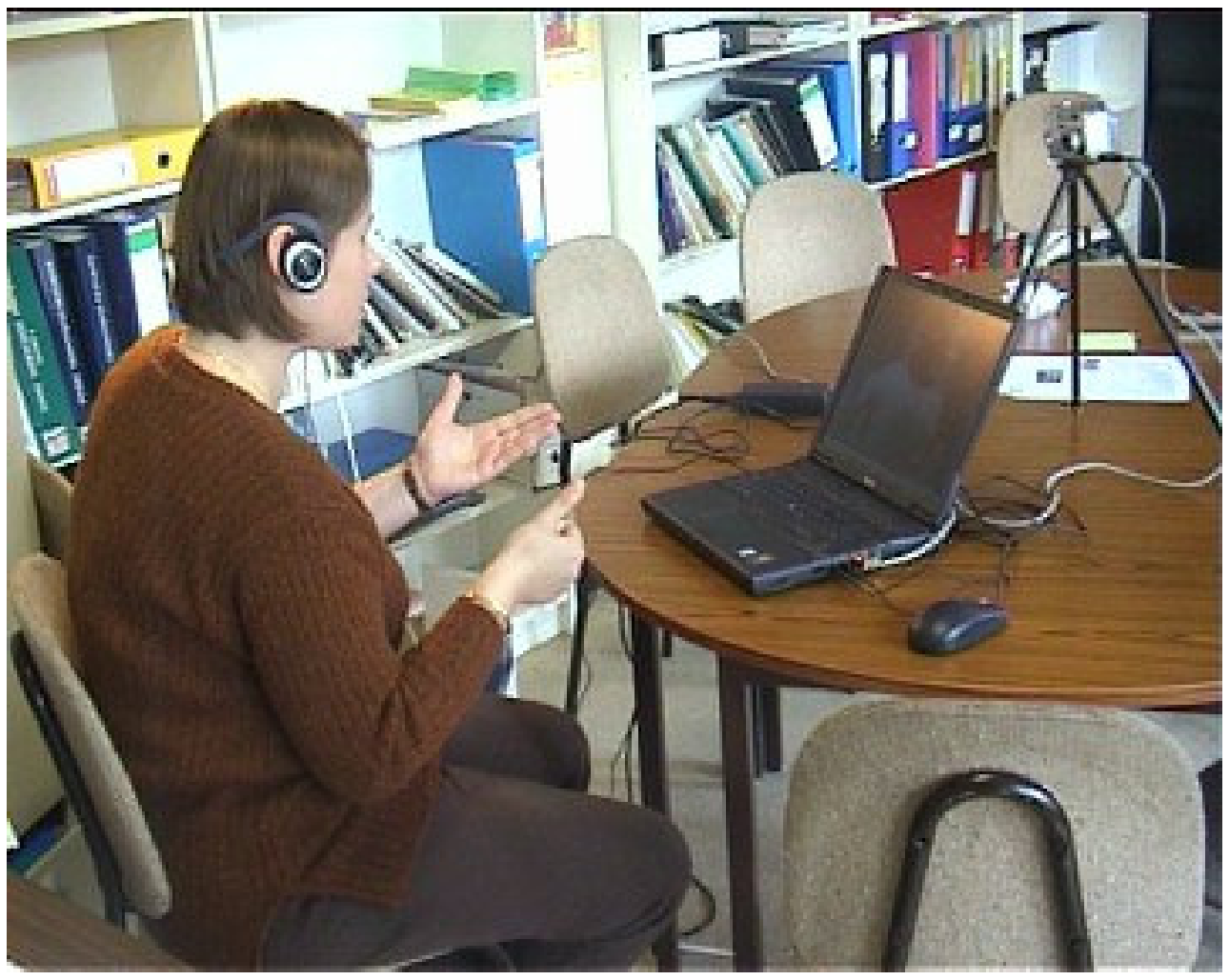

*Figure 40. Interpreter's workstation.*

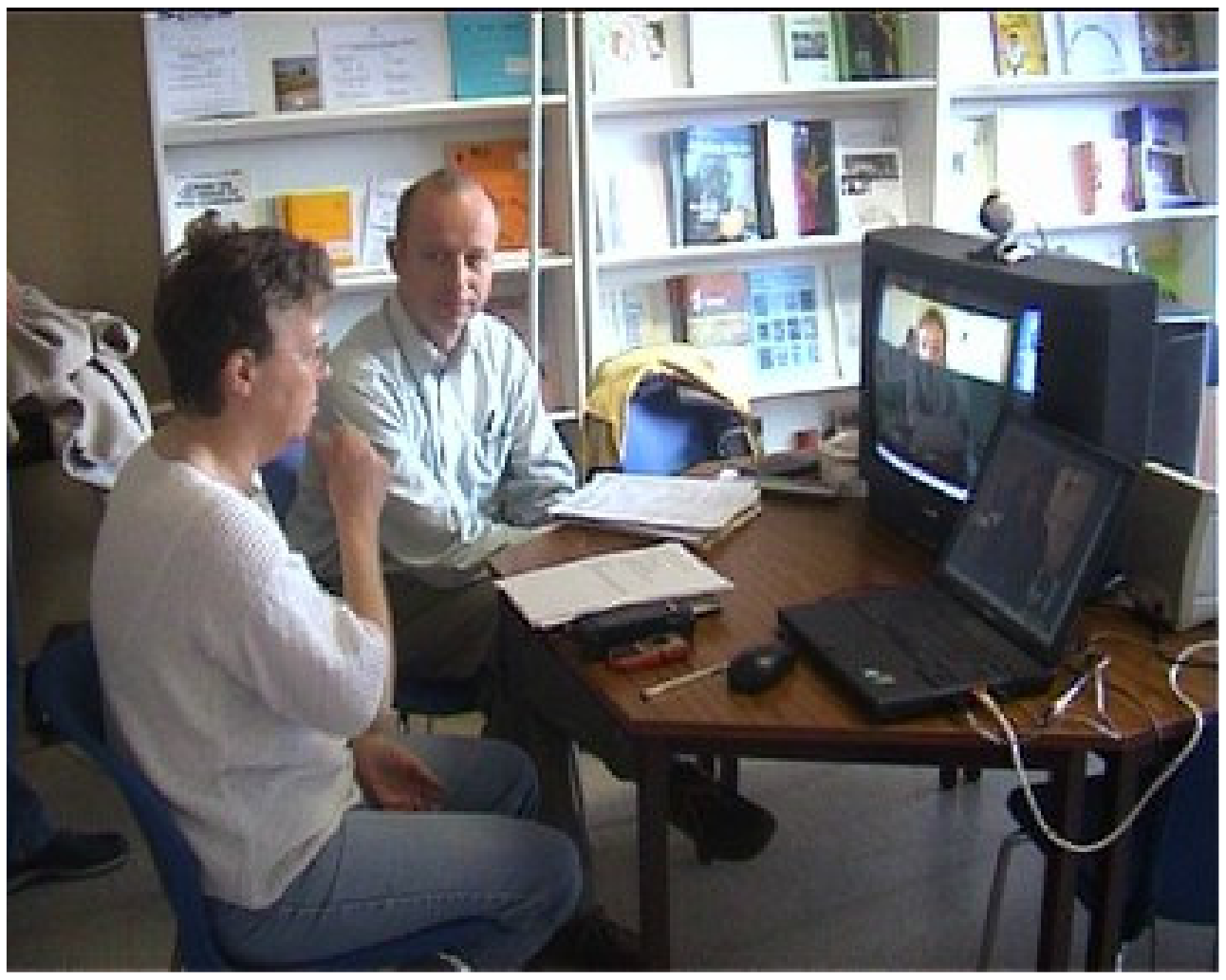

*Figure 41. Workstation in the public administration.*

The videoconferencing software chosen was France Telecom's "eConf". It was selected for its good performance, user-friendly interface and full-screen mode with inset. The software is not specific to deaf people but has a development kit for its interface, which is used by a Swedish firm to adapt the software for deaf people. Collaborative tools are available but are not enabled by default. We did not test them in depth.

The bandwidth throttling software used was "bandwidth controller" (available from http://bandwidthcontroller.com/). It makes it possible to limit upstream and downstream bandwidth and to monitor their instantaneous use.

Initially, the tests were to include a part on the real network, using the "cable" and ADSL connections available at the FFSB. However, they were not carried out because of technical installation problems and lack of time.

### 5.5.2 Test Scenarios

Two scenarios were planned:

1. a simple form-based scenario: issuing a passport

2. an open-ended discussion scenario, set in the context of recruitment and developed with the help of SARE

**A. Scenario 1: passport application**

The planned scenario was as follows:

- welcome by a hearing person with minimal training
- contact with the interpreter and preliminary welcome
- arrival of the employee with the form
- occurrence and resolution of a problem (example: no identity photo, so a digital photo is taken on the spot)
- end of the procedure and new appointment (date when the passport will be available)

**B. Scenario 2: job interview**

The general context is a company wishing to hire a deaf person. Several types of job are possible. The deaf person is asked in particular about his or her skills (studies, experience, use of office software) and expectations. The problems related to deafness and the means of overcoming them in daily work are obviously discussed, in particular through the use of the remote interpretation system.

**C. Implementation**

These scenarios were carried out with two deaf people under the same conditions in order to draw the most representative lessons. The employee was always the same. The first scenario was carried out with the SISW interpreter and the second with the SISB interpreter.

As for the technical parameters, the first scenarios used parameters ensuring excellent quality but unrealistic in terms of bandwidth. In the second part of the test, they were gradually reduced in order to determine the acceptability thresholds. For quality degradation, the trade-off between fluidity and resolution gave both parameters equal priority.

The stations were not quite identical (particularly regarding audio and TV display). Towards the end of the test, participants swapped places in order to give each party the opportunity to try each station and to confirm the differences in perception.

The main moments of the tests were filmed with a digital camcorder so that they could be studied. An edited version is available on the companion CD-ROM.

### 5.5.3 Participants

**A. Deaf people**

Martine Fraiture (FFSB)
Thierry Adnet (FFSB)

**B. Interpreters**

Laurence Beer (SISB)
Anne-Marie Wauquierre (SISW and president of ABILS)

**C. Employee:**

Jean-Francois Molderez (CETIC)

**D. Observers**

Jean-Louis Brassiène (FFSB)
Véronique Gernay (APEDAF)
Max Rasquinet (CFLS)

### 5.5.4 Lessons Learned

Here we give a general evaluation of the test results and a set of remarks on specific points that arose during the tests.

**A. General overview**

During and after the test, several evaluation discussions took place. The main remarks collected are summarised in Table 6.

| | **Interpreters** | **Deaf people** | **Observers** |
|---|---|---|---|
| Image quality | *3* | *4* | *3* |
| Quality of perception of signs | *3* | *3* | *3* |
| Role-play | *4* | *4* | *N/A* |
| Environment | *3* | *3* | *3* |
| Audio quality | *2 (echo)* | *N/A* | *3* |
| Overall assessment | *3* | *3* | *3* |

*Table 6. General evaluation of videoconferencing. Key: N/A = not applicable, 0 = none, 1 = poor, 2 = acceptable, 3 = average, 4 = good, 5 = excellent*

Overall, it was a success: all the scenarios could be completed, with no immediate rejection and no insurmountable difficulties. Two indicators are revealing:

- the deaf participants quite quickly took ownership of the means of communication to have fun among themselves
- part of the debriefing discussion was improvised by videoconference.

Image quality was judged average to good. The deaf participants' station had a dual setup (TV+LCD). The main improvement needed concerns lighting conditions: the spotlights used were not powerful enough.

Perception of signs was judged average to good. One interpreter complained of an annoying delay (about 1 second). It did not affect the deaf person's station. On further investigation, it seems that this delay was due to the bandwidth limiter running on that machine.

Image blur was present during fast movements (see Figure 42). It seemed to affect the image received by the deaf person more than that received by the interpreter. The problem seems to lie with the camera used and the lighting.

Both interaction scenarios were appreciated and judged fairly realistic. On the other hand, the general environment could be improved, notably lighting, background and finish (this was a prototype with many visible cables).

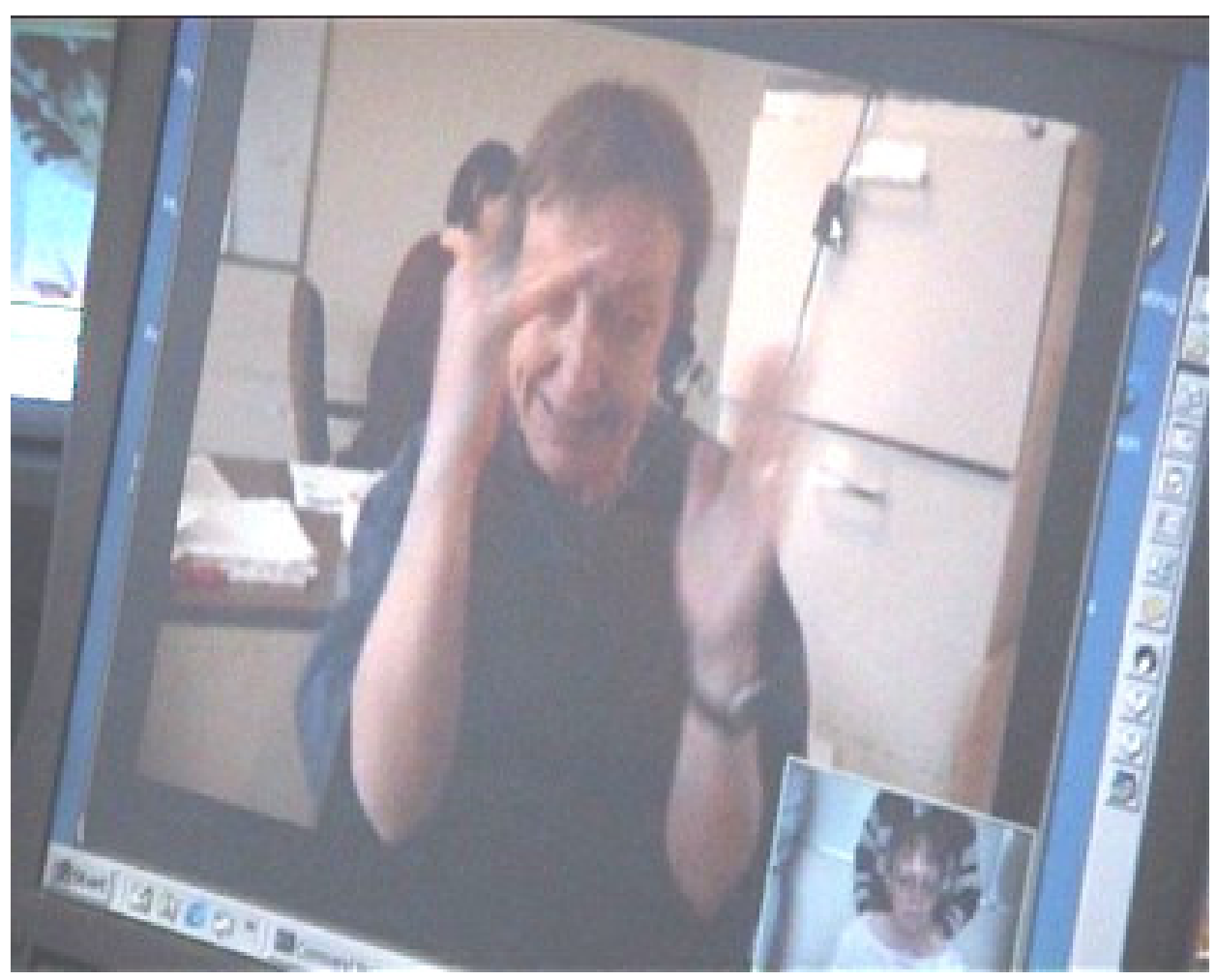

*Figure 42. Example of blur.*

**B. Display preference**

The preference between TV and LCD display varies from person to person. The TV seems brighter and offers richer colours, and it also has a larger display area. On the other hand, its image "flickers" more and is therefore tiring for long periods of use. For interpreters, a good-sized, good-quality LCD screen is essential. The main advantage of the television is its price and probably its availability in many public services

**C. Echo problem in the interpreter's headset**

One of the first surprises during the tests was the discovery of an echo in the interpreter's headset. It was caused by the voice played through the loudspeakers at the public service being picked up again by the microphone. It did not cause a major problem but was somewhat disturbing. Possible improvements are:

- reduce the loudspeaker gain (while remaining sufficiently audible)
- move the loudspeakers away from the microphone
- use a more directional microphone
- use an echo cancellation filter

**D. Minimum quality required**

For a videoconference using the standard, compatible protocol, the tests confirm the need for a CIF image format and a bit rate above 256 Kbps, preferably 384 Kbps. Attempts to reduce the size or the bandwidth quickly led to communication problems

**Holding a conversation at 128 Kbps limited to the QCIF format is not recommended for interpretation. It could be suitable for deaf people communicating directly with each other over the Internet, provided their connection gives them these bit rates. At present, these bit rates correspond to the upper limit offered by "mass-market" commercial solutions: the use of ADSL or cable connections is clearly ruled out at the time of writing. However, these technological obstacles to "mass-market" availability are likely to be removed in the short term (within a year or two) with connections offering greater upstream bandwidth and guaranteed quality of service, probably at first only between subscribers of the same Internet service provider.**

# 6 Virtual Interpretation

In this chapter, we analyse in more detail the possibilities for automating the translation of communication between a hearing person and a signing deaf person. We will first present the general structure of the communication flows, as well as the processes that can be put in place to carry out transfers between them. The feasibility of the approach in a general context is utopian: we will give a brief state of the art of the various technologies to be used, in order to highlight the most mature and the most problematic components. We will nevertheless end this chapter by illustrating several real systems, or systems under development, that meet partial needs. We will show how useful they are within public services.

## *6.1 Deaf ↔ Hearing Communication Chain*

Figure 43 illustrates the communication flows involved in transferring messages between a signing deaf person and a hearing person. Each party has different skills and relies on different main modalities.

Hearing people mainly use the sound channel (voice) and make very little use of the visual channel (a few gestures, body language). They can also use writing (letters, SMS, e-mail, etc.). The message is transmitted in French, either spoken or written. The average hearing person generally does not know sign language.

Deaf people mainly use the visual channel (gestures, facial expressions) and make very little use of the sound channel (sometimes breathing and clicking sounds are perceptible to the hearing person). The message is transmitted in sign language. The text channel cannot be used for sign language, which has no written representation in common use. The average deaf person's lack of command of French generally prevents them from using written French. The sound channel is used by deaf people who can speak.

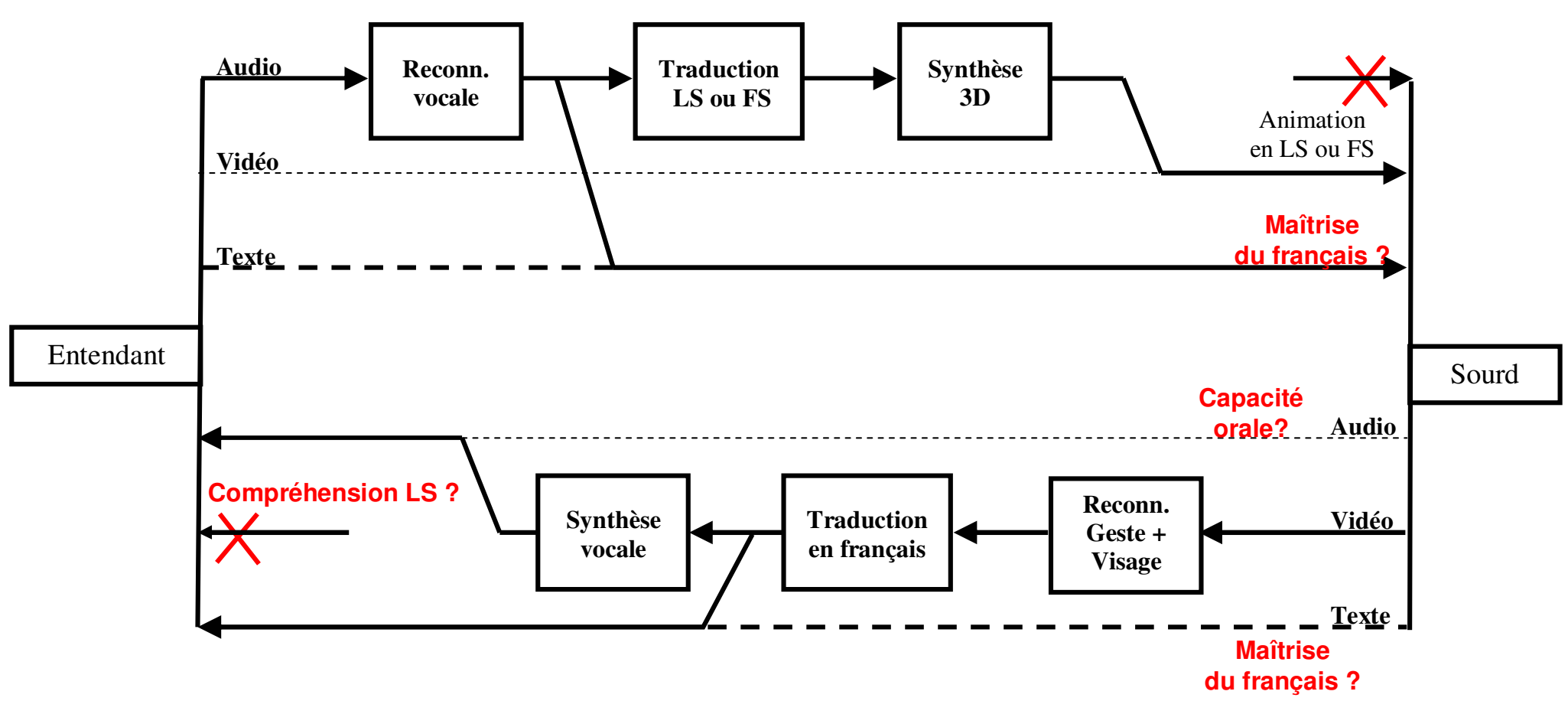

*Figure 43. Automating the translation of communication flows between deaf and hearing people.*

Let us examine both directions of communication to analyse how translation could be carried out. We will take the sentence as the semantic unit.

From the hearing to the deaf person, the following can be used:

- **speech recognition**, or direct text input.
- **translation** is carried out by semantic analysis of the sentence based on French grammar and identification of an equivalent in sign language (also based on knowledge of its grammar). A priori, the feasibility of this translation phase may be doubted; we will see, however, that there are solutions to simplify or avoid it
- the linguistic description in sign language can then be passed to a **synthetic character animation** component, which will express it visually.

In the opposite direction, from the deaf to the hearing person, a complementary chain is used:

- it starts with **automatic sign language recognition**. This requires the simultaneous identification of many elements, such as handshapes, arm movements, facial expressions and direction of gaze. Unlike French, symbolic input is not possible here (or else directly in French, and there is no more work to do).

- **translation from sign language into French** transforms the grammatical structures of sign language into French structures conveying an equivalent meaning.
- The result can be expressed as text or possibly sent to a **speech synthesiser**.

These various processes are very complex and difficult to implement because they deal with the meaning of the message. Note that the videoconferencing solution simply transmitted the raw messages, taking care to degrade them as little as possible.

## *6.2 State of the Art of Technologies*

### 6.2.1 Synthesis of Sign Language Content

Two approaches are being explored to generate sign language animation: one based on video and the other on the animation of synthetic characters.

The principle of **using video** is simple: videos of elementary signs are put in sequence to create a sentence in "sign language". The approach obviously has many limitations:

- a bank of elementary signs (a dictionary) signed by the same person is needed.

- joining the signs smoothly is difficult (see in particular [Kra99]).
- the approach is too rigid to really reproduce sign language. Problems arise, for example, with facial expressions, setting the amplitude or speed of certain signs, etc.

Despite these limitations, the approach is usable, but in the more restricted context of signed oral language. An example of a product is iCommunicator, which will be described later in this chapter.

The other approach is based on **the use of avatars**. These have become popular through their use in video games (for example in "First Person Shooters", "persistent virtual worlds", etc.). They are found widely in advertising, in television programmes (even live) and in the cinema (some films are now made entirely with computer-generated images). Compared with video, avatars have many advantages:

- their movements and speed of execution are fully controllable.

- movements can be fully described in a very compact language made up of control commands. The volume of data to be transmitted is several orders of magnitude smaller than that of the corresponding video.

- they are fully customisable (change of skeleton, "skin", etc.). The description of the character itself is also small.
- the viewing angle is chosen by the user

However, the characters we are interested in have particular requirements that make it impossible to use existing avatars as they are:

- **real-time animation**: the characters must be able to be animated "live". This rules out all very high-quality synthetic characters rendered offline.

- **sign language animation**: most virtual characters are simplified models that lack finesse in hand animation and in rendering facial expressions. Some virtual characters are animated by puppeteers.

Several systems specific to sign language have been developed; they are briefly presented in Table 7. Apart from the products based on Vcom3D and Televirtual technology (Figure 44 and Figure 45), these are mainly prototypes (see Figure 46, Figure 47 and Figure 48), which have their limitations.

The most commonly used techniques and algorithms for arm animation are:

- motion capture, which ensures realistic rendering since it reproduces a real movement. One drawback is that a single instance of a movement is captured, whereas it may have many variants that are not described. These must be captured separately or introduced by means of transformation operators.

- inverse kinematics, which makes it possible to calculate the "natural" movement of an articulated body between two points and avoids having to give the full set of angular descriptions at every instant. It is more compact and better suited to describing movements on a linguistic basis (see [Los00a] and [Los00b]). However, the dynamics of the movements may lack some naturalness. Complementary algorithms can correct this (see [Leb98], [Leb99], [Pot00]).

For facial animation: the technique of displacement nodes makes it possible to deform the mesh describing the face and thus to simulate frowning, smiles, etc. (see [Los02]).

Implementation technologies are generally based on one of the following 3D development standards:

- **VRML**: a standard language for describing 3D worlds in a text format suited to the Internet. Viewers exist for most computer platforms: Cosmos, Blaxxun, Cortona. In principle they are interchangeable, but in practice their compatibility is limited by proprietary extensions, particularly for interfacing with external control languages (Java, C++, script). These extensions are necessary because of the limited animation capabilities within VRML

- **Java3D** (SUN): a 3D programming interface for Java. Its use ensures good portability of the application. Performance is also good, as the implementation takes advantage of the underlying hardware acceleration. The programming interfaces offer a good level of abstraction.

- **Direct3D** (Microsoft): Microsoft's technology for 3D programming under Windows. It is part of the DirectX technology, which allows efficient graphics programming under Windows. The programming interfaces are, however, complex.

| **Name** | **Type** | **Technology** | **Remarks** |
|---|---|---|---|
| Vcom3D (USA) | Commercial | VRML (Blaxxun) | Full modelling of SL: hands and facial expressions<br>Several avatars available<br>**Design studio** adapted to SL<br>http ://www.vcom3.com |
| Visia (ViSiCAST) | Proto | ?? | http ://www.televirtual.com/euro_3.htm |
| I3D signer (Lille) | Research prototype (PhD thesis) | VRML C++ (cosmos) | Flat (2D) and 3D avatar<br>Good finish<br>Limited facial expressions<br>http ://www-cal.univ-lille1.fr/ |
| Virtual Signer (UCL) | Research prototype (MSc thesis) | Java3D | Inverse kinematics animation<br>Partial finish: no joints<br>No collision detection<br>No facial expressions<br>http ://www.touva.be/virtualsigner/ |
| Signer Paris XI | Research prototype (PhD thesis) | ?? | Very simplified character<br>No facial expressions<br>Complex sign (e.g. Crowd) |
| Signe3D (Aptitude) | Research demo | Direct3D (VisualBasic) | No facial expressions<br>Inverse kinematics animation |

*Table 7. List of the main signing avatar technologies.*

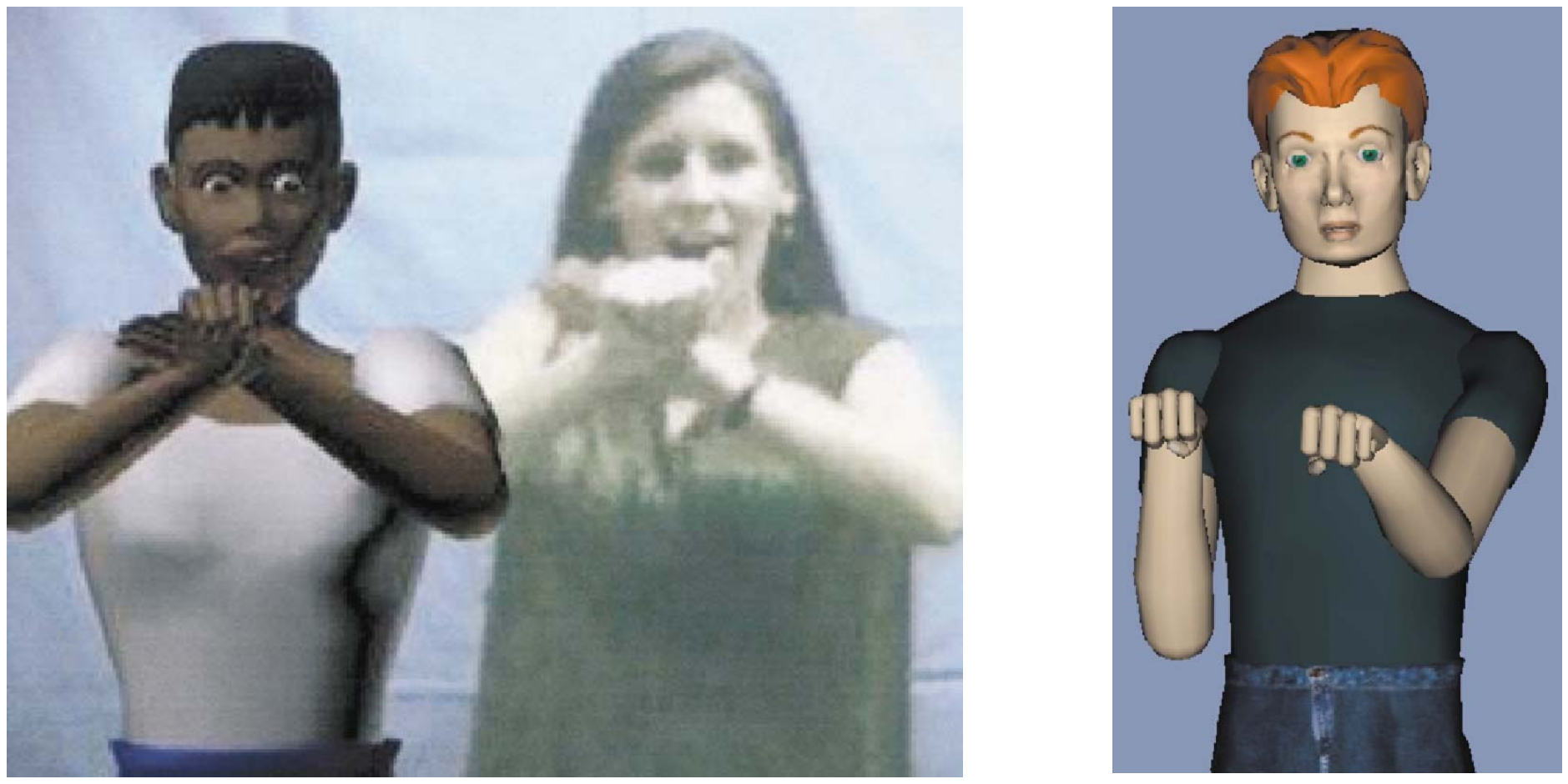

*Figure 44. Vcom3D signing avatar.*

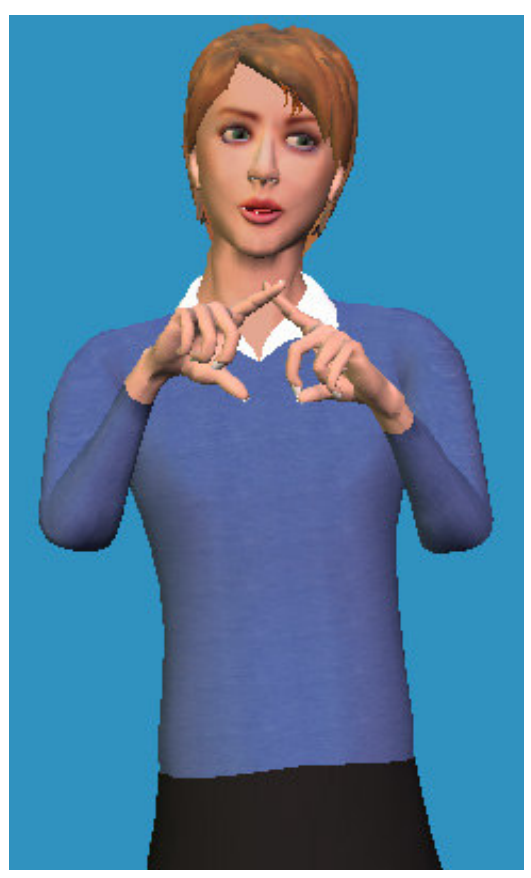

*Figure 45. Visia avatar (Televirtual)*

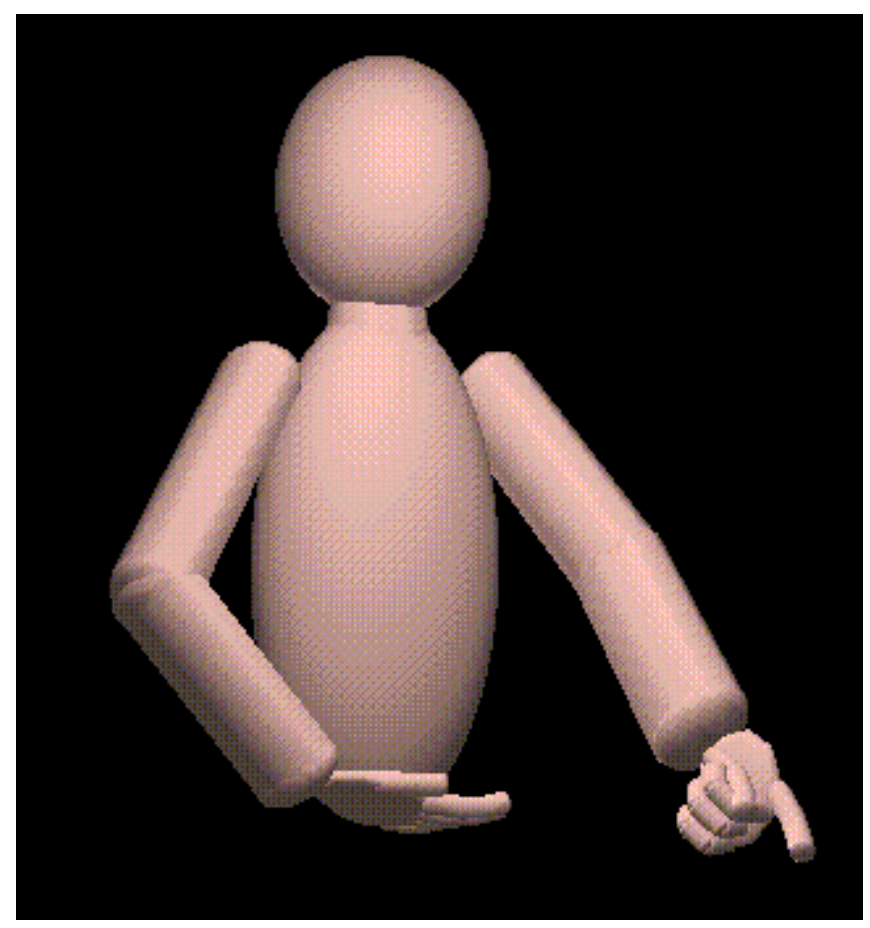

*Figure 46. Paris XI avatar*

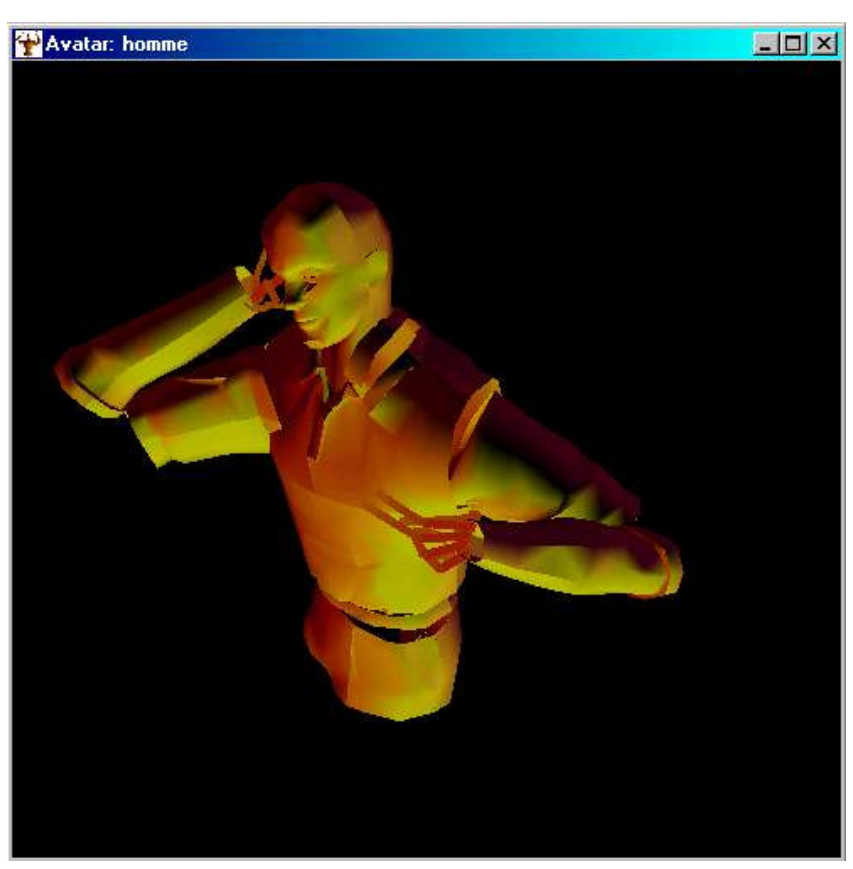

*Figure 47. Virtual Signer (UCL)*

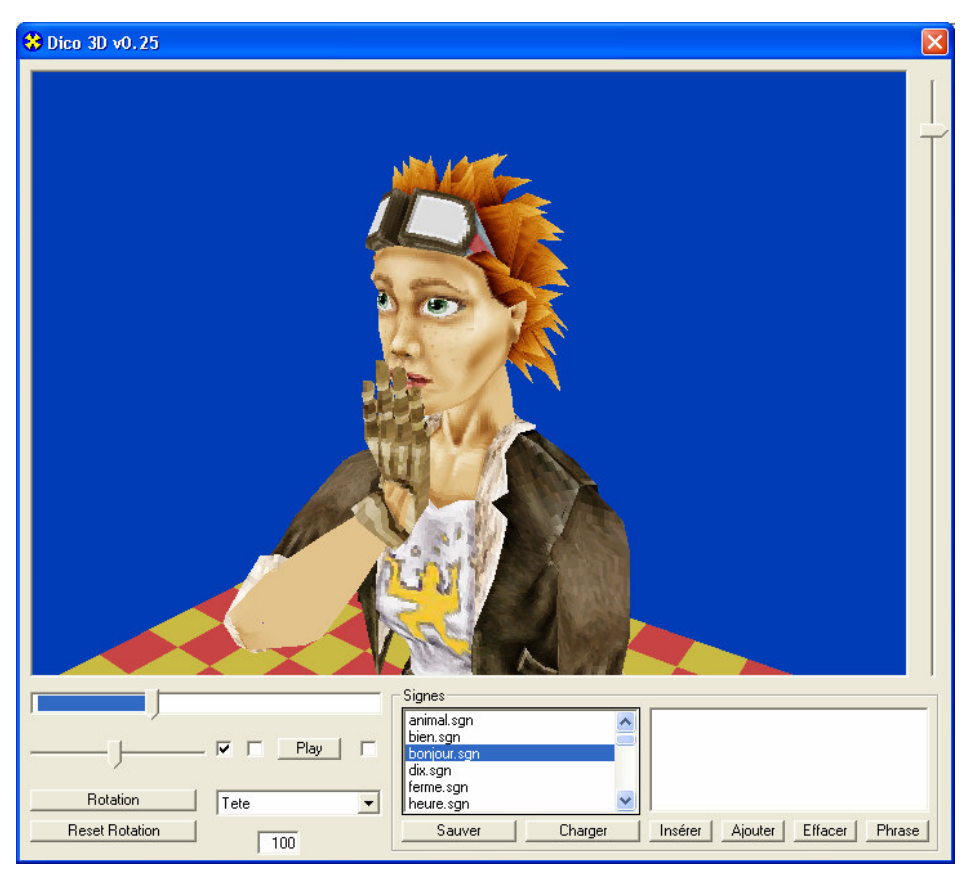

*Figure 48. Signe3D avatar.*

### 6.2.2 Speech Processing Technologies

Speech processing is a very broad and complex subject, of which we give only a very brief overview drawn from [Tut00] and from our interview at BabelTech.

The field is very active and has seen many developments linked to the GSM revolution, research (large databases, pooling of laboratory resources) and pure engineering tools (algorithmic in nature).

There are 3 types of techniques:

- **recognisers**, whose task is to decode the information carried by the speech signal from the data provided by the analysis. The aim may be to recognise the speaker or the speech. For speech recognition, a distinction is made between single-speaker, multi-speaker and speaker-independent cases. A distinction is also made between isolated words, connected words and continuous dictation.
- **synthesisers**, which have the opposite function: they produce artificial speech. They can be driven by a digital or symbolic representation (plain text or linguistically enriched text)
- the role of **coders** is to allow speech to be transmitted or stored at a reduced bit rate, which requires judicious use of the properties of speech production and perception.

Of these techniques, only the first two concern us and are detailed below.

**A. Speech synthesis**

Though it may seem trivial, speech synthesis is a complex process, both in terms of the cognitive processes involved (contextualisation, phonetisation, intonation, rhythm, etc.) and of the generating apparatus (configuration of the articulatory muscles, vibration of the vocal cords, etc.).

The system is organised as a series of software modules reproducing general human processing: pre-processing, morphological, contextual and syntactic-prosodic analysis, then the synthesiser itself.

The synthesis phase has recently made impressive progress through unit selection in a large database. Based on several hours of speech**, it is now possible to synthesise voices that are almost indistinguishable from human voices**. An example of a product based on this technique is BrightSpeech from BabelTech (http://www.babeltech.com)

In the context of this study, synthesis seems to be reserved for the hearing person. Note, however, that for late-deafened people, the system can be modified by replacing the final synthesis component with a component that animates the sound units as lip movements on a synthetic head. For deaf people with residual hearing, the synthetic voice can be a useful complement thanks to its clarity and regular rhythm.

**B. Speech recognition**

Various speech recognition systems have been developed, covering fields as wide as recognising a few keywords over telephone lines, voice dictation systems, command and control systems on PCs, and even natural language understanding systems.

The first technologies were based on word recognition. They were limited by vocabulary size and better suited to a single-speaker context. Under these conditions, almost perfect results are obtained today.

For large-vocabulary, multi-speaker systems, one has to go down to the phoneme level, and the aim is to derive a voice-independent model. Implementation is generally based on statistical acoustic models trained on a large amount of speech data. They use hidden Markov models or artificial neural networks, sometimes in combination. Current performance is still far from human performance: the error rate varies with vocabulary size and with whether the system is speaker-dependent or independent. Robustness is also a major challenge: the presence of noise quickly reduces recognition performance.

Products are sold for various types of applications (industrial control, voice dictation, etc.). **For voice dictation, the solutions are speaker-dependent and require an initial training phase** of about ten minutes. Examples include ViaVoice (IBM), Dragon Naturally Speaking (ScanSoft) and Microsoft Speech.

We had the opportunity to try out Dragon Naturally Speaking and even to use it in an application. We will come back to this in the applications part of this chapter.

### 6.2.3 Sign Language Recognition

Here, techniques can be used that rely on a **motion capture system** (based on a complex system of harnesses, gloves and recording of certain points on the face) or that **analyse a video stream** (using various image processing techniques: segmentation, extraction of discriminating features, etc.).

Sign language recognition mainly involves recognising manual and facial features. These are being researched in various laboratories: at UCL [Erg98], Aachen (RWTH, as part of WISDOM) [Ers03], in Japan [Sag97], in Australia [Kad95] and in the USA [Yan01], [Gri99], [Gri02]. Since 1996, an annual workshop, "Gesture and Sign Language in Human-Computer Interaction", has reviewed developments in the field [Wac97].

In terms of concrete achievements, the most striking progress has been made in the "edutainment" sector (combining entertainment and education). Figure 49 shows a system by the Belgian company Alterface (http://www.alterface.com). Typically, a filmed person is "immersed" in an application with which he or she interacts through gestures. Similar applications will soon be available on mass-market game consoles. However, gesture recognition is not fine enough to be applied to sign language recognition.

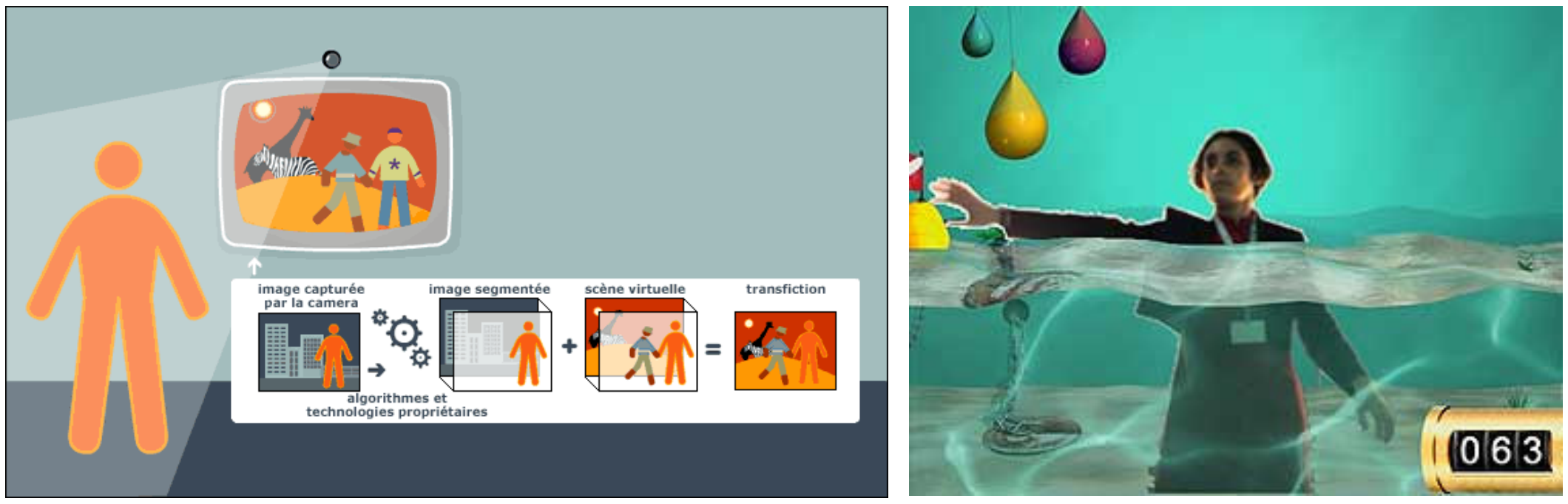

*Figure 49. Gesture recognition and "immersion" in a virtual environment (AlterFace).*

In facial recognition, convincing real-time applications are beginning to emerge, notably from EyeMatic (see Figure 50). A person's facial expressions (lip movements, smiling, surprise, frowning, etc.) are recognised and drive an avatar. This makes visual communication possible over narrow bandwidths. In the context of sign language, semantic information could be extracted to complement the information contained in the gestures.

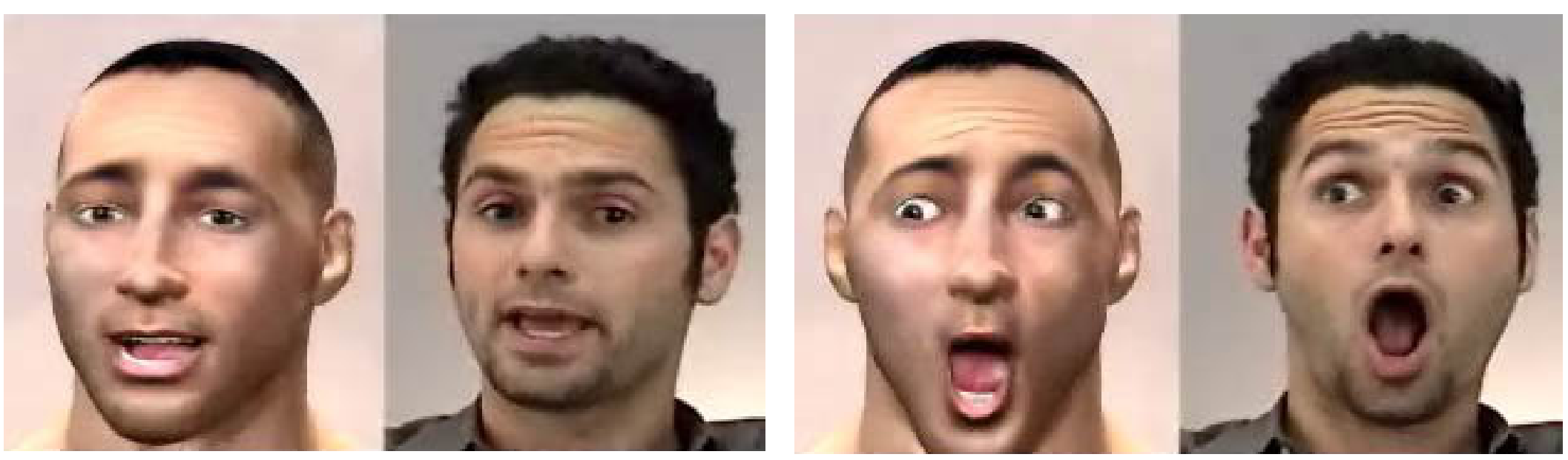

*Figure 50. EyeMatic synchronised facial avatars.*

### 6.2.4 Machine Translation

Machine translation between two languages is a difficult problem, still largely unsolved for oral languages, even though they have been studied for a long time and are well described. Sign language has the additional difficulties of differing considerably from oral languages and of still being very poorly described.

In addition to the work of linguists to establish its grammar more precisely, various work is under way on suitable computer representations. Research explores Liddell & Johnson's "Movement-Hold" paradigm ([Lid89], [Vog99], [Spe01], [Zij03]), HPSG notation (Head-driven Phrase Structure Grammar) [Bun02][Mar02], SiGML [Ell00] and TAG rewriting systems [Zha00].

To get around the translation problem, two options are possible:

- stay in the same language by using signed French
- restrict oneself to standard sentences whose translation is known (the so-called "formulaic language" approach)

We will see below practical systems that adopt these simplifications.

## 6.3 Presentation of Feasible Systems

### 6.3.1 General Possibilities

One of the main conclusions of the state of the art in the previous section is that the techniques supporting communication from the deaf to the hearing person are much less mature than those from the hearing to the deaf person. Neither visual recognition of sign language nor its translation is currently feasible. We will not explore them further in the rest of this chapter.

The direction of communication from the hearing to the deaf person suffers from several problems, but they can be overcome:

- the speaker-dependent nature of speech recognition is acceptable for certain classes of applications where the number of speakers is known and stable over time: for example in a school or an administration.

- translation into sign language can be avoided by using signed French (considered a second-best solution by deaf people but nevertheless accepted) or by using pre-encoded phrases.

In the rest of this section, we will detail some applications that use these working assumptions. We will also examine the broader possibilities that could suit late-deafened people.

### 6.3.2 Partial Interactive Communication Systems

These systems focus mainly on transmitting information from the hearing to the deaf person. Regarding the return channel, we will assume:

- either that it poses no problem: this is the case if the deaf person is able to speak
- or that it is of little importance because of the asymmetric nature of the communication. This is the case of a student attending a class. If interaction is needed, a manual "magic slate"-type system or an automatic system based on speech synthesis can be considered.

We must also consider the constraint that the speech recognition system must first be trained to the speaker. Basic training takes only a few minutes but already makes it impossible to address a passing person. Training that ensures more reliable and therefore more comfortable recognition may require many adjustments during use. We will therefore restrict ourselves to classes of systems that repeatedly involve the same speakers.

Two concrete classes can be mentioned:

- education: during the school year, the deaf student attends classes with the same teachers (about ten). In this case, the system is managed by the student, who makes sure to record the voice profiles of his or her teachers.

- routine procedures within an administration: for example, sending a registered letter at the post office or renewing a passport at the municipality. In this case, the system is managed by the administration, which makes sure to record the voice profiles of the employees who will interact with deaf people.

We will now illustrate an example of each of these systems.

**A. TESSA: Text and Sign Support Assistant (ViSiCAST project)**

Tessa is the main achievement of the ViSiCAST consortium, funded by the European 5th Framework Programme [Ban00]. The general objective of this project is similar to that of our study: to improve the integration of deaf people. The project is coordinated by the University of East Anglia (UK). Various European partners are involved, notably Televirtual (UK) for its avatar technology, the University of Hamburg (linguistic aspects), the Institut National des Télécommunications (France), and deaf organisations: the Royal National Institute for Deaf People (United Kingdom) and the Instituut voor Doven (Netherlands)

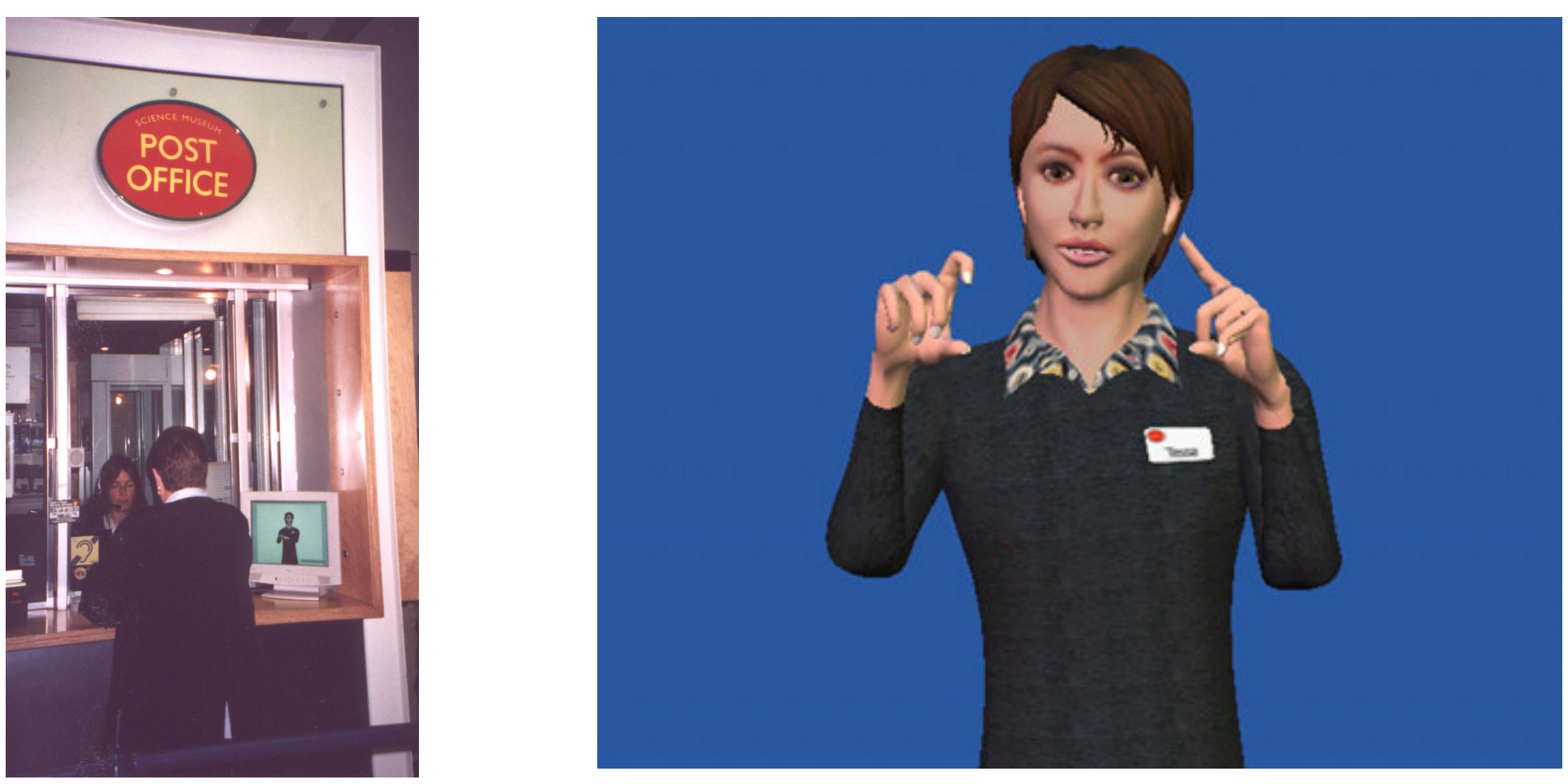

*Figure 51. Visia signing an explanation at the British Post Office.*

The objective of TESSA is to help post office staff in their interactions with deaf people (see Figure 51). The aim is to help complete simple, predictable transactions such as cashing a cheque or buying stamps.

The principle of the interaction is illustrated in Figure 52:

- *The clerk speaks into the microphone "Your parcel will be delivered on {date}"*
- Speech recognition determines the sentence spoken.
- The computer locates and assembles the corresponding animation

- The customer's screen shows the animation of the avatar (called Visia) together with a text version.

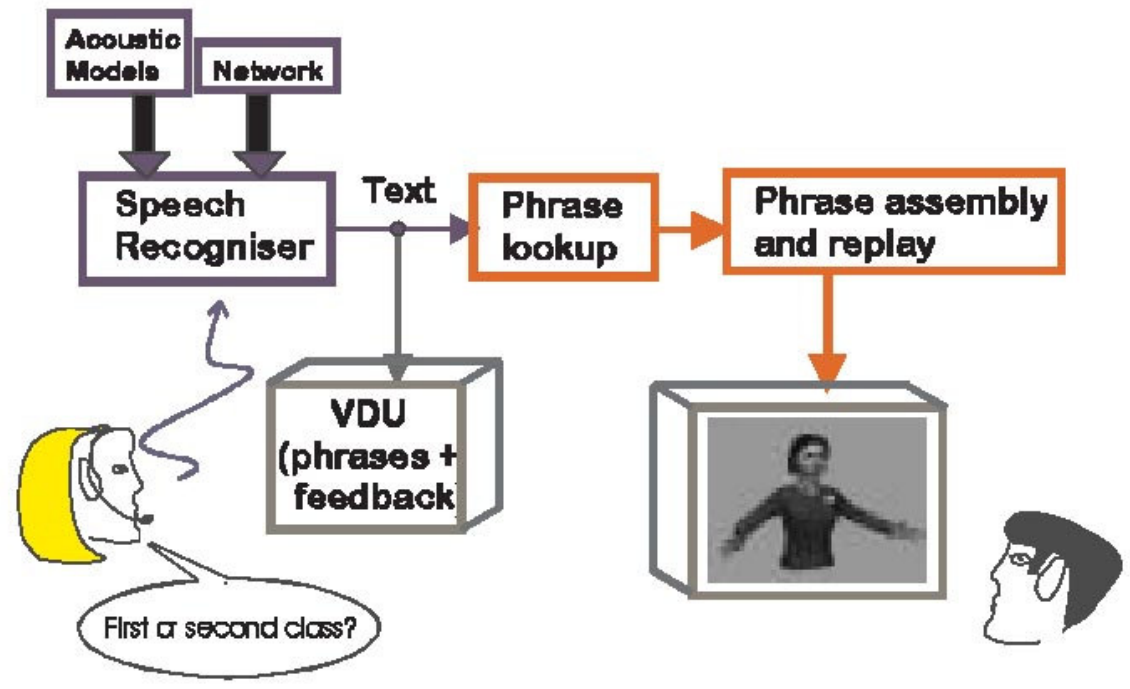

*Figure 52. Processing chain in the TESSA system.*

The interactions may seem very rigid, as they are limited to a set of key sentences. The animations can even be pre-computed in files. This amounts to playing video, and in the end little advantage is taken of what a synthetic character offers. The approach used to animate TESSA is based on replaying data captured with a motion-capture harness.

One may question the realism of such a system, which seems at the mercy of the slightest unforeseen event. The system was evaluated in depth in [Cox02][Cox03]. The study shows that the system makes it possible to complete transactions but significantly lengthens the time needed (time doubled) compared with a situation where the deaf person is left alone without an interpreter. The system was nevertheless judged useful overall, in particular as an alternative means to be used when the procedure gets bogged down.

To make the system more flexible, extensions are being considered, in particular greater use of speech recognition to go beyond standard sentences. The risk is then having to switch to signed English rather than BSL. Another extension under consideration concerns the communication channel from the deaf to the hearing person. Initially, this would involve recognising a limited number of signs.

The approach seems above all pragmatic and has the merit of being part of a broader accessibility programme for anyone who does not master English well: in addition to sign language, the system could also display subtitles in oral languages such as French or German. Such an extension could be useful in very touristic places. The synthetic character can also be used with late-deafened people by making it move its lips in sync with the audio and displaying subtitles.

We will come across other applications of this avatar later in this chapter, in particular on the web and on television.

**B. iCommunicator**

iCommunicator is a product sold in America by Interactive Solutions and presented on their website (http://www.myicommunicator.com). It is a product combining several functions (see Figure 53):

- **speech recognition** (based on Dragon Dictate). The recognised text is displayed in the bottom area and can be reworked later, for example to compile lecture notes or write up a conference report.

- a **presentation in signed American English**, with a video display of the meaningful words (nouns, verbs, adjectives, etc.). The framing can vary depending on the word: either a classic medium shot or a view of the hand when that is enough (examples: numbers, letters)
- an **input line** for the speech synthesiser, allowing a deaf person who is unable to speak to express himself or herself.

The product seems mainly aimed at teenagers in the context of their school activities and is therefore outside the strict scope of this study. The system fits well with the usage constraints mentioned above: training for the teacher, communication mainly from the teacher to the student. The "note-taking aid" dimension is also well worth noting. Indeed, an interpreter or the ability to lip-read requires the deaf person to maintain constant visual contact, and he or she must therefore choose between taking notes and "listening". Other alternatives are of course to use other students' notes or even the teacher's.

The system may also be suitable for use within a public service, provided the effort is made for the training phase. A demonstration video is available on the company's website as well as on the CD-ROM in the appendix.

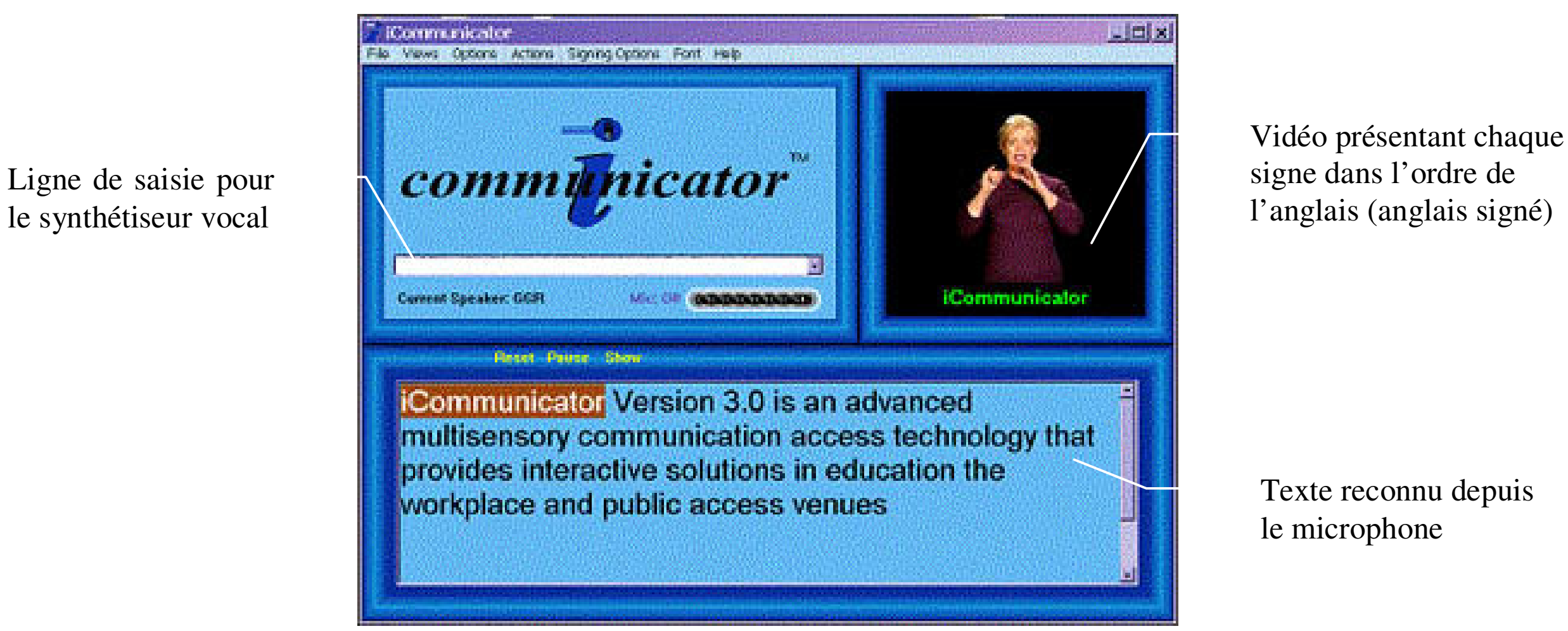

*Figure 53. iCommunicator interface*

Note that developing such an application is not particularly complex, since good-quality "off-the-shelf" commercial components can be assembled. In this case, the product is based on the speaker-dependent speech recognition product Dragon Naturally Speaking. We were able to test it separately. It has a powerful development kit that can be used in standard programming environments such as Visual Basic and Visual C++. We quickly used it to build a modest prototype limited to speech recognition, without any sign language visualisation technique. It is therefore not suitable for signing deaf people but could suit late-deafened or hard-of-hearing people.

The application simply displays the last sentence spoken in a very large font. Typically, this allows the deaf person to follow on the lips and glance at the screen to check understanding. Previous sentences are also stored and can be viewed in a text area in a smaller font. This text can be saved and possibly reworked in a word processor. This type of application has been studied in more detail in European projects such as VOICE (http://voice.jrc.it/).

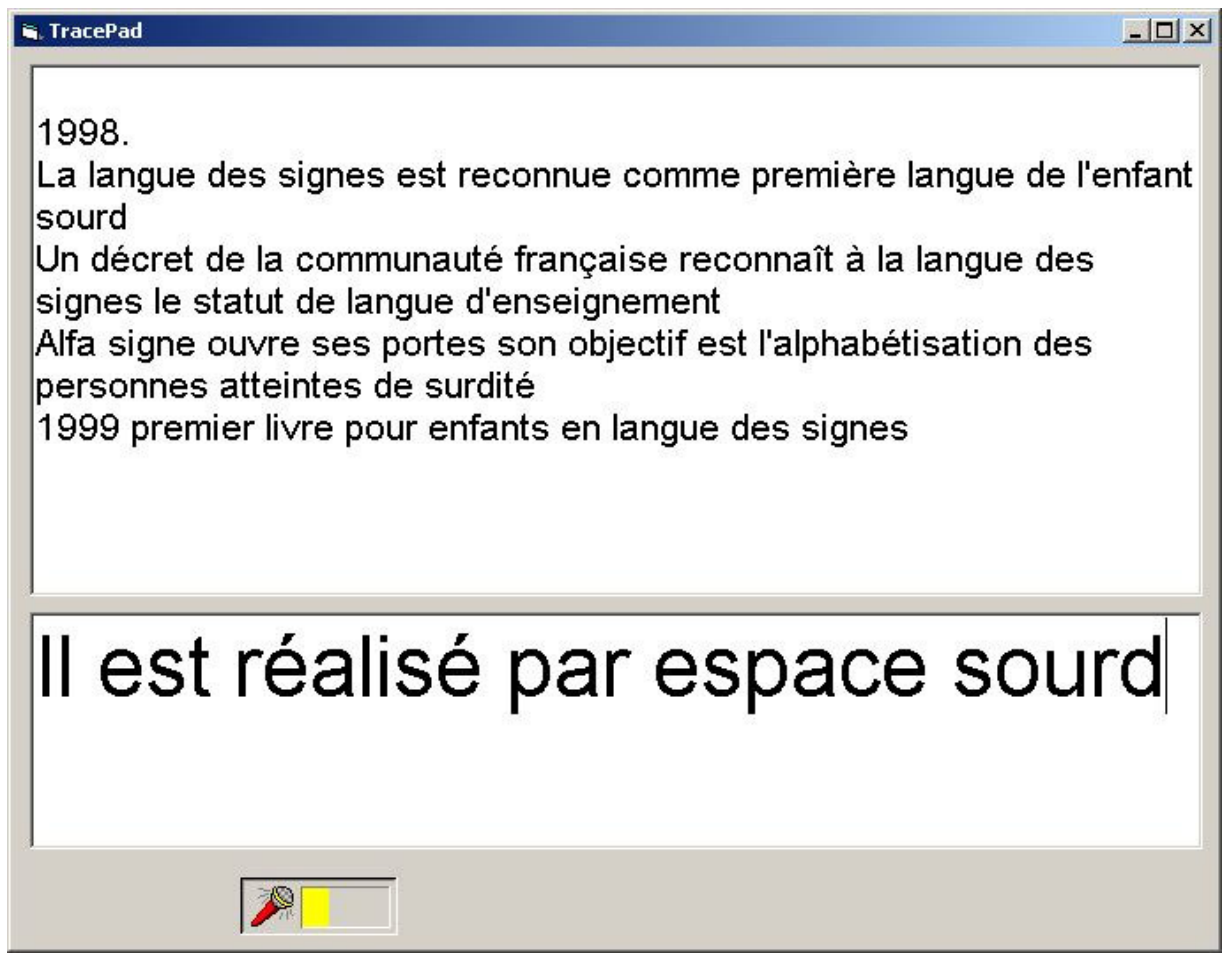

*Figure 54. Prototyping a simple interactive system based on text only.*

The figure above illustrates the application after a few lines of the collection "L'histoire des sourds" [Gerday] were dictated to it. The quality of recognition is very good here: the meaning is fully preserved and the spelling almost perfect. However, this test was carried out under good conditions:

- standard training of about ten minutes (carried out a few weeks earlier, without any particular adaptation)
- quiet environment
- very high-quality microphone
- recalibration before the start of the session
- particular attention to diction and articulation

The quality of the results degrades quickly and produces meaningless text as soon as one of these conditions is no longer met. We also encountered other causes of degradation:

- related to the speaker: for example, when he or she has an illness that changes the voice (cold, sore throat)
- related to the algorithms used: they favour the most probable word sequences (established on the basis of linguistic "corpora"). Poetic texts, for example, can cause problems because of their unconventional use of words. Homophones are also problematic (for example "ces chalets", "ses chalets", "ces chats laids", "sécha les"), because the use of context is limited and probabilistic.

### 6.3.3 Information Systems in Sign Language

Here, the communication chain is limited to rendering sign language material and poses no major problem. However, good tools are needed to generate this content easily. Recommendations on the use of sign language on the web are described in [Hel02].

We will review three classes of products: dictionaries, editors for hypertext content enriched with sign language, and real-time sign language broadcasting systems.

**A. Sign language dictionaries**

We have already mentioned the DILS, Mano and DicoLS dictionaries. All three are based on video. Vcom3D's "Sign Smith Illustrated Dictionary" illustrates the contribution of the synthetic character. In Figure 55, one can see in particular the controls for changing the speed (tortoise-hare), as well as the sliders for changing the viewing angle.

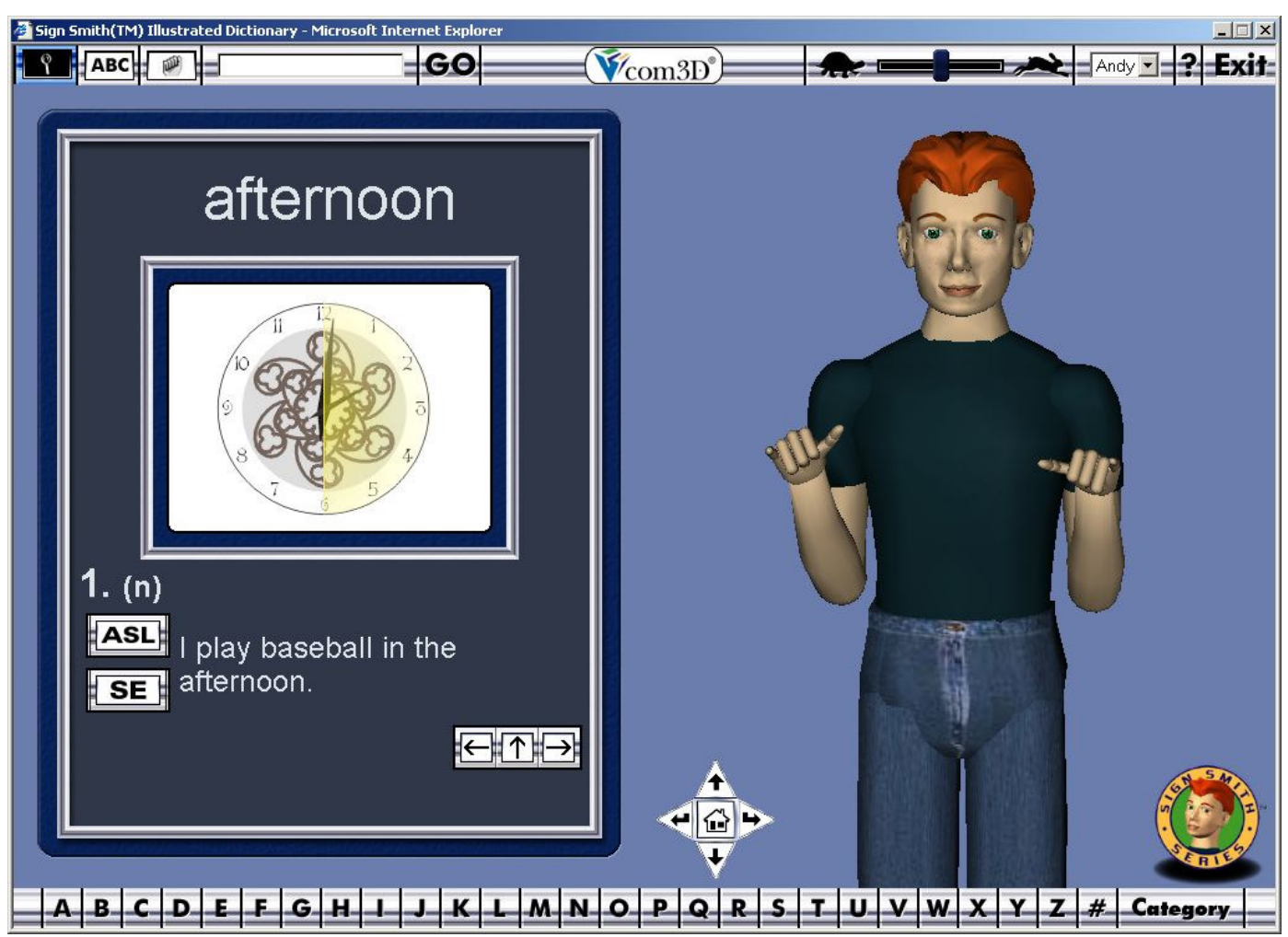

*Figure 55. Vcom3D avatar used in the "Sign Smith Illustrated Dictionary".*

**B. Sign language content management systems**

For sign language content management, the Swedish product LinkIt makes it possible to create and distribute the equivalent of a website combining text and sign language (with display synchronised with the text). However, it uses a proprietary format requiring a special viewer program. Figure 56 illustrates its use for children's stories. The product uses video and is intended rather for producing content to be distributed on CD-ROM.

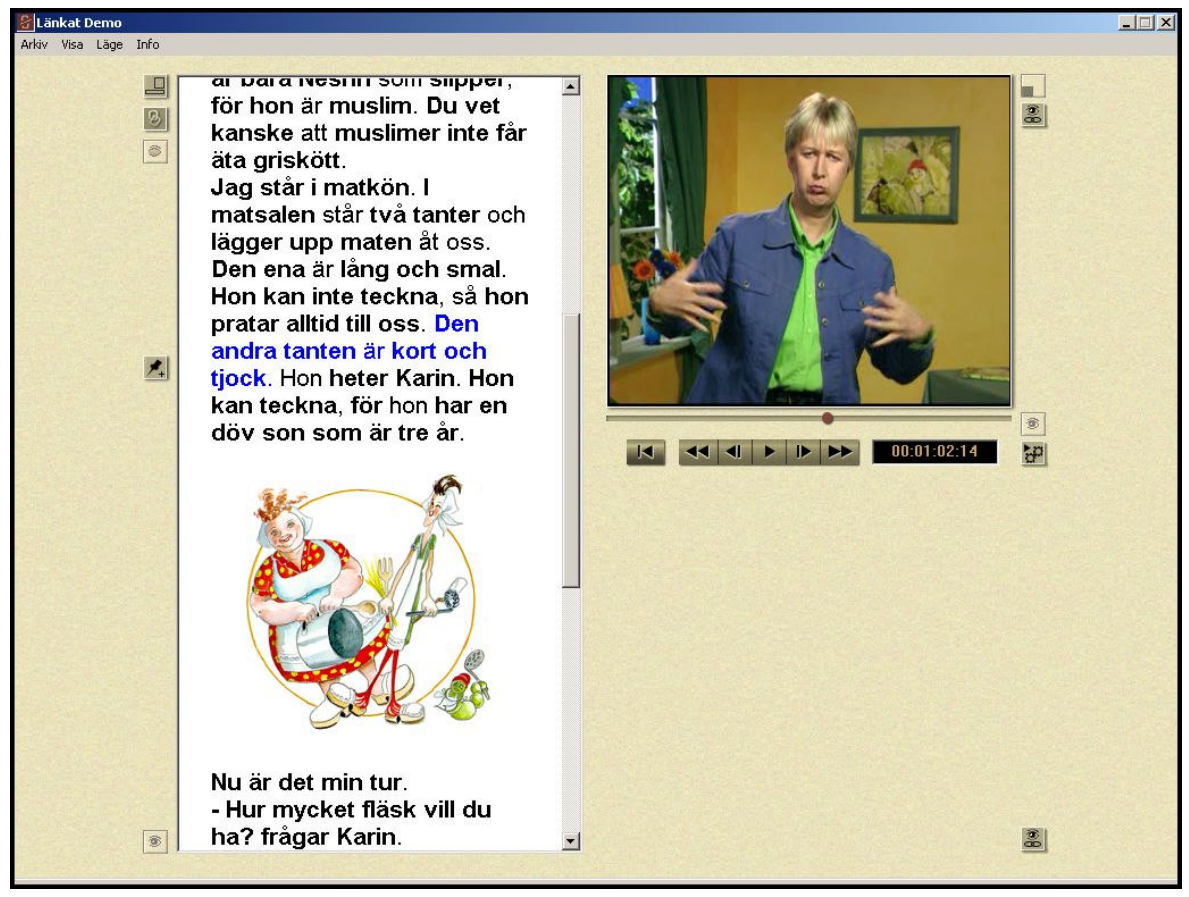

*Figure 56. LinkIt used to tell a children's story.*

Another system developed in France is LAC (multimedia software for learning communication techniques for deaf people). This software initially worked with videos. Experiments were successfully carried out to replace them with an avatar [Los00a] (see Figure 57).

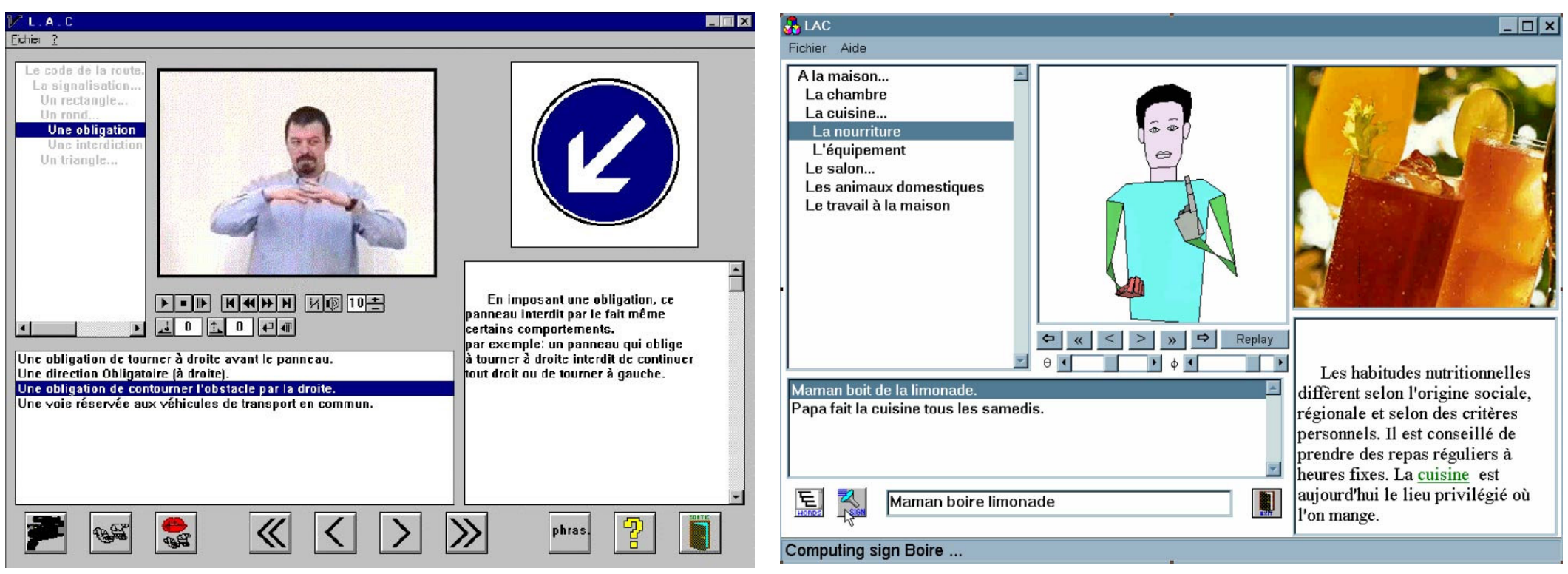

*Figure 57. LAC interface, in video mode and "flat" avatar mode.*

The Visia avatar developed by Televirtual as part of the ViSiCAST project has also been tested in a system for translating weather forecasts on the web. As with TESSA, the domain can be described by a fairly limited set of generic phrases.

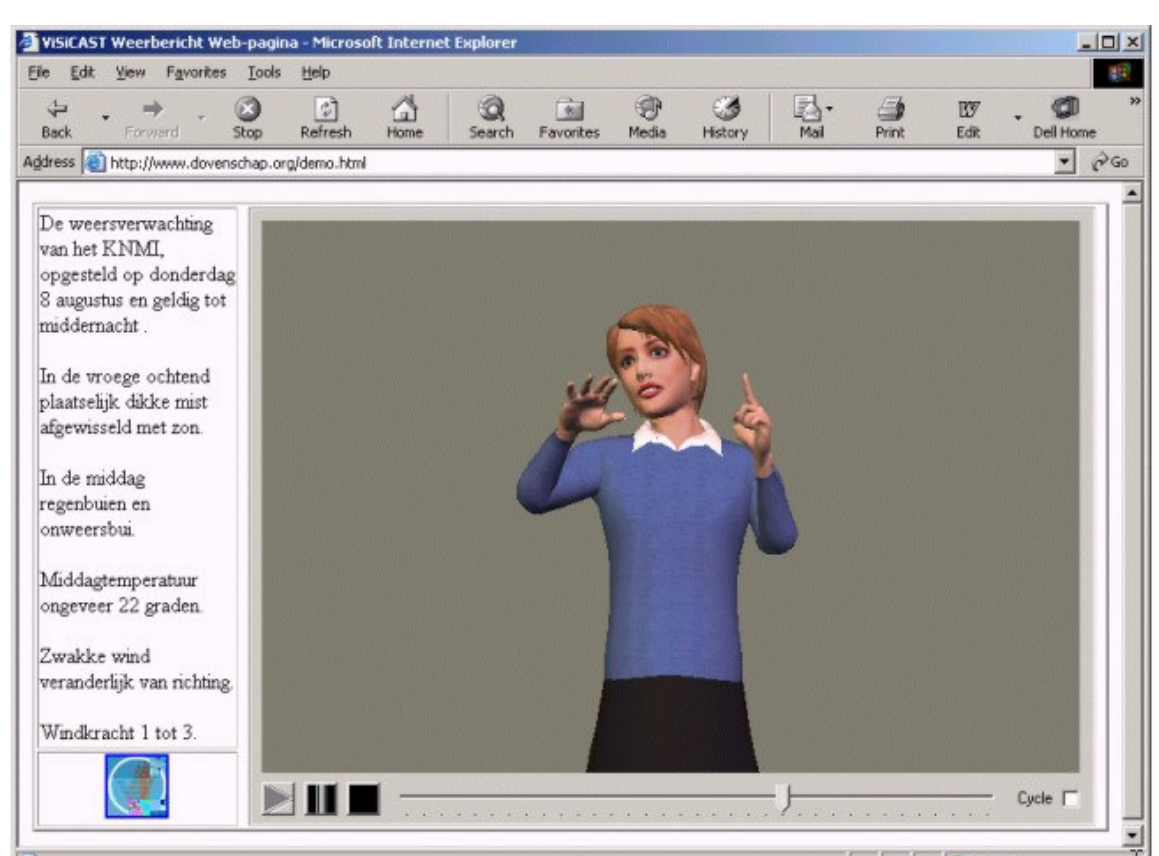

*Figure 58. Visia avatar signing the weather forecast on the web.*

**C. Real-time sign language broadcasting system**

The Visia avatar can also be used in a television broadcasting system. The principle is illustrated in Figure 59: subtitles would be extracted from Teletext by a decoder belonging to the deaf person. These subtitles would be fed into a translation component that would generate the animation commands for the avatar. The avatar image would then be inset on the screen with the picture and sound. Figure 60 and Figure 61 illustrate this with two different avatars.

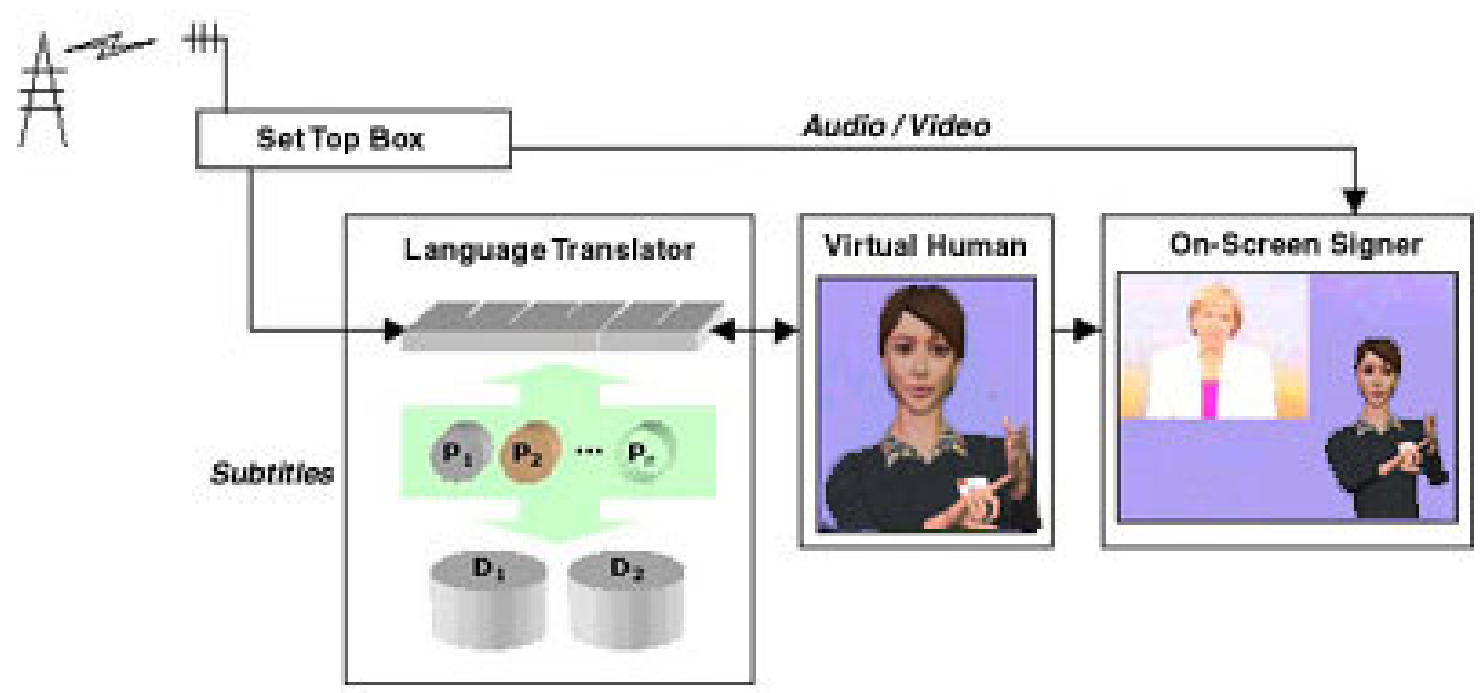

*Figure 59. Using Visia for television interpretation.*

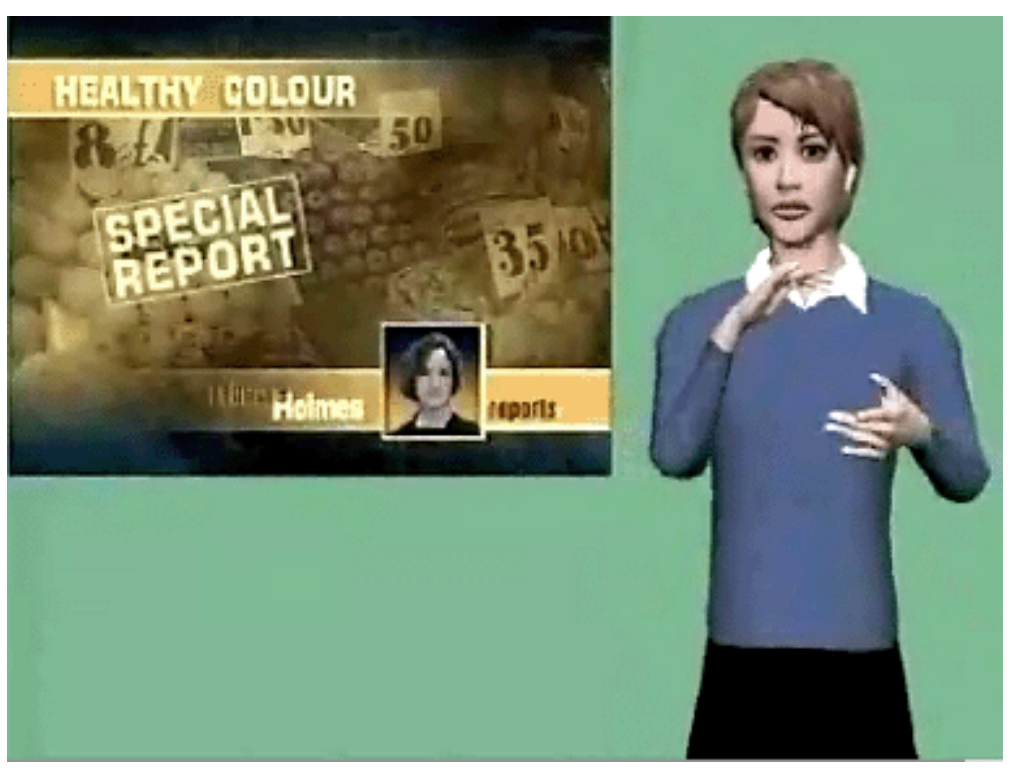

*Figure 60. Visia signing the TV news*

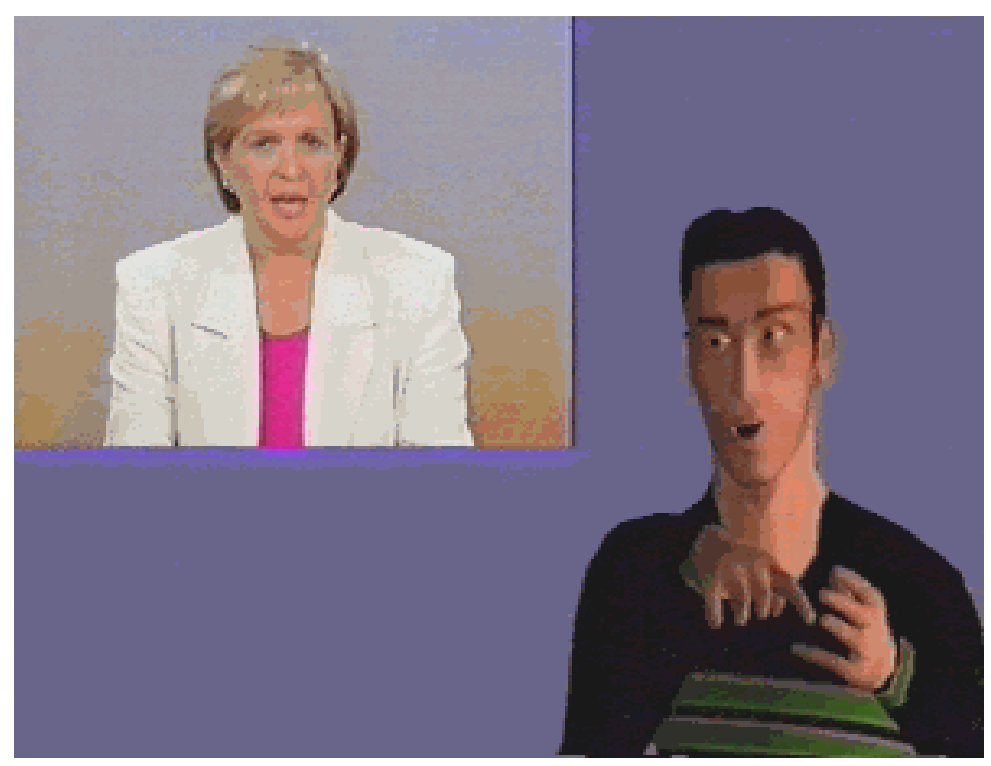

*Figure 61. Simon signing the TV news.*

As described, the translation component is not currently realistic for generating sign language: at most, signed language can be generated. An alternative system that could work in a first phase eliminates this component and generates the animation commands directly by capturing the movements of a real interpreter (at the production studio). The amount of data to be transferred remains small, around 15 Kbps.

### 6.3.4 Other Systems for Late-Deafened People

We briefly describe a few solutions suitable for late-deafened people who express themselves in French. These solutions make a somewhat different use of the technologies described above.

### A. Automatic lip synthesis

Lip synthesis directly combines speech recognition and the animation of a synthetic character. The character is reduced to a head or bust: only the movement of the lips needs to be rendered precisely, so as to allow lip reading.

Technically, the recognition problem is simpler because it is limited to discriminating phonemes. Such a system works in a **speaker-independent** way. Two implementations exist:

- the LipCell system (see Figure 62), produced by the Israeli company Speechview (http://www.speechview.com). It should be on sale soon (about $125). One of its aims is to make it easier for deaf people to receive GSM calls. It currently runs on PCs but should be ported to PDAs and GSM phones. We have no details on the technology used, except that it is based on phoneme recognition. A system of coloured dots is used to discriminate ambiguous phonemes.

- the SynFace system (see Figure 63), a European project led by Sweden, the United Kingdom and the Netherlands. The system uses phoneme recognition via 2 methods (HMM and ANN). It was evaluated in [Sic02]: the use of this technology seems to improve communication by 20% but remains somewhat below a human face (18%). Curiously, the system does not seem to add information to discriminate ambiguous phonemes.

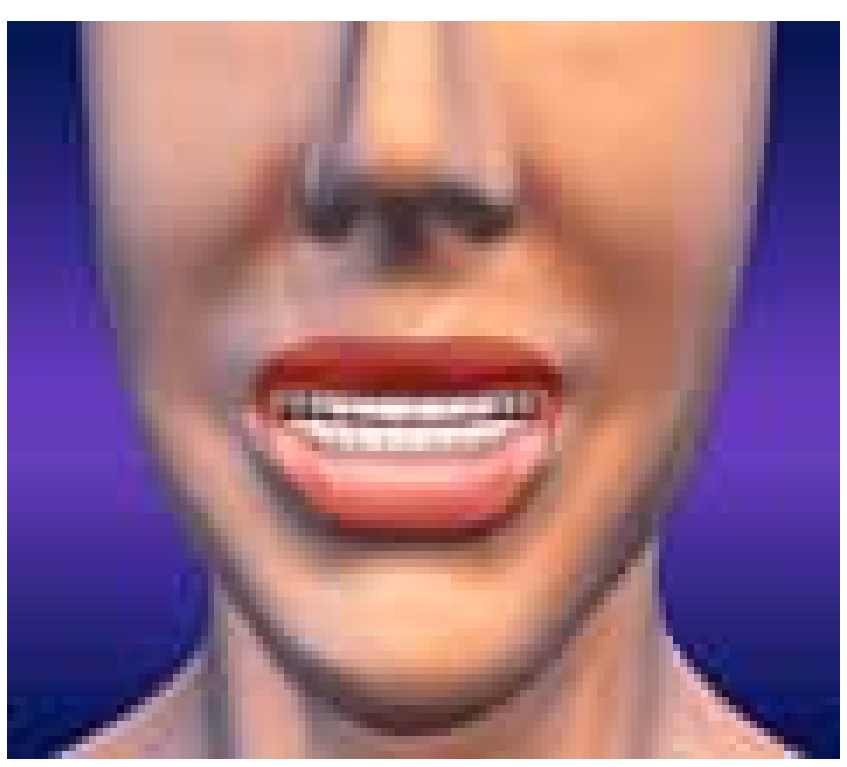

*Figure 62. Lip animation synthesis by Speechview.*

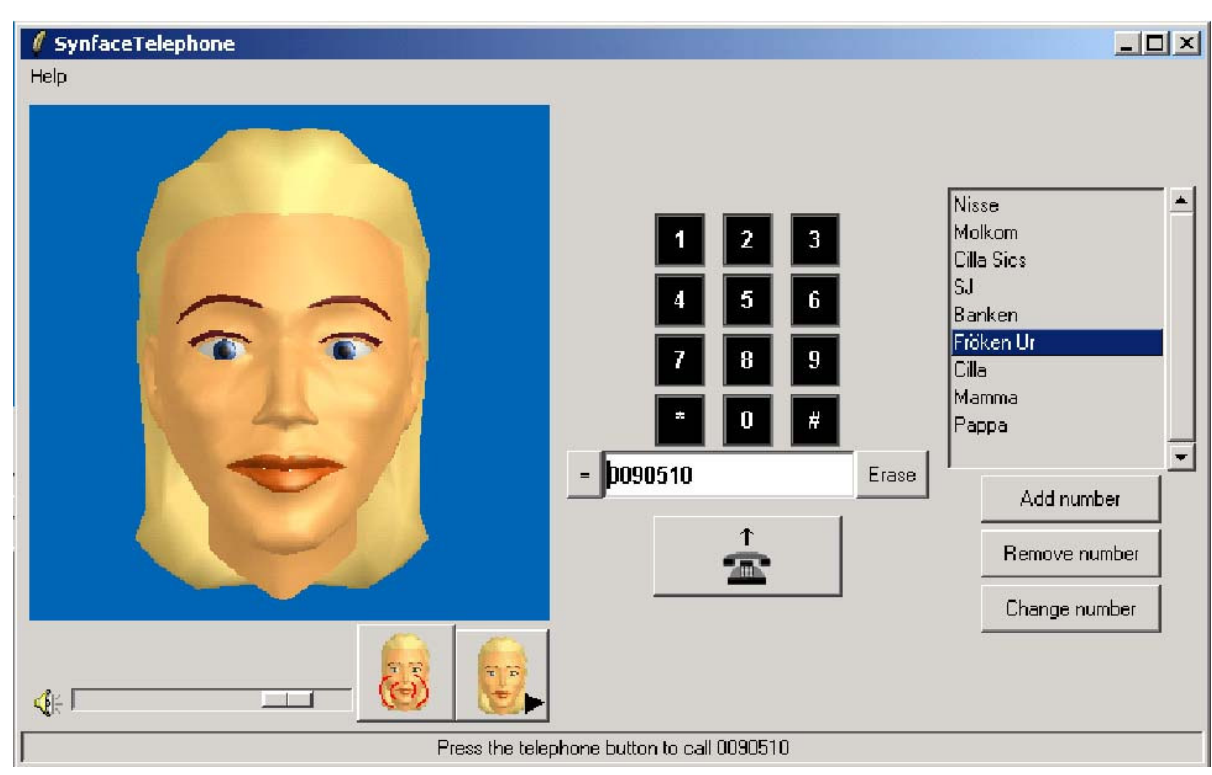

*Figure 63. Lip animation synthesis by SynFace.*

### B. Semi-automatic subtitling

Another possible use of speech recognition is semi-automatic subtitling for content that is not broadcast immediately (recall the 40:1 ratio between preparation and broadcast time) and above all for purely live content.

Speech recognition cannot be applied directly to the original soundtrack for two main reasons:

- the first is related to speech recognition technology, which cannot be multi-speaker over a large vocabulary.

- the second is related to subtitling requirements: a maximum rate of 150 words/ minute must be respected (110 for children). This implies a more concise reformulation of the content.

A solution used notably at VRT is to have a person listen to the live interview and reformulate it into a speech recognition system trained to his or her voice (see Figure 64). The subtitles are very quickly checked and possibly corrected by a second person before being sent to Teletext. The system works with an acceptable error rate of a few percent. Clearly, this system requires great concentration and a changeover after a certain time.

This system is useful for TV news, debates or football matches.

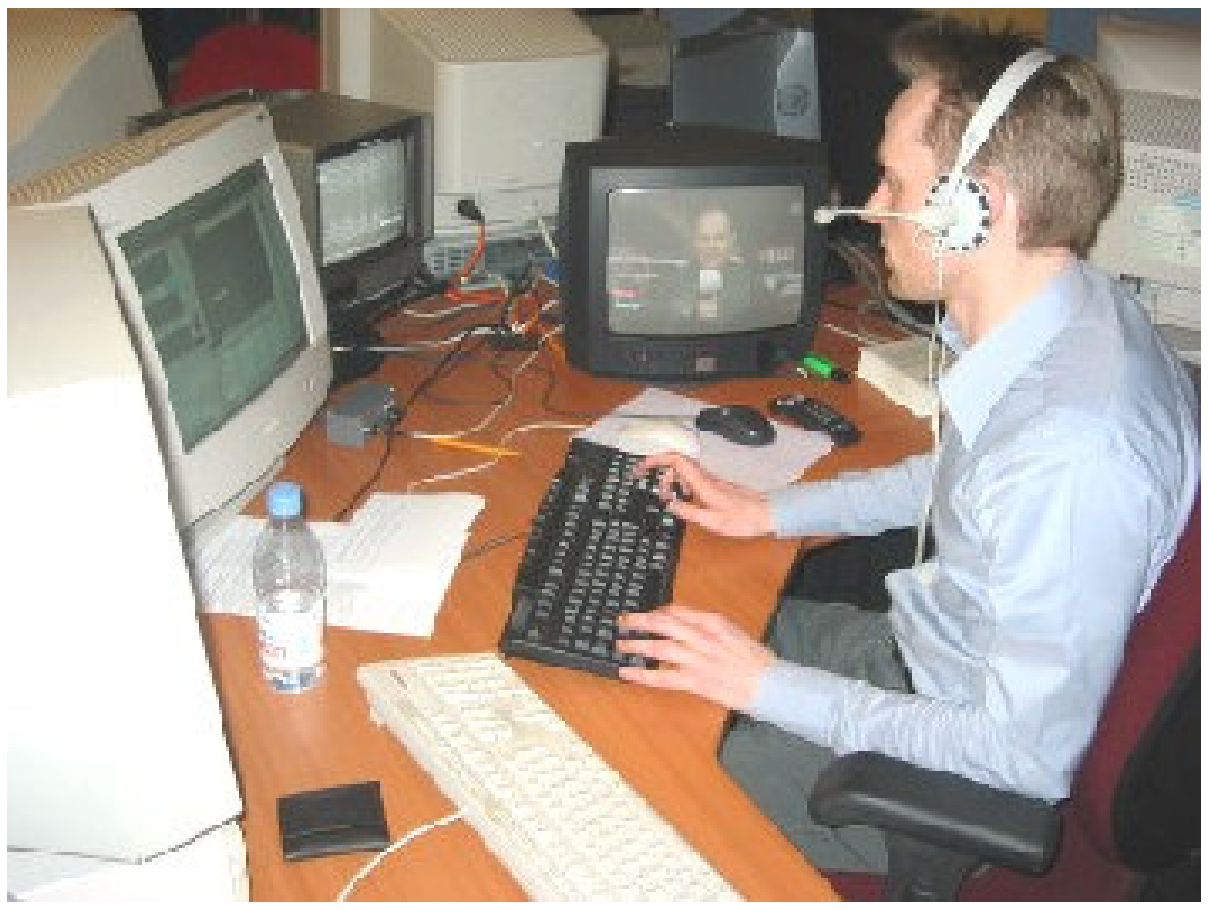

*Figure 64. Subtitling at VRT.*

## 6.4 Possible Uses in Public Services

Of the systems described, several are of definite interest for implementation within a public service:

- Sign language content management systems could be used to give explanations before the interaction, for example about the course of the procedure, the documents required, how to fill in forms, etc. These sign language explanations could be accessible at local kiosks or directly online on the Internet.

- A system similar to TESSA could help complete simple procedures that are well known to the deaf person.
- Subtitling could be used with videoconferencing for late-deafened people using an oral interpreter. The interpreter would clearly reformulate the employee's words, so that the deaf person would have both lip and text information, possibly supplemented by signed French.
- The lip animation system could also be used for late-deafened people directly at the counter. It could, for example, be integrated into the TESSA virtual character, which would express itself either in sign language or in signed and articulated French.

We also noted an interesting application with one of the partners we met: the CFLS, which has a database of about 9,000 unified signs. It is not exploited to the extent that current possibilities allow: it mainly serves as an archive of the signs published in the paper lexicons. This content could be put to use in various ways:

- it could be published on CD-ROM or directly on the Internet, coupled with an avatar animation system. This could be based on linguistic representations.

- the unification process could be carried out collaboratively on the Internet: the various proposed signs could be put online by several means (video, linguistic encoding or through a suitable interface) and discussed by the unification committee before the weekly meetings. This would speed up the process. It could also be given more visibility by allowing other deaf people to take part in the process (suggestions, opinions, etc.).

- the organisation of the linguistic knowledge base could be enriched by experiments carried out in research centres. This can be envisaged for aspects specific to sign language, such as [Bon98] and [Los00], but also in the more general context of the "semantic web". In the long term, this could lead to search engines capable of indexing content in sign language.

# 7 Towards Greater Mobility

In this chapter we examine the new possibilities opened up by the development of high-speed mobile communications. We will first examine their main technological characteristics, then mobile terminals and finally the services useful to deaf people. This part is largely based on [Tis99], [Hol01] and [Sus00].

## *7.1 Infrastructure for Mobile Communications*

In recent years, mobile telephony has become part of the daily lives of most Belgians, including deaf people, for whom SMS has become an indispensable everyday communication tool. It gives them a degree of autonomy and an ability to communicate with other people, deaf or hearing, that they did not have before. This system offers very low bit rates (14 Kbps), which limit it to transmitting text messages or very restricted Internet access (WAP).

In the world of computer networks, meanwhile, high-speed access has become widespread and reaches a large number of homes and workplaces. This is also true for deaf people, though less widely than the craze for SMS. Some people are trying out applications such as videoconferencing over the Internet.

The future that is emerging will see these two worlds converge: mobile telephony networks and computer networks:

- the lack of high speed in the former will be remedied by the planned development of the **UMTS** standard ($3^{rd}$ generation GSM), promising bit rates of around 2Mbps. In the meantime, an intermediate standard called **GPRS** (2.5 generation GSM) offers theoretical bit rates of 140 Kbps (35 Kbps in practice, comparable to a traditional modem) and allows somewhat more user-friendly Internet access: "iMode"
- the lack of wireless connection in the latter is already being solved by the deployment of **WiFi**, a wireless local area network standard. It runs at an appreciable speed, around 11 Mbps.

We will briefly describe these technologies in order to compare them.

### 7.1.1 UMTS

UMTS is organised into two distinct domains, the "circuit" domain and the "packet" domain, intended for different applications:

- the **circuit domain** will handle real-time services corresponding to telephone conversations, video telephony, video games, streaming (sending a continuous flow of information that is processed instantly, with the possibility of displaying the data before the whole file has been downloaded) and multimedia applications. These applications require short transfer times and quality of service: only the circuit domain will be able to handle these flows when UMTS is introduced. The bit rate supported by this mode will be 384 Kbps (compatible with good-quality videoconferencing)

- The **packet domain** will handle non-real-time services corresponding to web browsing, network games and e-mail. These applications are less sensitive to transfer time, and the data can travel in packet mode. The supported bit rate may reach 2 Mbps. This domain will of course be connected to the Internet and to other public or private data networks.

A fundamental feature of UMTS is the ability to negotiate the properties of the radio bearer according to the **quality of service** required for the type of data to be transmitted. A negotiation protocol initiated by the application or by the network makes it possible to request the required level, check its availability under the user's subscription, consult the available resources and make an offer. Four traffic classes corresponding to different qualities of service are identified in UMTS:

- **Conversational**: the best-known application in this class is voice transmission over a switched circuit; with the Internet and multimedia, new applications, notably video telephony, will be part of this class. It is the only one of the 4 quality of service classes whose properties are directly imposed by human perception.

- **Streaming**: this quality of service class covers multimedia applications in which the client, receiving data as a steady, continuous flow, converts it directly into sound or images; such applications are asymmetric and tolerate a longer transmission delay than conversational ones.

- **Interactive**: this class covers applications in which an end user (human or software) requests data online from a remote site; response time and transparent packet transfer are the key properties of this class.

- **Background**: this class covers applications such as electronic mail (e-mail), SMS or database downloads. This quality of service class is characterised by the fact that the destination does not set a deadline for receiving the data.

The "universal" UMTS deployment model does not aim to guarantee high speed everywhere, but according to urban density: in the countryside, base stations will cover longer distances at lower bit rates, while in towns they will be closer together and offer high speed. One problem is coverage inside buildings.

Other features of the UMTS design are support for many simultaneous connections, handover from cell to cell, support for identification and billing systems, monitoring of quality of service, and security.

UMTS will not be deployed for several years, because the system is complex, the infrastructure large and the investments colossal. There was much speculation when UMTS licences were sold to operators; since then, the global economic situation has not been conducive to committing the necessary expenditure and has instead favoured a simpler technology that already works in computer networks: WiFi.

### 7.1.2 WiFi

WiFi technology is being deployed as high-speed Internet access points in busy places (called "Hot Spots") such as hotels, airports, companies, etc. Unlike UMTS, which aims at global mobility, WiFi settles for the concept of nomadism: the person must be in a "Hot Spot" to access the network.

Technically, this technology relies mainly on the 802.11a, 802.11b and 802.11g revisions of the IEEE 802.11 standard:

- the 802.11a standard defines links in the 5 GHz band for bit rates of up to 54 Mbps.
- the 802.11b standard defines links in the 2.4 GHz band for bit rates of up to 11 Mbps and is currently the most widespread. The 802.11a and 802.11b standards are not compatible with each other.
- the 802.11g standard defines links in the 2.4 GHz band for bit rates of up to 54 Mbps. This standard, which has not yet been ratified, is compatible with 802.11b.

Deployment works well in buildings. The power levels involved are fairly low, around 10MW, so ranges are fairly short: around 50 m indoors and 100 m outdoors. Several access points may be needed to cover a given place.

As for quality of service, WiFi of course suffers from the limitations of the Internet (of which it forms the physical and data link layers). In addition, it was not designed for a large number of simultaneous connections.

In practice, for a computer to use WiFi, all that is needed is a wireless network card costing less than €100. Setting up a WiFi network requires a wireless access point, priced from €100. Note also that installing a WiFi network in one's private home is completely unrestricted.

Security mechanisms providing access control exist but are not always well implemented, which can result in possible intrusions into the network and the connected computers.

### 7.1.3 Comparison

Table 8 compares the main characteristics of each solution at present. However, these standards should not be seen as competitors. Figure 65 shows that each solution can find a niche depending on the coverage offered: "WLAN" for WiFi and "WWAN" for UMTS.

The WiFi approach is more gradual and more driven by real applications but faces major problems, particularly regarding quality of service. This is very important in the conversational applications that concern us in this study. Solutions can be envisaged (through "prioritised" flows, for example) but require the current standards to evolve. In the long term, when these problems are solved, the core network is likely to switch to "all IP".

| Technologie | WiFi – 802.11 | UMTS – 3G |
|---|---|---|
| **Débit théorique** | **11** Mbit/s (802.11b); 54 Mbit/s (802.11a) | **1** mbit/s |
| **Portée des cellules** | 50 mètres | 300 mètres (et plus) |
| **Qualité de service** | **Mauvaise** : le WiFi a été conçu pour un petit nombre de connections simultanées. Au-delà de quelques dizaines d'utilisateurs, les transmissions se dégradent singulièrement. | **Bonne** à priori : l'UMTS a été conçu d'emblée pour prendre en charge un grand nombre de connexions. |
| **Prix d'une borne** | A partir de **1500** € | A partir de **100 000** € |
| **Etat de la législation (France)** | **Défavorable** pour le moment : Autorisation de test valable pour les lieux privés. Un prestataire n'a pas le droit d'installer une borne WiFi dans la rue, sauf s'il obtient une dérogation. | **Favorable** |
| **Scalabilité** | Moyenne | Très bonne |
| **Facturation** | Solutions propriétaires | Prévue |
| **Suivi de cellule** | Pas prévu | Protocole de changement d'émetteur |
| **Sécurité** | Mal maîtrisée | Spécifiée |

*Table 8. WiFi vs UMTS comparison (source JDNet)*

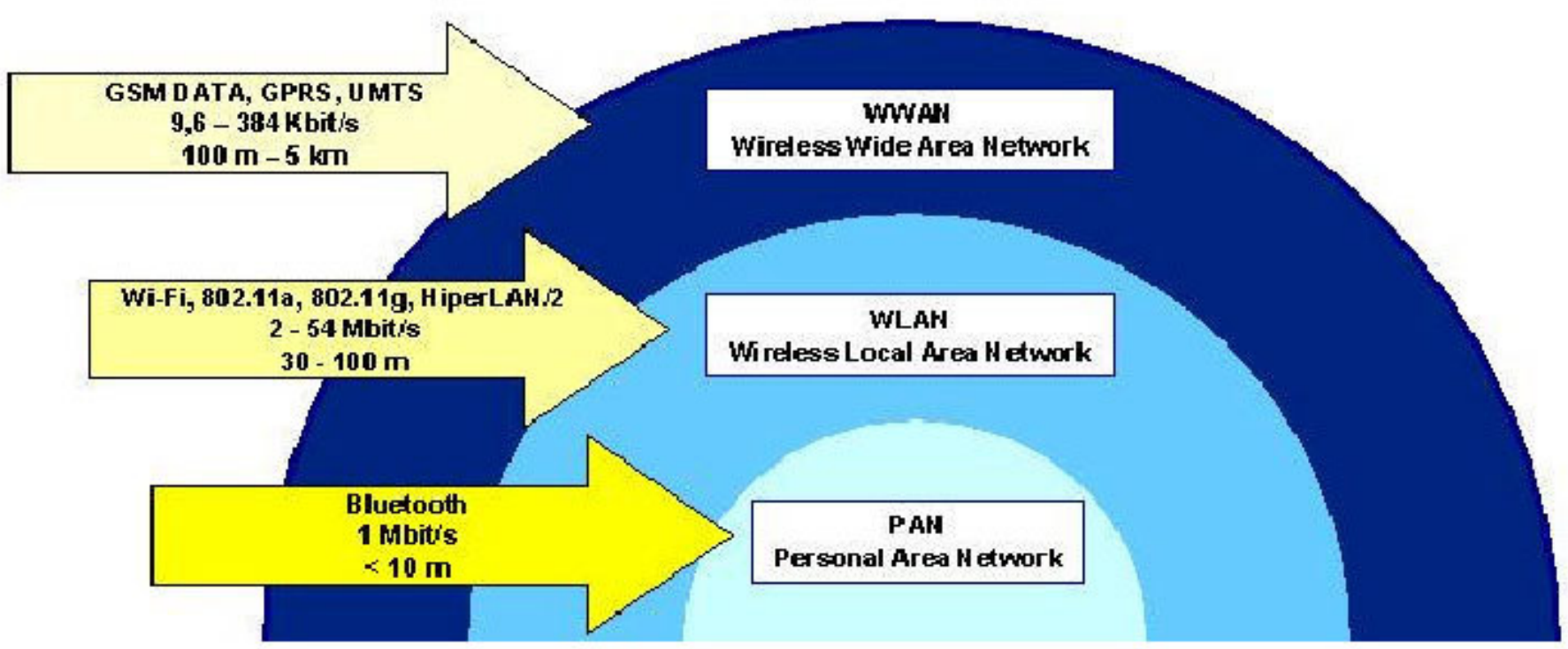

*Figure 65. Complementarity of wireless solutions (source [Ces02])*

## 7.2 Portable Terminals

Mobile use will require suitable mobile terminals. Here too the trend is towards convergence between personal computer terminals and GSM terminals: personal digital assistants (PDAs) and high-end GSM phones are already looking more and more alike (see Figure 66).

While GSM phones are currently limited to GPRS (slow but accessible), some PDAs have built-in or optional wireless access. GSM phones, however, are more advanced in multimedia capture, notably through small digital cameras for creating MMS messages. Some models, such as the Nokia 3650 (see Figure 66), can also capture a few seconds of video and send it.

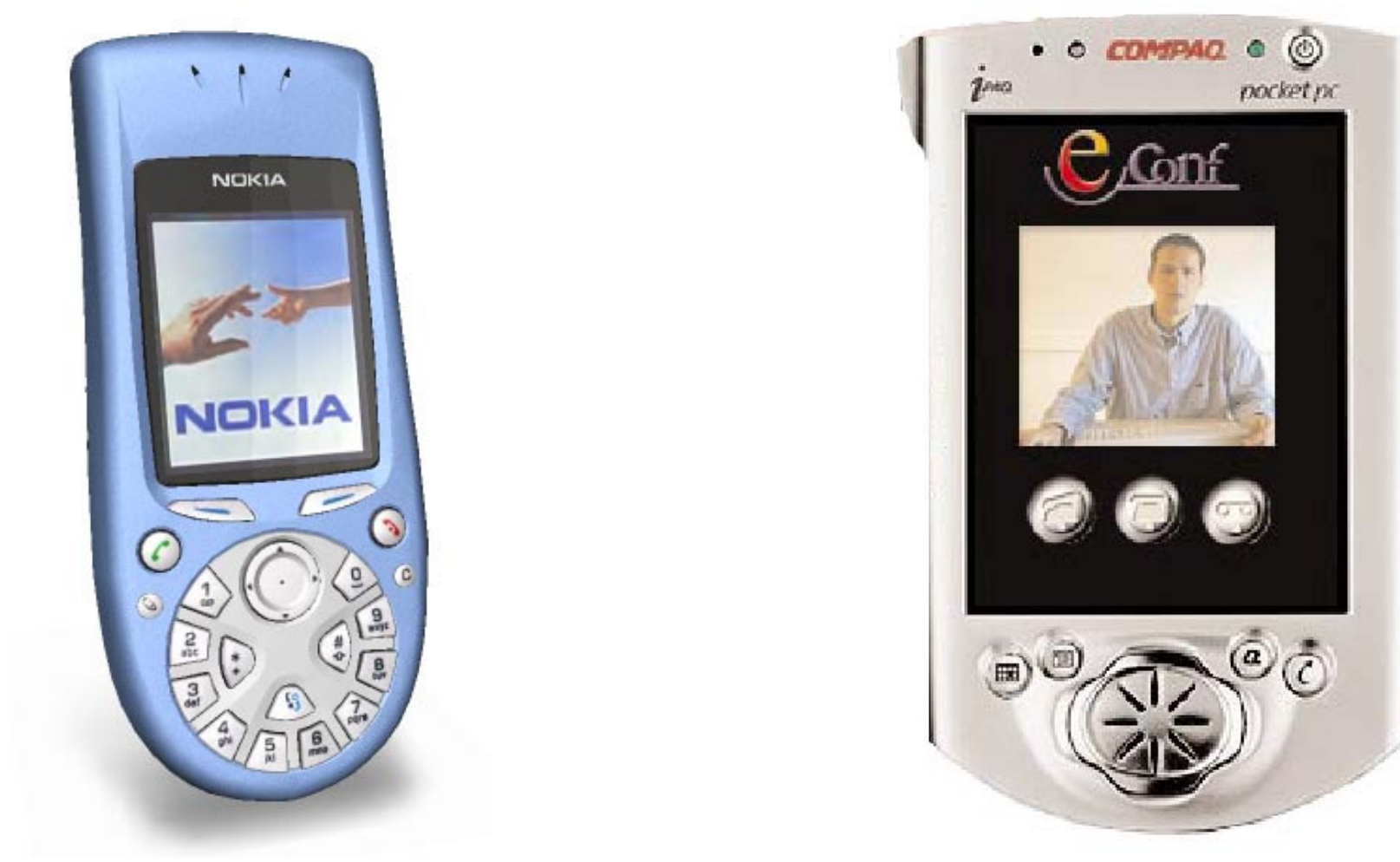

*Figure 66. Nokia 3650 and iPaq.*

In another product range, note also the Tablet PC (see Figure 67), which is a hybrid between a laptop (for its display size, computing power and type of operating system) and a personal assistant (for its interface and input method).

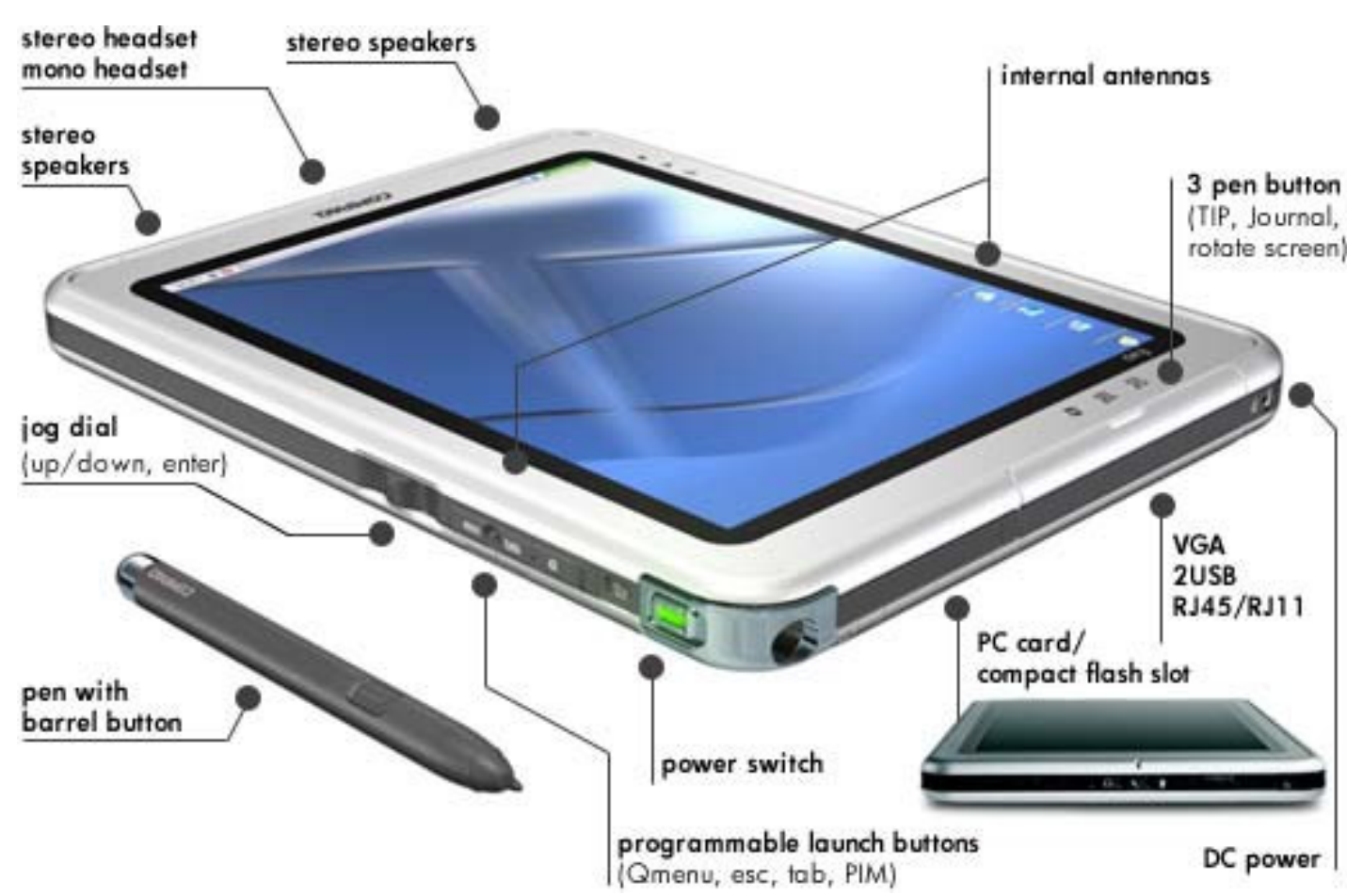

*Figure 67. General structure of a tablet PC*

Regarding the specific needs of deaf people, one can imagine displaying a sign language video, whether for a real-time call or a deferred message:

- screen size is an important factor. The Nokia 3650 above offers 208x176 and the iPaq 320 x 240. The most suitable standard format is

QCIF format. Figure 66 shows such an image displayed on an iPaq with the eConf videoconferencing software

- cameras are not yet very common on PDAs. They typically have a photo resolution of 640x480. In video mode, the Nokia 3650 captures in the H.263 videoconferencing format (QCIF resolution) and stores it in a 3GP file. We carried out a short test that confirmed the limits of the QCIF format; the frame rate was also fairly low (estimated at between 10 and 15).

- note also that cameras are not always well placed for filming oneself (example: lens on the back of the display). Moreover, signing fully in principle requires both hands, which calls for a stand or outside help.

## 7.3 Service Offering

High-speed mobility opens up new application prospects for both hearing and deaf people. For the latter, the main hope is to have a means of communication in sign language. This can take several forms:

- **Real-time conversational access**: "the GSM for deaf people", which would use videoconferencing. The main obstacle is quality of service: UMTS must support it according to its specification. WiFi, at present, offers no guarantee in this respect.
- **Exchanging messages in sign language**: here real time is no longer necessary. "Streaming"-type access or simply a download without time constraints followed by viewing the message is enough

- **Access to an information server in sign language**, possibly taking into account the caller's geographical location (example: finding the nearest garage). Here, the appropriate access mode would probably be streaming.

- **Access to a relay service or remote interpretation service in sign language, possibly with an on-call service for emergencies.**

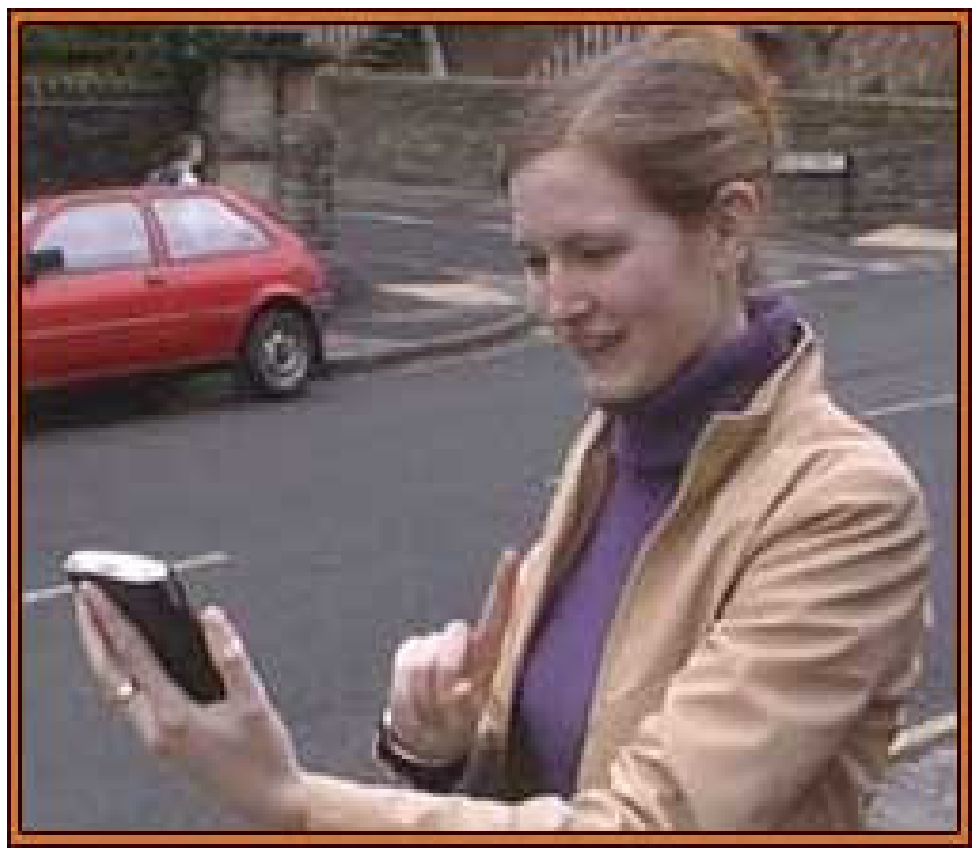

*Figure 68. "Mobile" conversation in sign language.*

The main objective of the European WISDOM project (Wireless Information Service for Deaf People On the Move: http://www.mobilewisdom.org) is to explore these possibilities. Among the particular requirements:

- the terminal must have visual or vibrating alerts for incoming calls

- the input methods considered will be text-based but also in sign language, for example to navigate the video server. Current research has already produced a system that accepts continuous sentences in a restricted vocabulary of 152 signs, with a recognition accuracy of 91.1%.

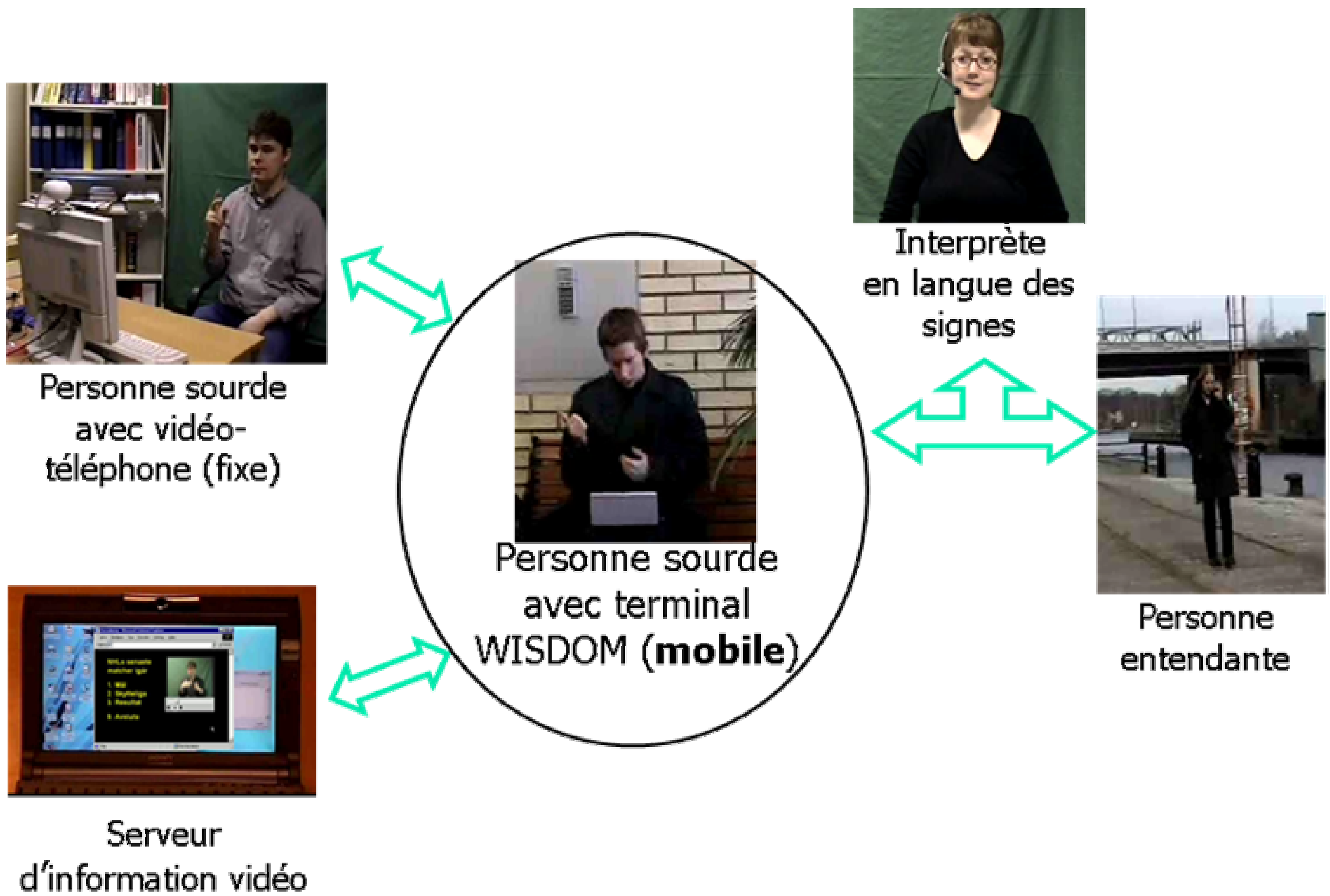

*Figure 69. WISDOM project.*

# 8 Conclusions and Recommendations

At the end of this study, we have highlighted various ways of improving **the services offered to deaf people using sign language within administrations and public services**. These solutions make more or less extensive use of information and communication technologies. They were motivated by a preliminary domain study and an analysis of the goals to be achieved.

We first considered the solution closest to the current one and also the most realistic: implementing a **videoconferencing system for remote interpretation**. Interpretation is provided by a professional interpreter. With a suitable connection, this system can be deployed with current technology. The requirements to be met were described in detail in section 5.4. The system is certainly not suited to all situations, but it helps to improve the accessibility of public services for deaf people and also increases the availability of interpreters by reducing their tedious travel. Finally, it makes it possible to set up an emergency interpretation service.

We then explored the option of **machine translation** and studied the many processing steps involved. We showed that the approach was not realistic as a whole, but that useful systems were feasible for particular classes of applications, mainly combining speech recognition and the animation of synthetic characters. Among these applications, those that can be implemented in a public service are:

- consulting explanatory information in sign language
- translating into signed French the communication of employees known to the system
- translating simple, well-known procedures into sign language All these solutions can usefully rely on synthetic characters to present information in sign language to the deaf person[3].

Toutes
de synt
sourde

In functional terms, these two solutions therefore overlap very little: the first is mainly focused on interactive communication and the second on consulting information available on a medium. They can therefore be implemented in a complementary way: for example, a kiosk or a website could provide information in sign language on how to carry out a procedure in a public service. The procedure itself could then be carried out using the remote interpretation system.

Finally, we discussed the opportunities offered by the development of mobile systems and the promise of additional communication and autonomy they hold for deaf people.

---

At this level, one could therefore still speak of sign language interpretation, but in the narrower sense of the avatar's "acting performance".

Recall that this study focused mainly on deaf people using sign language. However, we have also pointed out where a system lent itself to extensions that could benefit late-deafened or hard-of-hearing people: in particular the use of a remote oral interpreter, subtitling, or automatic lip synthesis. We have also mentioned some uses in contexts other than public services, such as education or private use.

We will end with a list of recommendations concerning the main possible actions and how to implement them. We first make some recommendations related to the general context of the study. These partly overlap with those made in [Rec03].

- **Raising deaf people's awareness of information** and communication technologies. Interacting with a videoconferencing terminal or consulting a multimedia website with avatars requires a minimum familiarity with basic computing: using a web browser, instant messaging, e-mail, specific software such as sign language dictionaries, etc. Initiatives to train deaf people in these techniques should be encouraged, in particular through deaf associations. Financial support for buying equipment and for Internet connection is another element that goes in this direction.

- **Raising hearing people's awareness** of the problems faced by deaf people (and disabled people in general). Hearing people who come into contact with deaf people must be informed of the current and future possibilities for communicating and must be able to use them more easily. Appropriate training for public service staff should be provided. "Mass-audience" actions such as the Surdimobil should be continued. The introduction of the passe-partout index and the European Year of People with Disabilities are also good opportunities to highlight the problems faced by deaf people.

The second set of recommendations is the most important and concerns remote interpretation by videoconference.

- **Deploying high-speed connections** between videoconferencing centres is a prerequisite. To be realistic, it should probably be implemented in a context broader than services for the target minority, for example as part of the modernisation of IT infrastructure or the development of e-government.

- **Recognising the work of interpreters**. The videoconferencing system is not intended to do away with interpreters; on the contrary, it relieves their workload by reducing tiring travel. Complementary measures to improve their training and recognise their status should be encouraged. Their number should also be increased significantly.

- **Carrying out a pilot experiment in remote interpretation**. Remote interpretation increases the autonomy of deaf people and the

availability of interpreters. However, it is not applicable in all situations, and before considering a large and costly deployment, a **small-scale pilot experiment under real conditions** should be carried out. This would make it possible to validate and refine the proposed organisation. Awareness-raising actions are also needed to familiarise the deaf public with the possibilities of this means, for example on World Deaf Day.

- **Cost estimation and deployment.** Several deployment models are possible, in terms of phasing and of the type of partnership with the equipped administrations and public services (full equipment, provision of subsidies, provision of access to the interpretation centre, etc.). These points have not been addressed in this document but can be developed, in particular on the basis of the requirements described in section 5.4.

Finally, the last set of recommendations concerns the dissemination of content in sign language.

- **Dissemination of content in sign language.** Websites aimed at deaf people should be able to offer more sign language material than at present. Initiatives in this direction should be encouraged. The French WebSourd project, which should soon become operational, should be followed closely, and collaborations could be considered.

- **Encouraging the development of an avatar animation system**. Sign synthesis based on an animated virtual character overcomes many limitations of current systems: ease of storage and transmission, viewing options, adaptability. Avatars specific to sign language already exist (European ViSiCAST project); however, they suffer from limitations in expressiveness and content creation. Current approaches make too little use of the structure of sign language, which can be exploited successfully, as various studies show. The diversity of skills required and the research dimension still involved call for a multidisciplinary project bringing together:

  - computer scientists, preferably with experience in this field (such as the non-profit Aptitude, CETIC, I3D)
  - deaf and hearing linguists (such as the CFLS, Marie-Haps, Liège)
  - deaf associations, to highlight needs and validate results.

  Implementation can be envisaged at the Walloon level but could benefit from wider collaboration: since the software is largely reusable for other sign languages, its value could extend well beyond the deaf community of French-speaking Belgium:
  - at the Franco-Belgian level, very positive contacts have taken place with the Flemish federation (Fevlado), the French federation (IVT), the University of Lille, France Telecom (Paris) and WebSourd (Toulouse).
  - at the European level, the second call for proposals of the 6$^{th}$ Framework Programme also offers interesting opportunities to extend the work carried out in the ViSiCAST project. Preliminary contacts with the Bristol centre and the University of East Anglia show this willingness.

## Non-bibliographic references

Les médias suivant (CD-ROM, vidéo), plus adaptés à la langue des signes ont
The following media (CD-ROM, video), better suited to sign language, were also consulted:

**CD-ROM :**

**CD-ROM:** S

- DILS
- Les signes de Mano
- Apprentissage interactif de la LSF
- Sign Smith Studio Demo
- Sign Smith Illustrated Dictionary Demo
- Lankät text teckensprak (en suédois)

**Vidéo :**

**Video:** es Enfants du Silence (1986)

- Le pays des Sourds (1992)
- Dites-moi : Emmanuelle Laborit (1995)
- Danse des mains – La renaissance de la langue des signes en Europe(1997)

# 10 Appendices

## *10.1 Glossary and Acronyms*

| | |
|---|---|
| Tinnitus | Auditory sensation (buzzing, whistling, etc.) that is not caused by external stimulation of the ear. |
| AKA | Assisted kineme alphabet. Lip-reading aid based on easily recognisable gestures that convey both the movements of the sentence and variations in intonation. More complex to learn than LPC but more flexible and allows greater fluency of expression. |
| ADSL | Asymmetric Digital Subscriber Line. High-speed but asymmetric means of connecting to an Internet access provider: 1 to 3 Mbps downstream to the subscriber and around 128 Kbps upstream to the access provider. |
| ASL | American Sign Language |
| Auslan | Australian Sign Language |
| BSL | British Sign Language |
| CS | Cued Speech. See LPC. |
| CODEC | Coder-decoder: component responsible for compressing and decompressing digital video or audio signals in order to reduce the bandwidth needed to transmit them, possibly at the expense of their quality. |
| Decibel (dB, dBA) | Unit expressing the ratio between 2 sound intensities on a logarithmic scale. More precisely, it is expressed mathematically as a logarithmic function of the pressure ratio: 20 log (p2/p1).<br>The reference level (0 dB) corresponds to the faintest sound audible to the human ear. The average pain threshold for the human ear is 130 dB.<br>To take into account the ear's sensitivity, which varies with frequency and attenuates low (< 1 kHz) and high (> 10 kHz) sounds, measuring instruments are fitted with a filter (A filter) whose frequency response simulates that of the human ear. The corresponding unit of measurement is then the dBA. |
| FPS | Frames per second. Video frame rate. Rendering movement requires more than 10 FPS. Excellent quality requires more than 25 FPS. |
| Signed French (SF) | Spoken French translated into gestures borrowed from LSFB (for Belgium). The word order follows that of French. |
| GPRS | Global Packet Radio Service |
| h.261 | ITU video transmission standard designed for low-speed ISDN lines |
| h.263 | Improvement of the ITU video transmission standard, also for low-speed lines. |
| h.264 | New emerging ITU standard |
| h.323 | ITU standard specifying the components, protocols and procedures that provide real-time audio, video and data communication services over packet networks, notably IP-based networks. The H.323 standard in fact groups together a large number of protocols that define the different stages of an H.323 communication. |
| IEEE 802.11 | Official name of the wireless transmission standards, known to the general public as WiFi. |
| Cochlear implant | Device made up of 2 parts (an external part and an internal implanted part) which converts sounds into electrical signals that are sent directly to the auditory nerve. |
| ISDN | "Integrated Services Digital Network". Digital network offering 2 independent full-duplex 64kbps digital channels, which can be combined into a 128kbps link, as well as a 16kbps signalling link. |
| ISL | International Sign Language (aka Gestuno) |
| ISP | Internet Service Provider. Service provider that connects an individual or a company to the Internet. |
| ITU | International Telecommunication Union. Standardisation body in the field of telecommunications. |
| KAOS | Goal-oriented requirements acquisition and specification method; see references [Dar93] and [Lam00]. |
| Kbps | Kilobit per second. Transmission rate of one thousand bits per second. |
| Kineme | Mouth and facial movement produced when voicing a phoneme. Several phonemes may correspond to the same kineme, which makes lip reading ambiguous. |
| Lip reading (LR) | Technique that consists of following only the movement of the lips, mouth and facial expression to recognise a word. |
| LPC | Langage Parlé Complété (Cued Speech). Technique that consists of using articulated speech accompanied by meaningful hand gestures. |
| Sign language | Language made up of a set of gestures, facial expressions and body movements, with its own grammar and syntax (varying between countries and regions) |
| LSBF | French Belgian Sign Language, synonym of LSFB. |
| LSF | French Sign Language (France) |
| LSFB | French Belgian Sign Language |
| Mbps | Megabit per second. Transmission rate of one million bits per second. |
| MPEG | Digital video compression standard used notably for DVDs. |
| Oralism | Educational philosophy and methods aimed at teaching deaf children to use spoken language. |
| Late-deafened person | Person who could hear previously and has completely lost the sense of hearing following an illness or accident. |
| Hard-of-hearing person | Person whose hearing is more or less reduced (quantified between 35 and 70dB) and who, with or without technical aids, uses residual hearing to understand speech. This person mostly uses French to communicate and has communication habits closer to those of hearing people. |
| Deaf person | Person more or less completely deprived of hearing, usually since birth or early childhood. Very often, this person regularly uses sign language to communicate and shares, to varying degrees, the culture, values, communication habits, rules of behaviour, goals and traditions of the members of the deaf community. |
| Linguistic register | Variants within the same linguistic system corresponding to different uses of the language (formal, familiar, etc.) |
| Residual hearing | Remaining hearing ability in a person who is not totally deaf. |
| RNIS | Réseau Numérique à Intégration de Services. See ISDN. |
| SDSL | Symmetric Digital Subscriber Line. |
| Signer | Person communicating by means of sign language |
| UML | Unified Modelling Language: graphical modelling language for describing the static and dynamic aspects of organisations and computer systems. |
| UMTS | Universal Mobile Telecommunication System |
| VCD | Video CD. System for distributing digital video on CD-ROM using the MPEG-1 format. |
| WiFi | Wireless Fidelity: wireless LAN. |

## *10.2 Results of the Survey Conducted by the FFSB*

**A. Which of the administrative or public services listed below should be equipped with this system? (3 choices maximum)**

| | |
|---|---|
| Hospitals | 24 |
| Gendarmerie (Police) | 24 |
| Municipal offices | 16 |
| Law courts | 12 |
| Bank | 11 |
| ORBEM / FOREM | 6 |
| AWIPH | 4 |
| Youth contact services | 3 |
| Health insurance fund | 3 |
| Brussels service for the integration of disabled people | 3 |
| Family planning | 2 |

**B. Absence of the interpreter:**

| | |
|---|---|
| Communicating with an interpreter via a screen does not bother me | 28 |
| Having an interpreter at my side is essential | 15 |
| I have no confidence in an online sign interpretation system | 4 |

**C. Preconceptions about the system**

| | |
|---|---|
| Yes | 20 |
| Sometimes | 10 |
| No | 5 |

**D. Details of any concerns**

| | |
|---|---|
| Poor image quality | 19 |
| I am not afraid... | 16 |
| A breakdown | 5 |

## *10.3 Contents of the Companion CD-ROM*

The companion CD-ROM contains the material cited during this study: videos, study reports, research articles and the software used (demo version). It also includes an electronic version of this report.

The CD-ROM can be viewed with a web browser simply by opening the index.html file at the root of the CD-ROM. On Windows, it should start automatically.

The files are mainly in the following formats:

- *rm format* (RealVideo) for videos; the viewer is available free of charge at http://www.real.com
- *pdf format* for articles and reports; the viewer is available free of charge at http://www.adobe.com/products/acrobat/readstep2.html